\documentclass[reprint,amsmath,amssymb,aps,prb,floatfix]{revtex4-1}

\usepackage{graphicx}
\usepackage{dcolumn}
\usepackage{bm}
\usepackage{multirow}
\usepackage[table,xcdraw]{xcolor}
\usepackage{enumerate}
\usepackage{physics}
\usepackage{bigints}
\usepackage{hyperref}

\begin{document}

\preprint{APS/123-QED}

\title[ACBN0 Bulk]{Improved Description of Perovskite Oxide Crystal Structure and Electronic Properties using Self-Consistent Hubbard $U$ Corrections from ACBN0}

\author{Kevin J. May}
 \email{kmay@mit.edu}
 \altaffiliation[Present address: ]{Department of Materials Science and Engineering, Massachusetts Institute of Technology, Cambridge, MA 02139 USA.}
\author{Alexie M. Kolpak}%
 \email{kolpak@mit.edu}
\affiliation{%
 Department of Mechanical Engineering\\
 Massachusetts Institute of Technology, Cambridge, MA 02139 USA
}%

\date{\today}

\begin{abstract}
The wide variety of complex physical behavior exhibited in transition metal oxides, particularly the perovskites A$B$O$_3$, makes them a material family of interest in many research areas, but the drastically different electronic structures possible in these oxides raises challenges in describing them accurately within density functional theory (DFT) and related methods. Here we evaluate the ability of the ACBN0, a recently developed first-principles approach to computing the Hubbard $U$ correction self-consistently, to describe the structural and electronic properties of the first-row transition metal perovskites with $\left(B=\textrm{V}-\textrm{Ni} \right)$. ACBN0 performs competitively with hybrid functional approaches such as the Heyd-Scuseria-Ernzerhof (HSE) functional even when they are optimized empirically, at a fraction of the computational cost. ACBN0 also describes both the structure and band gap of the oxides more accurately than a conventional Hubbard $U$ correction performed by using $U$ values taken from the literature.
\end{abstract}

\maketitle

\section{\label{sec:intro}Introduction}
Density functional theory (DFT) is one of the most often-used computational approaches for modeling the electronic structure of complex molecules and solids. However, the approximate exchange-correlation (XC) term in the total energy functional, informed by early work on the homogeneous electron gas\cite{dirac1930,ma1968,perdew1992}, leads to significant inaccuracies in DFT. Notable examples are the underestimation of fundamental gaps in the electronic structure, or the prediction of metallic ground states in transition metal oxides in cases where the true ground state is insulating. Transition metal oxides are materials of interest in a wide variety of applications, including renewable energy and catalysis. In certain cases the trends captured by DFT are sufficient, but when quantitative predictions (e.g. location of a catalyst on a ``volcano'' plot) or band gaps are needed, ``beyond-DFT'' methods are required. This is especially important in perovskite oxides with formula unit $\text{ABO}_3$ (where $A$ is usually a lanthanide or alkaline earth metal and $B$ is usually a transition metal), which include band insulators\cite{okamoto2004,ozgur2005,pentcheva2007}, Mott-Hubbard insulators\cite{imada1998}, charge transfer insulators\cite{zaanen1985}, and correlated metals\cite{gou2011}. Perovskites and other related structures have found interest in a wide variety of applications, ranging from fundamental physics phenomena (metal-to-insulator transitions\cite{imada1998}, topological insulators\cite{hasan2010}, magnetism\cite{stoneham2010}, superconductivity\cite{anderson1987,dow2009}, ferroelectricity\cite{cohen1992}) to catalysts\cite{suntivich2011,hwang2017}, battery materials\cite{kendrick2013}, and oxide electronics\cite{ramesh2008,bibes2011}. Being materials where electron correlations play an important role in determining the properties, they are challenging to describe universally using current theoretical approaches.

Approximate XC functionals such as the local density approximation (LDA) or the various flavors of the generalized gradient approximation (GGA) do not cancel out the self-interaction energy in the Coulomb (Hartree) functional, leading to excessive delocalization\cite{perdew1981}. This is an important reason for qualitatively incorrect predictions in systems where charge is strongly localized, such as in many transition metal oxides. In addition, the total DFT energy for a given system as a function of electron occupation is smooth for approximate XC functionals, whereas for the exact Kohn-Sham (KS) potential the energy is piece-wise linear, with derivative discontinuities at integer occupation numbers\cite{perdew1982}. This is one of the reasons for the significant underestimation of fundamental gaps by approximate XC functionals\cite{sham1983,sham1985,perdew1986,cohen2008}. It is therefore unsurprising that several beyond-DFT methods introduce derivative discontinuities in the total energy vs. electron occupation. Hybrid functionals, where a fraction of the exact Hartree-Fock (HF) exchange acting on the KS orbitals is used, intuitively reduce delocalization via the cancellation of self-interaction in the Hartree energy, but also introduce discontinuity into the XC potential\cite{seidl1996}. While the most commonly used mixing fraction of 25\ exact exchange (75\ approximate DFT exchange) was justified for atomization energies of molecules\cite{perdew1996}, in practice the mixing fraction is often used as an empirical parameter in order to optimize the description of a desired material property, as has been done with perovskite oxides\cite{he2012}. A self-consistent hybrid functional (sc-hybrid) based on the PBE0 functional has also been reported\cite{skone2014}, which avoids empiricism in the mixing fraction by setting it equal to the inverse of the static dielectric constant of the material. This relationship can be justified from a comparison of the hybrid functional exchange-correlation functional to the expression for self-energy in the Coulomb hole plus screened exchange (COHSEX) approximation\cite{hedin1965}, a static version of the $GW$ approximation. In the original paper\cite{skone2014}, the dielectric constant is calculated using the coupled perturbed KS equations (CPKS) within first-order perturbation theory\cite{ferrero2008,johnson1993}, the mixing fraction is updated, and this process is repeated until self-consistency is achieved. In plane-wave calculations, the static dielectric constant may computed the Berry phase technique\cite{king-smith1993,resta1994}, as has been demonstrated in recent work\cite{fritsch2017}. This approach has improved upon fixed and empirical mixing fraction hybrid functionals for a variety of semiconductors and insulators\cite{skone2014,fritsch2017,he2017}, though the method is computationally expensive owing to the multiple iterations of hybrid functional calculations needed for each material.

DFT+$U$, inspired by the Hubbard model\cite{hubbard1963}, is another approach to improving the description of correlated materials. In DFT+$U$, a corrective term is added to the total DFT energy functional that energetically favors orbitals in the chosen Hubbard manifold (typically $d$ or $f$ electrons but not exclusively) being either completely empty or full\cite{himmetoglu2014} via screened HF-like Coulomb ($U$) and exchange ($J$) interactions that act only on this set of localized orbitals, usually between orbitals on a single site but potentially between neighboring sites as well\cite{campo2010}, and removing a ``double-counting'' term from the DFT energy functional. Unfortunately, there is no unique choice for the set of localized orbitals onto which to project the KS orbitals, nor for the double-counting term or the method of calculating the values of $U$ and $J$ themselves. Atomic-like orbitals (e.g. from the pseudopotentials) are often used as a basis\cite{anisimov1997,dudarev1998,cococcioni2005,himmetoglu2011}, as are Wannier functions\cite{schnell2002,oregan2010,franchini2012,ma2016}. The correction is often applied in a simplified scheme\cite{dudarev1998} using a single interaction parameter which is often assumed to take both $U$ and $J$ interactions into account, an ``effective'' $U_{\textrm{eff}}=U-J$. In this work, $U$ refers to such an effective value unless $J$ is explictly mentioned. The value of $U$, similarly to the fraction of exact exchange in hybrid functionals, is often used as an empirical parameter that is varied to produce the desired results. First-principles approaches to calculating $U$ do exist, however. The linear response (LR) method defines $U$ in such a way that the curvature of the total energy as a function of electron occupation is canceled out for non-integer occupations, giving rise to a derivative discontinuity in the energy\cite{cococcioni2005}. A frequency-dependent, screened $U$ can also be calculated via the constrained random phase approximation (cRPA)\cite{aryasetiawan2004,karlsson2010,sasioglu2011,vaugier2012}. Self-consistent values of $U$, meaning that $U$ is calculated iteratively until a convergence criteria is met, have been shown to improve on empirical choices of $U$ for structural properties and defect formation energies\cite{hsu2009,ricca2019}. The downside of some of these approaches (which will be briefly described further in the next section) is that they can be computationally demanding for large cells when there are many unique sites that warrant treatment with DFT+$U$.

Recently, a new approach to calculate the value of the Hubbard $U$ has been reported\cite{agapito2015}, inspired by previous work computing $U$ via unrestricted HF orbitals\cite{mosey2007,mosey2008}. The ACBN0 method defines $U$ based on the bare Coulomb and exchange interactions and a renormalized occupation matrix, where KS orbital occupations are reduced based on the Mulliken population of each KS orbital projected on the Hubbard manifold. The main advantages are flexibility with respect to unique Hubbard sites and extremely low computational cost, negligible compared to the main DFT calculation, making ACBN0 particularly well-suited for high-throughput applications\cite{supka2017}. Corrections are applied using $U$ values for both metal $d$ ($U_{dd}$) and oxygen $p$ ($U_{pp}$) sites. In principle, these values can be calculated for any Hubbard manifold on any given atomic site. Due to the renormalization of the occupation matrix, if few occupied KS states have strong character of the chosen Hubbard manifold, the magnitude of the correction will be drastically reduced. This has the effect of reducing computed values of $U$ for more covalent materials. This method was originally tested on several benchmark materials ($\text{TiO}_2$, MnO, NiO and wurtzite ZnO) and later on wide-gap semiconductors\cite{gopal2015} and several other binary oxides\cite{gopal2017}, showing improved agreement with more computationally-expensive beyond-DFT methods such as hybrid functionals and the $GW$ approximation. It is worth mentioning that also recently, a new method of computing $U$ that is equivalent to the LR approach has been demonstrated using density-functional perturbation theory, which, similarly to ACBN0, allows for computing self-consistent values of $U$ on individual atomic sites without the use of large supercells\cite{timrov2018,ricca2019,cococcioni2019}.

This work provides a further test of ACBN0 on the theoretically-demanding transition metal perovskites A$B$O$_3$, where $B$=Ti--Ni. Moreover, since there are few studies which look at DFT+$U$ on all of these materials (especially with first-principles calculations of $U$), we offer comparison with fixed values of $U$ chosen from values in the literature that were calculated using various methods such as cluster configuration interaction (cluster-CI) calculations\cite{david1999} fit to experimental photoemission and X-ray absorption spectra, or the LR method mentioned previously. This is referred to in the text as ``PBE+$U_{dd}^{\text{Lit.}}$''. We chose these cluster-CI values since they often lie in the median range of $U$ values reported, and are all determined in the same way from experimental results. It is important to mention that in general, values of $U$ are not transferable; factors such as the method of calculating $U$, the XC functional, the pseudopotentials (also the XC functional used in generating them), the Hubbard manifold to which $U$ is being applied and the implementation of the DFT+$U$ method all affect the value of a first-principles $U$ and the effect it has when applied. We apply these values of $U$ from the literature in an intentionally ``na\"{i}ve'' way and ignore the non-transferable nature of $U$ to illustrate how using either empirically-chosen or first-principles values of $U$ in this way can affect the results of a DFT calculation. This is not intended as a comment on the ``correctness'' of those values of $U$ or on the results obtained using these corrections in their respective studies. We examine the prediction of magnetic ground state, lattice geometry, and electronic structure for the 1st-row transition metal perovskites, and compare with higher theory and experimental data when possible, providing a necessary test of ACBN0 as well as a guide for treating these materials with computationally-inexpensive first-principles methods.

\section{Theoretical Background}
\subsection{\label{ssec:hubU}The Hubbard Correction in DFT: DFT+$U$}
Here we briefly cover an introduction to several formulations of DFT+$U$, omitting the derivations that are covered in the original works. We refer to a thorough review of DFT+$U$ for more details and discussion on the development of the method.\cite{himmetoglu2014} The basic idea of DFT+$U$ is that a correction to the DFT energy is applied by separately treating electronic interactions between \emph{localized} electrons, usually considered on a single atomic site. It frequently may be the only feasible choice for large systems, owing to its negligible cost compared to the base DFT calculation. Various implementations of the correction are possible\cite{anisimov1993,liechtenstein1995,dudarev1998,pickett1998,cococcioni2005,himmetoglu2011,oregan2012}, but the basic form is:\cite{himmetoglu2014}
\begin{align}\label{eq:DFTUgen}
E_{\text{DFT}+U}[n(\textbf{r})] = & E_{\text{DFT}}[n(\textbf{r})] \nonumber \\
& + E_{\text{Hub}}[n_{mm^{\!\prime}}^{I\sigma}] - E_{\text{dc}}[n^{I\sigma}]
\end{align}
where $E_{\text{DFT}}$ is the energy of a DFT calculation (LDA or GGA), $E_{\text{Hub}}$ is the Hubbard correction term, $E_{\text{dc}}$ is a double-counting correction (to remove the energy calculated in $E_{\text{Hub}}$ from $E_{\text{DFT}}$) and $n_{mm^{\!\prime}}^{I\sigma}$ is the occupation number of a localized orbital in the set $\varphi_{m}$ (the Hubbard basis) on atomic site $I$, with spin index $\sigma$. This last term is often computed by projecting the KS orbitals onto a set of localized orbitals, such as pseudo-atomic orbitals or Wannier functions, expressed as:
\begin{equation}
n_{mm^{\!\prime}}^{I\sigma} = \sum_{k,i} f_{ki}^{\sigma} \braket{\phi_{ki}^{\sigma}}{\varphi_{m^{\!\prime}}^{I}} \braket{ \varphi_{m}^{I}}{\phi_{ki}^{\sigma}}
\end{equation} 
where $f_{ki}^{\sigma}$ is the occupation of $\phi_{ki}^{\sigma}$, the $i^{th}$ KS orbital labeled by k-point and spin index. 

While the early work on DFT+$U$ was done using a form for Eq. (\ref{eq:DFTUgen}) that is more reminiscent of the Hubbard model, it was not invariant upon rotation of the localized orbitals.\cite{anisimov1993} A rotationally-invariant formulation was developed by Liechtenstein \emph{et al.} that is similar to a HF (HF) calculation:\cite{liechtenstein1995}
\begin{align}
E_{\text{Hub}}[n_{mm^{\!\prime}}^{I\sigma}] = & \frac{1}{2}\sum_{\{m\},\sigma\!,\,j} \left\{ \mel{m,m^{\!\prime\!\prime}}{V_{ee}^{s}}{m^{\!\prime},m^{\!\prime\!\prime\!\prime}}n_{mm^{\!\prime}}^{I\sigma}n_{m^{\!\prime\!\prime}m^{\!\prime\!\prime\!\prime}}^{I-\sigma}\right. \nonumber \\
& + \left( \mel{m,m^{\!\prime\!\prime}}{V_{ee}^{s}}{m^{\!\prime},m^{\!\prime\!\prime\!\prime}}\right. \nonumber \\
& - \left. \left. \mel{m,m^{\!\prime\!\prime}}{V_{ee}^{s}}{m^{\!\prime\!\prime\!\prime},m^{\!\prime}}\right)\left( n_{mm^{\!\prime}}^{I\sigma}n_{m^{\!\prime\!\prime}m^{\!\prime\!\prime\!\prime}}^{I\sigma}\right) \right\} \label{eq:fullU}
\end{align}
\begin{align}
E_{\text{dc}}[n^{I\sigma}]= & \sum_{I} \bigg\{ \frac{U^{I}}{2}n^{I}(n^{I}-1) \nonumber \\
 & - \frac{J^{I}}{2}\left[ n^{I\uparrow}(n^{I\uparrow}-1)+n^{I\downarrow}(n^{I\downarrow}-1)\right]\bigg\} \label{eq:fullJ}
\end{align}
where $V_{ee}^{s}$ are the screened Coulomb interactions between electrons. The rotation invariance comes from the quadruplet integrals and the dependence on the trace of the occupation matrices in the Hubbard and double-counting terms, respectively. The Coulomb integrals $\mel{m,m^{\!\prime\!\prime}}{V_{ee}^{s}}{m^{\!\prime},m^{\!\prime\!\prime\!\prime}}$ are expressed as:\cite{himmetoglu2014}
\begin{equation}\label{eq:HubInt}
\bigintssss\!\!\!\!\bigintssss \varphi_{Im}^{\dagger}(\textbf{r})\varphi_{Im^{\!\prime}}(\textbf{r}) \frac{e^2}{\absolutevalue{\textbf{r}-\textbf{r}^{\prime}}}\varphi_{Im^{\!\prime\!\prime}}^{\dagger}(\textbf{r}^{\prime})\varphi_{Im^{\!\prime\!\prime\!\prime}}(\textbf{r}^{\prime}) \, d\textbf{r}^{\prime} d\textbf{r} 
\end{equation}
In the case of spherically symmetric Hubbard basis orbitals (i.e. atomic orbitals), these integrals can be separated into radial and angular parts. The parameters $U$ and $J$ can be expressed as atomic averages of the Coulomb integrals over the states with the same quantum number $l$:\cite{himmetoglu2014}
\begin{gather}
U = \frac{1}{(2l+1)^2} \sum_{m,m^{\prime}}\expval{V_{ee}^{s}}{m,m^{\prime}}=F^{0} \\
J = \frac{1}{2l(2l+1)} \sum_{m\neq m^{\prime},m^{\prime}}\expval{V_{ee}^{s}}{m^{\prime},m}=\frac{F^2+F^4}{14}
\end{gather}
where $l$ is the angular quantum number of the Hubbard orbitals (e.g. 2 for atomic-like $d$ states), and $F^{\nu}$ is a Slater integral from the radial part of Eq. (\ref{eq:HubInt}). The above equations are strictly valid only for unscreened Coulomb kernels and spherically-symmetric Hubbard basis sets, but they are often used to evaluate Slater integrals in screened systems by working backwards: first computing $U$ and $J$ for the system then assuming the ratios between the Slater integrals have the same values as those for symmetric atomic orbitals, and solving for $F^{\nu}$ and $V_{ee}^{s}$. Further details are in the original paper\cite{liechtenstein1995} and discussed in later work\cite{vaugier2012,himmetoglu2014}.

Eqs. (\ref{eq:fullU}) and (\ref{eq:fullJ}) can be significantly simplified.\cite{dudarev1998} By retaining only the lowest order integrals in Eq. (\ref{eq:HubInt}), only $F^{0}$ remains and the functional can be rewritten as:
\begin{equation}\label{eq:simpU}
E_{\text{Hub}}\left[ n_{mm^{\!\prime}}^{I\sigma}\right] - E_{\text{dc}}\left[ n^{I\sigma}\right] = \sum_{I,\sigma} \frac{U_{\text{eff}}^{I}}{2} \textrm{Tr}\left[ \textbf{n}^{I\sigma} (1-\textbf{n}^{I\sigma})\right]
\end{equation}

This results in the calculation of forces, stresses etc. being greatly simplified, and hence it is likely the most widely-implemented form of DFT+$U$. This version gives extremely similar results to that of the full rotationally-invariant formulation, with the possible exception of some materials such as Fe-pnictides, heavy fermion or non-collinear spin materials, and multi-band metals\cite{himmetoglu2014}. This is due to the loss of the explicit higher-order interaction $J$ in materials where it is especially important, via the parameter $U_{\text{eff}}=U-J$ mentioned earlier. A correction to the simplified version that takes an explicit $J$ term into account without losing the simple form of the Hubbard energy (DFT+$U$+$J$) has been derived from the second-quantized form of the total electronic interaction potential, and verified on the CuO system\cite{himmetoglu2011}. A similarly-derived expansion of DFT+$U$, termed DFT+$U$+$V$, that takes into account inter-site Coulomb interactions, has been shown to improve the treatment of materials that exhibit a higher degree of covalency\cite{campo2010}. These expanded schemes are not used in this work and therefore will not be expounded upon. 

To interpret the effects of applying a $U$ correction, we can examine Eq. (\ref{eq:simpU}) and recognize that we can choose a representation of the Hubbard basis that diagonalizes the occupation matrix\cite{cococcioni2005}:
\begin{equation}
\textbf{n}^{I\sigma}\textbf{v}_{j}^{I\sigma}=\lambda_{j}^{I\sigma}\textbf{v}_{j}^{I\sigma}
\end{equation}
with the constraint that $0\leq \lambda_{j}^{I\sigma} \leq 1$ (since they are eigenvalues of the occupation matrix), the DFT+$U$ correction can be rewritten as
\begin{equation}
E_{\text{Hub}}[n_{mm^{\!\prime}}^{I\sigma}] - E_{\text{dc}}[n^{I\sigma}] = \sum_{I,\sigma} \sum_{j} \frac{U_{\text{eff}}^{I}}{2} \lambda_{j}^{I\sigma}(1-\lambda_{j}^{I\sigma})
\end{equation}
where we see that the $U$ parameter imposes an energetic penalty for partial occupation of the localized orbitals, favoring fully occupied or fully empty orbitals. This introduces a difference in the potential seen by occupied and unoccupied states and gives rise to a discontinuity in the potential as a function of occupation. 

When comparing DFT+$U$ to HF or hybrid functionals, there is some resemblance in the functional form, as seen in Eqs. (\ref{eq:fullU}) and (\ref{eq:fullJ}). It could therefore be considered as a substitution of a HF-like Hamiltonian for part of the density that is normally treated by the Hartree and approximate XC functionals. In this way it acts as a correction to the self-interaction energy of localized states (where self-interaction error is the largest). DFT+$U$ differs from hybrid functionals in that only a subset of states projected onto localized orbitals are treated, as opposed to interactions between all the KS states. The interaction is often done in an orbital-averaged way for simplicity (i.e. the same $U$ for all $d$ states). In addition, the discontinuity in the energy as a function of the number of electrons only arises from the subset of Hubbard orbitals. In other words, the potential is linearized with respect to occupations of the localized states--not the number of electrons in the whole system--with discontinuities at integer occupations, considering the atomic states as isolated and in contact with the ``bath'' of the rest of the crystal. 

\subsection{Self-Consistent Determination of \textit{U}}
In order to improve confidence in the predictions made by theory, it is desirable to minimize empirical parameters and perform fully \emph{ab initio} calculations whenever possible. It is common to consider (incorrectly) that values of $U$ are transferable between systems where the chemical environment differs significantly, and there is a need for methods of calculating the interaction parameters self-consistently for a given atom, crystal, magnetic ordering and localized basis set. There have been several approaches reported for calculating self-consistent values of $U$ for use in DFT+$U$, the two most common being the LR approach\cite{cococcioni2005} and the constrained random phase approximation (cRPA)\cite{aryasetiawan2004}. Here, we will briefly describe these and related methods.

\subsubsection{\emph{U} from Linear Response}
The LR method developed by Cococcioni \emph{et al.} interprets $U$ as a correction that counteracts the curvature of the total energy as a function of the fractional number of electrons present in the system, which is a result of the unphysical self-interaction energy present in approximate-XC DFT\cite{cococcioni2005}. This is a reformulation of the so-called "constrained DFT" approach\cite{dederichs1984}, where one explicitly calculates the change in energy as a function of localized orbital occupation. This was typically not done in plane wave DFT but via other methods (such as muffin tin methods) where orbital occupation could be easily defined and fixed. A value of $U$ could then be chosen that eliminates the unphysical curvature in the exchange-correlation energy by calculating this curvature but then also calculating and subtracting the curvature in the total energy that arises from hybridization, from a non-interacting KS formulation of the same system. In LR, instead of varying the orbital occupation, a small perturbative potential is applied and varied, and the resulting change in the total energy and localized orbital occupation is used to determine the curvature in the energy/occupation relationship. This method implicitly takes screening into account via a self-consistent method where the procedure is repeated until the calculated value of $U$ converges. The method also requires that a supercell structure be used, which prevents interactions between periodic images of Hubbard sites, which may be computationally prohibitive, and may not work as well for closed-shell systems where the response of the system to linear perturbation is very weak.

\subsubsection{\emph{U} from the Constrained Random Phase Approximation}
The cRPA method finds use in both DFT and DFT+DMFT as it yields a frequency-dependent $U$. The idea is that by separating the polarization of a system into localized (e.g. from $d$ orbitals) and delocalized states, the inverse dielectric function can be factorized and the effective interaction acting on the localized states can be calculated while including the screening from the extended states. The basis for the localized states can be constructed from just the manifold of interest for applying $U$ (e.g. a $d$--$d$ model), or be expanded to include other localized or itinerant states (e.g. a $d$--$dp$ model). Calculating interactions between multiple subsets of states is also possible, such as when considering $U_{dd}$, $U_{pp}$ and an inter-site term $V_{dp}$ (e.g. a $dp$--$dp$ model). While the dielectric function should be calculated using both Hartree and XC kernels, in cRPA only the Hartree term is included for simplicity (hence ``constrained''). The calculated value of $U$ is an expectation value of $W_r$, the Hartree kernel divided by the delocalized part of the dielectric function. The calculation of the localized and delocalized parts of the polarization requires some care in materials where the bands are entangled\cite{miyake2009}.

\subsubsection{\label{ssec:acbn0}Self-Consistent DFT+\emph{U} with ACBN0}
Recently, Agapito \emph{et al.} proposed a new scheme for self-consistently determining $U$, based in part on work by Mosey and Carter\cite{mosey2007,mosey2008} that utilized unscreened Coulomb and exchange interactions between Hubbard orbitals (taken as unrestricted HF orbitals obtained for a given system) to explicitly calculate $U$. While the authors have termed the approach a "pseudohybrid Hubbard density functional", we wish to clarify that \emph{in its current implementation} it is not a variational scheme or applied at each self-consistent step in the DFT calculation, but rather consists a post-processing method on an existing self-consistent DFT+$U$ calculation that is then repeated iteratively until the value of $U$ is converged. Therefore it is \emph{not} currently a functional. The general outline of this ACBN0 method (named for the authors) is as follows\cite{agapito2015}:

\begin{enumerate}[i.]
    \item The Hubbard orbitals are chosen as the pseudo-atomic orbitals (PAOs) present in the pseudopotential files.
    \item The PAOs are expressed as a combination of three Gaussian atomic orbitals, fit to the PAOs from a starting point of a Slater-type orbital (STO-3G) basis. This greatly improves the computational efficiency of calculating HF-like interaction integrals.
    \item The KS orbitals are projected onto the PAO basis via a scheme developed by the same authors\cite{agapito2013,agapito2016a,agapito2016b,damico2016}.
    \item The occupation of the each KS orbital is ``renormalized'' by the Mulliken population of that orbital projected on the Hubbard basis (including all atoms with states that have the same quantum numbers in the unit cell).
    \item The Hartree ($U$) and exchange ($J$) terms are explicitly calculated using the HF Coulomb and exchange kernels divided by the occupation numbers of the Hubbard orbitals.
\end{enumerate}

The Coulomb and exchange energies are calculated as follows. First, the KS orbital ($\phi_{i}$) occupations are renormalized according to the Mulliken charge on the extended basis $\{\bar{m}\}$ (including all orbitals on all sites that have quantum numbers equal to the Hubbard orbitals in the single-site basis $\{ m \}$):
\begin{equation}
\bar{n}_{\phi_{i}}^{\textbf{k}\sigma} \equiv \sum_{\mu \in \{\bar{m}\}} \sum_{\nu} c_{\mu i}^{\textbf{k}\sigma \dagger}S_{\mu\nu}^{\textbf{k}} c_{\nu i}^{\textbf{k}\sigma}
\end{equation}
where $\textbf{k}$ is an vector index for a specific k point in the Brillouin zone, $c_{\nu i}^{\textbf{k}\sigma}$ is the expansion coefficient of the localized state $\varphi_{\nu}$ as a component of KS state $\phi_i$, and $S_{\mu\nu}$ is the overlap matrix element for localized states $\varphi_{\mu}$ and $\varphi_{\nu}$. A renormalized density matrix is then defined by adding the contributions of every KS state to the relevant Hubbard orbitals:
\begin{equation}
\bar{P}_{\mu\nu}^{\sigma} = \frac{1}{\sqrt{N_{\textbf{k}}}} \sum_{\textbf{k},i} \bar{n}_{\phi_{i}}^{\textbf{k}\sigma} c_{\mu i}^{\textbf{k}\sigma \dagger} c_{\nu i}^{\textbf{k}\sigma}
\end{equation}

Meanwhile, the occupations of the Hubbard orbitals are:
\begin{equation}
n_{m}^{\sigma} = \frac{1}{\sqrt{N_{\textbf{k}}}} \sum_{\textbf{k},i,\nu} c_{mi}^{\textbf{k}\sigma \dagger}S_{m\nu}^{\textbf{k}} c_{\nu i}^{\textbf{k}\sigma}
\end{equation}
where $N_{\textbf{k}}$ is the number of k points in the Brillouin zone. By using the HF expression for the energy, we then get
\begin{equation}
\begin{split}
E_{\text{hub}} = & \frac{1}{2} \sum_{\{m\},\sigma} \left[ \bar{P}_{mm^{\!\prime}}^{\sigma} \bar{P}_{m^{\!\prime\!\prime}m^{\!\prime\!\prime\!\prime}}^{\sigma^{\prime}} \right]\mel{m,m^{\!\prime\!\prime}}{V_{ee}}{m^{\!\prime},m^{\!\prime\!\prime\!\prime}} \\
& + \frac{1}{2} \sum_{\{m\},\sigma} \left[ \bar{P}_{mm^{\!\prime}}^{\sigma} \bar{P}_{m^{\!\prime\!\prime}m^{\!\prime\!\prime\!\prime}}^{\sigma} \right]\mel{m,m^{\!\prime}}{V_{ee}}{m^{\!\prime\!\prime},m^{\!\prime\!\prime\!\prime}}\end{split}
\end{equation}
which bears a clear resemblance to Eq. (\ref{eq:fullU}). By comparing this expression to another form of the energy given by Eq. (\ref{eq:simpU}),
\begin{equation}
E_{\text{hub}}\approx \frac{U}{2}\sum_{\{m\},\sigma} n_{m}^{\sigma}n_{m^{\!\prime}}^{-\sigma} + \frac{U-J}{2}\sum_{m\neq m^{\!\prime},\sigma} n_{m}^{\sigma}n_{m^{\!\prime}}^{\sigma}
\end{equation}
we get as expressions for $U$ and $J$:
\begin{gather}
U = \frac{\sum_{\{m\},\sigma} \left[ \bar{P}_{mm^{\!\prime}}^{\sigma} \bar{P}_{m^{\!\prime\!\prime}m^{\!\prime\!\prime\!\prime}}^{\sigma^{\prime}} \right]\mel{m,m^{\!\prime\!\prime}}{V_{ee}}{m^{\!\prime},m^{\!\prime\!\prime\!\prime}}}{\sum_{m,\sigma \neq m^{\!\prime},\sigma^{\prime}} n_{m}^{\sigma}n_{m^{\!\prime}}^{\sigma^{\prime}}} \\
J = \frac{\sum_{\{m\},\sigma} \left[ \bar{P}_{mm^{\!\prime}}^{\sigma} \bar{P}_{m^{\!\prime\!\prime}m^{\!\prime\!\prime\!\prime}}^{\sigma} \right]\mel{m,m^{\!\prime}}{V_{ee}}{m^{\!\prime\!\prime},m^{\!\prime\!\prime\!\prime}}}{\sum_{m \neq m^{\!\prime},\sigma} n_{m}^{\sigma}n_{m^{\!\prime}}^{\sigma}}
\end{gather}

The strengths of ACBN0 are that the parameters can be calculated for every atom in the unit cell, without needing supercells or costly additional calculations. The HF integrals are evaluated very quickly when the Hubbard orbitals are projected onto a three-gaussian (3G) basis, in a fraction of the time needed for the DFT calculation. The renormalization should, in theory, reduce the calculated values of $U$ and $J$ for systems where the KS orbitals are not well-represented by a localized basis.

\section{\label{sec:methods}Computational Methods}
DFT calculations were performed using Quantum ESPRESSO 6.1\cite{giannozzi2009,giannozzi2017}, using optimized norm-conserving pseudopotentials from the SG15 library\cite{schlipf2015} (La and Sr) and standard-accuracy (stringent for Cr) Pseudo-Dojo\cite{vanSetten2018} (transition metals and O), generated from the Optimized Norm-Conserving Vanderbilt Pseudopotential code\cite{hamann2013}. Plane wave cutoff, $k$-point mesh, and self-consistency convergence threshold were converged with respect to the total energy ($<$ 15 meV/atom), total force ($<$ $10^{-4}$ Ry/a.u./atom), and unit cell pressure ($<$ 0.3 kbar), versus a well-converged calculation with cutoff 250 Ry, a dense k-point mesh (9 $\times$ 9 $\times$ 9 Monkhorst-Pack\cite{monkhorst1976}) and threshold of $10^{-9}$ Ry. Convergence test results and k-point paths for band diagrams are shown in the Supplemental Material\cite{supp}. A plane wave cutoff of 100 Ry was used (except for Cr, which used 120 Ry) with a Monkhorst-Pack grid of 4 $\times$ 4 $\times$ 3 and convergence threshold of $10^{-6}$ Ry were used for all calculations. Variable-cell relax calculations decreased the convergence threshold to $10^{-9}$ Ry for the final relaxation steps.

DFT+$U$ was performed using $U$ values calculated with ACBN0, using Python scripts to both automate the self-consistent electronic structure calculations and determine the electron repulsion integrals. Detail on the calculated $U$ values, as well as first-principles values of $U$ taken from the literature (since several first-principles or empirical values of $U$ have been reported, the median value is used in this work), is shown in the Supplemental Material\cite{supp}.  The simplified rotationally-invariant implementation of Dudarev \emph{et al.}\cite{dudarev1998} and Cococcioni \emph{et al.}\cite{cococcioni2005} was used. Initial spin states and starting atomic magnetizations were set according to the experimentally reported electronic configurations for each transition metal in the associated perovskite structure (such as high-spin $\text{Fe}^{3+}$, with ${t}_{2g}$: $\uparrow\,\uparrow\,\uparrow$ and ${e}_{g}$: $\uparrow\,\uparrow$)\cite{he2012}. Antiferromagnetic (AFM) AFM-A, AFM-C, AFM-G, ferromagnetic (FM) and nonmagnetic (NM) magnetic orderings were calculated depending on what the primitive unit cell size and symmetry allow. ACBN0 is not fundamentally limited to a certain set of localized orbitals, but in the original paper and in this work, the atomic-like orbitals from the pseudopotentials are used for simplicity and for the convenience of fitting a minimal three-Gaussian (3G) basis set for rapid evaluation of the electron repulsion integrals. The ACBN0 $U$ correction was calculated and applied to transition metal $3d$ states ($U_{dd}$) and oxygen $2p$ states ($U_{pp}$). Literature $U$ values were only applied to the metal $3d$ states as is common practice (calculations labeled as PBE+$U_{dd}^{\text{Lit.}}$). We again mention that ACBN0 as implemented in this work is not a functional, but a method to calculate $U$. Comparing energies between different calculations using ACBN0 is not possible since presumably different values of $U$ would be used in each calculation. To deal with this issue, we choose to present both an average $U$ from the calculations of energetically-similar magnetic states, and the $U$ from the experimentally-determined ground state.

\section{Results and Discussion}
\subsection{Computed Values of \emph{U}}
\begin{figure}[!tpb]
   \centering
   \includegraphics{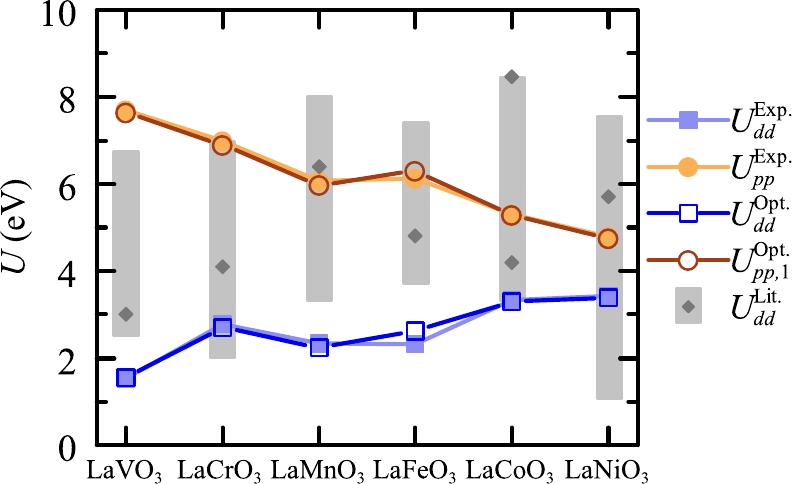}
   \caption{Values of $U$ obtained from ACBN0 for metal $3d$ ($U_{dd}$) and oxygen $2p$ ($U_{pp}$) for the perovskite oxides studied in this work. Filled symbols refer to calculations using the experimental structure, while open symbols represent calculations that have been optimized with several iterations of relaxation and ACBN0 until the values of U change by less than 0.01 eV. $U_{dd}^{\text{Lit.}}$ values from the literature used for comparing to conventional DFT+$U$ are included for comparison, with the range of literature $U$ values found and cited in this paper bounded by grey rectangles.}\label{fig:Uvals}
\end{figure}

Before discussing the main DFT+$U$ results, we first present the values of $U$ calculated by ACBN0 in this work and make some broad comparisons with other values of $U$ presented in the literature. The results of ACBN0 calculations of $U$ are presented visually in Fig.~\ref{fig:Uvals}. The magnitude of $U_{dd}$ is between 1.5 and 3.5 eV, with a trend of increasing $U_{dd}$ with $d$ shell filling. The increasing $d$--$p$ hybridization as the $d$ manifold is filled is apparent by the relatively small increase in $U_{dd}$ for the metal cations as well as the decrease in the $U_{pp}$ calculated for oxygen $2p$ states, which ranges from around 4.8 to 7.6 eV. These larger magnitudes of $U_{pp}$ on oxygen ions are in qualitative agreement with the observation of experimental $U_{pp}$ values from X-ray spectroscopy interpreted in the Zaanen, Sawatsky, and Allen (ZSA) scheme of oxide electronic structure\cite{sawatzky1979,ghijsen1988,bar-deroma1992,chainani1992,chainani1993,sarma1994,chainani2017}, though the values themselves correspond to an interpretation of $U$ which is different to that used in the present work and should not be quantitatively compared. The application of $U_{pp}$ to oxygen $p$ states is an important point of discussion and is covered in more detail in Sec.~\ref{subsec:Uox}.

The values of $U_{dd}$ calculated in this work are typically smaller than those which have been used in the literature to date for the same materials. We again caution that values of $U$ should not be considered transferable in general. Using the same DFT functional, code, and implementation of $U$, even a change in pseudopotential (especially from untested pseudopotentials, or those generated with a different XC functional than the DFT calculation) can result in changes on the order of eV. For example, the original ACBN0 paper\cite{agapito2015} found values of $U_{dd}=0.15~\text{eV}$ and $U_{pp}=7.34~\text{eV}$, using (untested) norm-conserving pseudopotentials from PSLibrary 1.0.0\cite{dalcorso2014}. With the pseudopotentials used in our current work (see Computational Methods, Sec.~\ref{sec:methods}), a test calculation on TiO$_2$ yielded $U_{dd}=0.12~\text{eV}$ and $U_{pp}=8.4~\text{eV}$. Aside from the pseudopotentials affecting the accuracy of the underlying DFT calculation, they also contain the pseudo-orbitals that are used for applying the $U$ correction--a significant change in the pseudo-orbitals can affect the calculated value of $U$ by up to several eV\cite{pickett1998}. That being said, the difference between ACBN0 and the various sources from the literature varies between 0 eV (for $U_{dd}$ of LaCoO$_3$ fitted to enthalpy of formation) and over 5 eV (for $U_{dd}$ of LaMnO$_3$ fitted to band gap, and LaCoO$_3$ from LR). It is worth noting that for the two most common forms of self-consistent $U$ (constrained DFT/LR and cRPA) that typically, the lowest values of $U$ (i.e. those in closest agreement with ACBN0) were those calculated from cRPA (introducing another difficulty in comparison, namely the type of interaction model used in the cRPA) and the highest were for constrained DFT/LR. More detailed tables listing our calculated values of $U$ (including the explicit $U$ and $J$ values), and values of $U$ sourced from the literature are available in the Supplemental Material\cite{supp}.

\subsection{Crystal Structure}
\begin{figure}[!tpb]
   \centering
   \includegraphics{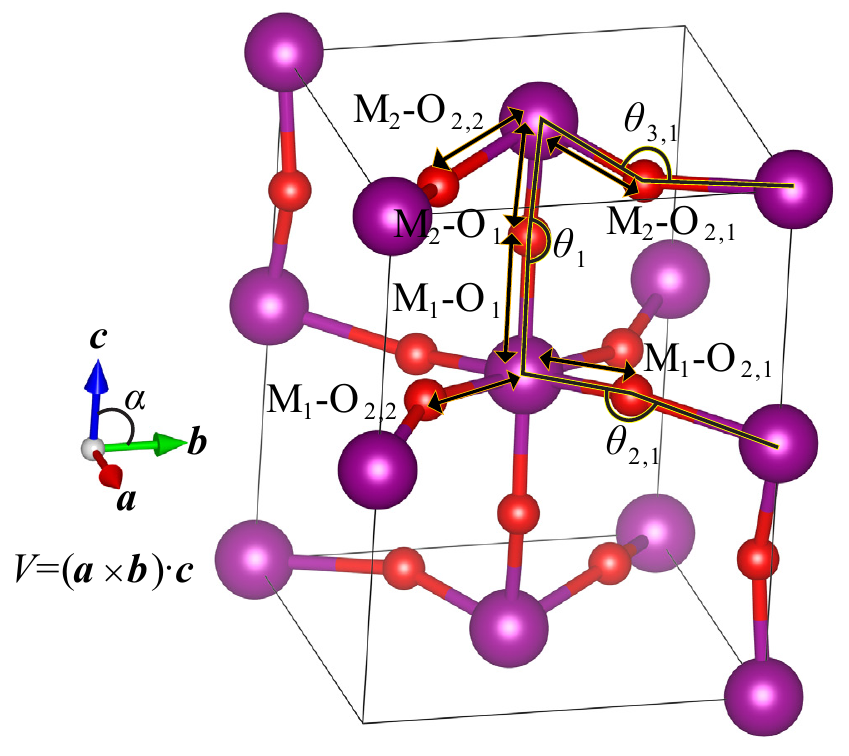}
   \caption{Perovskite structure parameters used in the determination of mean absolute relative error (MARE).}\label{fig:MAREcell}
\end{figure}

The structural parameters of perovskites LaMO$_{\text{3}}$ (M = V--Ni) have been reported according to the definitions shown in Fig.~\ref{fig:MAREcell}, similar to those reported by He and Franchini in their HSE hybrid functional study of first-row transition metal perovskites\cite{he2012}. HSE results mentioned refer to this work unless otherwise noted. Optimized structures are analyzed only for the calculations using the experimentally-observed magnetic ordering (except for paramagnetic LaNiO$_3$, where a non-magnetic state is used). These consist of the lattice parameters, unit cell volume, various metal-oxygen bond lengths and metal-oxygen-metal bond angles. Crystallographic representations have been chosen to be consistent among all perovskites (i.e., the space group unique axes are oriented in a such a way that allows a direct mapping of atomic site positions between different materials). While the He and Franchini also include the Jahn-Teller (JT) distortion modes Q2 and Q3 as parameters, their small magnitudes are not suitable for including in the mean absolute relative error (MARE) and will not be included in this analysis for simplicity. They can be still calculated from the information provided herein. One should ensure that the same experimental reference structures are used when comparing between different studies whenever possible. In the following discussion this is the case unless otherwise noted.

\subsubsection{LaVO$_{\text{3}}$}
LaVO$_3$ has monoclinic symmetry in the $P2_1/b$ space group, with two unique V sites in the unit cell. The structural parameters, presented in Table~\ref{table:structV}, reflect this by including bond lengths and angles for both V sites. The MARE for ACBN0 is 0.69\%, which compares favorably to the PBE value of 0.88\%, and especially to the MARE of 2.6\% obtained from PBE+$U_{dd}^{\text{Lit.}}$. The value of $U_{dd}$ used (3.0 eV) is both from cluster-CI model fits to experimental spectra\cite{nohara2009}, and also from empirical fits to band gap\cite{fang2004}. This value is also close to cRPA results\cite{kim2018}. Despite this, PBE+$U_{dd}^{\text{Lit.}}$ describes the structure of $\text{LaVO}_3$ quite poorly. Hybrid functional calculations, using both the commonly-used mixing fraction of 0.25 (HSE-25) and an empirically-chosen value of mixing to improve the overall structural and electronic properties (HSE-Opt), show improved structural agreement with experiment at 0.48\% and 0.35\%, respectively\cite{he2012}. It is interesting to note where the variation in MARE arises from in the different methods. The largest error values typically arise from the bond angles; however PBE+$U_{dd}^{\text{Lit.}}$ also results in a significantly overestimated cell volume, and also incorrectly predicts some relative bond lengths and angles such as M$_2$--O$_{2,1}$ $>$ M$_2$--O$_{2,2}$. He and Franchini's PBE results, while the MARE similar to that reported here (0.98 vs. 0.88 \%), differ significantly in some other parameters such as volume (0.2 vs. 1.08\%). This illustrates some of the difficulty in comparing different works that utilize different codes, computation parameters and pseudopotentials, even though in general DFT is becoming increasingly reproducible across various DFT implementations\cite{lejaeghere2016}. The main picture for LaVO$_3$ is that ACBN0 marginally improves in all areas vs. PBE (which still describes structure adequately with MARE $<$ 1\%), while hybrid functionals have shown lower overall errors by improving the accuracy of bond lengths and angles, despite similar errors in the lattice constant and volume as compared to PBE and ACBN0. 

\begin{table}[!tpb]
\centering
\caption{Structural parameters for AFM-C LaVO$_3$. Experimental data measured at 10 K is taken from Bordet \emph{et al.}\cite{bordet1993} $U_{dd}^{\text{Lit.}}=3.0$ eV\cite{nohara2009}. Relative absolute error is shown in italics (in \%), with the mean absolute relative error (MARE) listed at the bottom of the table. ACBN0 calculations are PBE+$U_{dd}$+$U_{pp}$.}
\label{table:structV}
\begin{ruledtabular}
\begin{tabular}{rcccc}
\multicolumn{1}{c}{} & \multicolumn{4}{c}{\textbf{LaVO$_3$}} \\
\multicolumn{1}{c}{} & \multicolumn{1}{c}{Expt.} & \multicolumn{1}{c}{PBE} & \multicolumn{1}{c}{PBE+$U_{dd}^{\text{Lit.}}$} & \multicolumn{1}{c}{ACBN0} \\ \cline{2-5}
V (\AA$^3$) & 241.10 & 242.28 & 250.75 & 242.32 \\
 &  & \textit{1.08} & \textit{4.00} & \textit{0.51} \\
$a$ (\AA) & 5.5623 & 5.575 & 5.602 & 5.545 \\
 &  & \textit{0.22} & \textit{0.71} & \textit{0.31} \\
$b$ (\AA) & 5.5917 & 5.637 & 5.726 & 5.609 \\
 &  & \textit{0.80} & \textit{2.41} & \textit{0.31} \\
$c$ (\AA) & 7.7516 & 7.710 & 7.817 & 7.791 \\
 &  & \textit{0.53} & \textit{0.84} & \textit{0.51} \\
$\beta$ ($^{\circ}$) & 90.13 & 90.02 & 89.84 & 90.40 \\
 &  & \textit{0.12} & \textit{0.32} & \textit{0.30} \\
$\text{M}_1$--$\text{O}_1$ (\AA) & 1.978 & 1.961 & 2.014 & 1.963 \\
 &  & \textit{0.88} & \textit{1.79} & \textit{0.76} \\
$\text{M}_1$--$\text{O}_{2,1}$ (\AA) & 1.989 & 2.025 & 2.101 & 1.990 \\
 &  & \textit{1.82} & \textit{5.63} & \textit{0.05} \\
$\text{M}_1$--$\text{O}_{2,2}$ (\AA) & 2.042 & 2.023 & 2.021 & 2.057 \\
 &  & \textit{0.94} & \textit{1.03} & \textit{0.74} \\
$\text{M}_2$--$\text{O}_1$ (\AA) & 1.979 & 1.961 & 2.010 & 2.018 \\
 &  & \textit{0.88} & \textit{1.58} & \textit{2.00} \\
$\text{M}_2$--$\text{O}_{2,1}$ (\AA) & 1.979 & 2.021 & 2.099 & 2.000 \\
 &  & \textit{2.12} & \textit{6.04} & \textit{1.07} \\
$\text{M}_2$--$\text{O}_{2,2}$ (\AA) & 2.039 & 2.025 & 1.996 & 2.028 \\
 &  & \textit{0.70} & \textit{2.14} & \textit{0.56} \\
$\theta_1$ ($^{\circ}$) & 156.74 & 158.80 & 152.51 & 156.15 \\
 &  & \textit{1.31} & \textit{2.70} & \textit{0.37} \\
$\theta_{2,1}$ ($^{\circ}$) & 156.12 & 156.64 & 152.53 & 154.08 \\
 &  & \textit{0.33} & \textit{2.29} & \textit{1.31} \\
$\theta_{2,2}$ ($^{\circ}$) & 157.83 & 156.84 & 152.97 & 156.53 \\
 &  & \textit{0.63} & \textit{3.08} & \textit{0.83} \\
MARE (\%) &  & \textbf{0.88} & \textbf{2.47} & \textbf{0.69} \\
\end{tabular}
\end{ruledtabular}
\end{table}

\subsubsection{LaCrO$_{\text{3}}$}
LaCrO$_3$ has an orthorhombic structure with GdFeO$_3$ (GFO) tilting distortions to the octahedra and space group $Pnma$ (represented here in the $Pbmn$ setting). As shown in Table~\ref{table:structCr}, ACBN0 (MARE 1.09\%) performs slightly worse than PBE (MARE 0.94\%), mostly due to the poor description of bond lengths, despite slightly improved accuracy with regard to the lattice parameters and bond angles. PBE+$U_{dd}^{\text{Lit.}}$ results with a cluster-CI value of $U_{dd}=4.1$ eV\cite{nohara2009} again result in a drastically poorer description of the structure. Other similar values of $U_{dd}$ from fits to enthalpy of formation\cite{wang2006}, band gap\cite{hong2012}, and HSE calculations\cite{yang1999} would likely provide the same general result. The HSE results of He and Franchini are referenced to a different (room temperature) experiment, but compared to the 11 K reference used here, PBE, HSE-25 and HSE-Opt (mixing 0.15) gave MARE values of 0.75\%, 0.43\% and 0.59\%, respectively\cite{he2012}. HSE-Opt improves on the lattice parameters and bond lengths but worsens the error on the bond angles. HSE-25 shows similar error on the bond lengths but drastically improves all other structure descriptors considered here. Again, the difference between the previously reported PBE results and the current work can likely again be explained by computational differences such as choice of pseudopotential or DFT input parameters.

\begin{table}[!tpb]
\small
\centering
\caption{Structural parameters for AFM-G LaCrO$_3$. Experimental data measured at 11 K is taken from Gilbu Tilset \emph{et al.}\cite{gilbutilset1998} $U_{dd}^{\text{Lit.}}=4.1$ eV\cite{nohara2009}. Relative absolute error is shown shaded in gray (in \%), with the mean absolute relative error (MARE) listed at the bottom of the table. ACBN0 calculations are PBE+$U_{dd}$+$U_{pp}$.}
\label{table:structCr}
\begin{ruledtabular}
\begin{tabular}{rcccc}
\multicolumn{1}{c}{} & \multicolumn{4}{c}{\textbf{LaCrO$_3$}} \\
\multicolumn{1}{c}{} & \multicolumn{1}{c}{Expt.} & \multicolumn{1}{c}{PBE} & \multicolumn{1}{c}{PBE+$U_{dd}^{\text{Lit.}}$} & \multicolumn{1}{c}{ACBN0} \\ \cline{2-5}
V (\AA$^3$) & 233.60 & 237.54 & 244.14 & 237.26 \\
 &  & \textit{1.69} & \textit{4.51} & \textit{1.57} \\
$a$ (\AA) & 5.4718 & 5.521 & 5.588 & 5.522 \\
 &  & \textit{0.90} & \textit{2.12} & \textit{0.92} \\
$b$ (\AA) & 5.5093 & 5.519 & 5.557 & 5.519 \\
 &  & \textit{0.18} & \textit{0.86} & \textit{0.18} \\
$c$ (\AA) & 7.7491 & 7.796 & 7.863 & 7.785 \\
 &  & \textit{0.61} & \textit{1.47} & \textit{0.47} \\
M--$\text{O}_1$ (\AA) & 1.968 & 1.987 & 2.016 & 1.990 \\
 &  & \textit{0.92} & \textit{3.29} & \textit{2.21} \\
M--$\text{O}_{2,1}$ (\AA) & 1.974 & 1.989 & 2.019 & 1.990 \\
 &  & \textit{0.73} & \textit{3.27} & \textit{1.52} \\
M--$\text{O}_{2,2}$ (\AA) & 1.968 & 1.987 & 2.018 & 1.990 \\
 &  & \textit{1.00} & \textit{2.42} & \textit{1.08} \\
$\theta_{1}$ ($^{\circ}$) & 159.59 & 157.67 & 154.33 & 156.07 \\
 &  & \textit{1.20} & \textit{2.28} & \textit{0.79} \\
$\theta_{2}$ ($^{\circ}$) & 160.04 & 158.02 & 154.81 & 157.60 \\
 &  & \textit{1.26} & \textit{2.56} & \textit{1.11} \\
MARE (\%) &  & \textbf{0.94} & \textbf{2.53} & \textbf{1.09} \\
\end{tabular}
\end{ruledtabular}
\end{table}

\subsubsection{LaMnO$_{\text{3}}$}
LaMnO$_3$ has the largest JT distortions among the 3$d$ perovskites studied here. The structural results are presented in Table~\ref{table:structMn}. This has important consequences for the calculated electronic structure, which is why a very high structural accuracy is required in this material for predicting electronic properties and ground states (discussed in the next section). PBE and ACBN0 provide almost identical error, with MARE values of 0.93\% and 0.99\%, respectively. This is in contrast to the work of He and Franchini, who report a large MARE for PBE (1.9\%), caused by large inaccuracy in bond lengths that describe the JT distortions, with the largest individual bond error being over 5\% (the largest PBE bond length error in this work is 1.62\%)\cite{he2012}. Their calculations using HSE-Opt and HSE-25 show some marginal improvement over the PBE and ACBN0 calculations here, but still give a similar overall picture. The PBE+$U_{dd}^{\text{Lit.}}$ calculations once again show a significantly larger error at 2.53\%, with over 3\% error on two of the three bond lengths, when using a value of $U_{dd}=6.4$ eV\cite{nohara2009}. A wide range of $U_{dd}$ values have been used in the literature, including 3.3 eV (from cRPA\cite{jang2018}) and 7.1 eV (from constrained DFT\cite{nohara2009}). Hashimoto \emph{et al.}\cite{hashimoto2010} reported that PBE+$U_{dd}^{\text{Lit.}}=2.0$ eV can improve the treatment of JT distortions in LaMnO$_3$ under full cell relaxation, but both ACBN0 and PBE+$U_{dd}^{\text{Lit.}}$ fail to improve over the PBE case in this work. We performed a quick test with $U_{dd}=2.0$ eV, which yielded a MARE of 2.06\%, with errors on the bond lengths still well above 1\%. The reason for this discrepancy in how PBE+$U_{dd}^{\text{Lit.}}$ describes the JT distortions in fully structurally-optimized LaMnO$_3$ is unknown. One thing to note is that in these studies\cite{he2012,hashimoto2010}, plane wave cutoffs between 30-40 Ry were used. We cannot claim that these results are unconverged with respect to calculating relaxed structures, but in this work, a cutoff of at least 100 Ry was found to be necessary to be converged with respect to cell pressure (within 0.5 kbar, see Supplemental Material\cite{supp}). For energy differences lower cutoffs may be adequate, but quantitative comparison of unit cell structure requires highly converged calculation parameters to get accurate forces and stresses. Unconverged calculations may provide a fortuitous improvement in describing structure--additional test calculations with both $U_{dd}=0.0$ eV (plain PBE) and $U_{dd}=2.0$ eV performed at 35 Ry plane wave cutoff energy yielded a MARE of 2.61\% and 1.46\% respectively for LaMnO$_3$, worsening accuracy for the plain PBE case but improving accuracy for the $U_{dd}=2.0$ eV case. This is entirely an artifact of unconverged geometry from low plane wave energy cutoffs. Another issue could be differing localized basis sets for applying the $U$ correction, and the specific implementation of DFT+$U$ used; the nature of the orbitals chosen and whether the $J$ exchange terms are included explicitly vs. in a combined effective $U$ can strongly affect calculated values of $U$ and the resulting material properties. This will be discussed further in the next section.

\begin{table}[!tpb]
\small
\centering
\caption{Structural parameters for AFM-A LaMnO$_3$. Experimental data measured at 4.2 K is taken from Elemans \emph{et al.}\cite{elemans1971} $U_{dd}^{\text{Lit.}}=6.4$ eV\cite{nohara2009}. Relative absolute error is shown shaded in gray (in \%), with the mean absolute relative error (MARE) listed at the bottom of the table. ACBN0 calculations are PBE+$U_{dd}$+$U_{pp}$.}
\label{table:structMn}
\begin{ruledtabular}
\begin{tabular}{rcccc}
\multicolumn{1}{c}{} & \multicolumn{4}{c}{\textbf{LaMnO$_3$}} \\
\multicolumn{1}{c}{} & \multicolumn{1}{c}{Expt.} & \multicolumn{1}{c}{PBE} & \multicolumn{1}{c}{PBE+$U_{dd}^{\text{Lit.}}$} & \multicolumn{1}{c}{ACBN0} \\ \cline{2-5}
V (\AA$^3$) & 243.57 & 248.11 & 264.51 & 247.44 \\
 &  & \textit{1.86} & \textit{8.59} & \textit{1.59} \\
$a$ (\AA) & 5.532 & 5.563 & 5.631 & 5.549 \\
 &  & \textit{0.56} & \textit{1.79} & \textit{0.31} \\
$b$ (\AA) & 5.742 & 5.806 & 5.994 & 5.819 \\
 &  & \textit{1.12} & \textit{4.39} & \textit{1.34} \\
$c$ (\AA) & 7.668 & 7.681 & 7.837 & 7.663 \\
 &  & \textit{0.17} & \textit{2.21} & \textit{0.07} \\
M--$\text{O}_1$ (\AA) & 1.957 & 1.972 & 2.048 & 1.970 \\
 &  & \textit{0.76} & \textit{4.61} & \textit{0.64} \\
M--$\text{O}_{2,1}$ (\AA) & 2.185 & 2.190 & 2.268 & 2.206 \\
 &  & \textit{0.23} & \textit{3.81} & \textit{0.99} \\
M--$\text{O}_{2,2}$ (\AA) & 1.904 & 1.934 & 1.995 & 1.924 \\
 &  & \textit{1.62} & \textit{4.82} & \textit{1.08} \\
$\theta_{1}$ ($^{\circ}$) & 156.69 & 153.63 & 146.21 & 153.04 \\
 &  & \textit{1.95} & \textit{6.68} & \textit{2.33} \\
$\theta_{2}$ ($^{\circ}$) & 154.34 & 154.20 & 149.31 & 153.45 \\
 &  & \textit{0.09} & \textit{3.26} & \textit{0.57} \\
MARE (\%) &  & \textbf{0.93} & \textbf{4.46} & \textbf{0.99} \\
\end{tabular}
\end{ruledtabular}
\end{table}

\subsubsection{LaFeO$_{\text{3}}$}
Orthorhombic $Pbnm$ LaFeO$_3$ has fully occupied $e_g$ and $t_{2g}$ manifolds (high spin) that suppress JT distortion. While PBE performs fairly well at describing the structural parameters (MARE of 1.20\%, Table~\ref{table:structFe}), ACBN0 improves the accuracy of every unit cell parameter (MARE 0.79\%). The PBE+$U_{dd}^{\text{Lit.}}$ structure ($U_{dd}=4.8$ eV\cite{nohara2009}) again shows significantly worsened structural accuracy with a MARE of 2.90\%. Hybrid functionals offer additional improvement vs. the ACBN0 results, with the empirically-optimized HSE-Opt yielding a MARE of 0.32\% and HSE-25 yielding a MARE of 0.30\% (note these MARE values have been adjusted from the original publication to correspond to the experimental data used here, which is very similar).

\begin{table}[!tpb]
\small
\centering
\caption{Structural parameters for AFM-G LaFeO$_3$. Experimental room-temperature data is taken from Etter \emph{et al.}\cite{etter2014} $U_{dd}^{\text{Lit.}}=4.8$ eV\cite{nohara2009}. Relative absolute error is shown shaded in gray (in \%), with the mean absolute relative error (MARE) listed at the bottom of the table. ACBN0 calculations are PBE+$U_{dd}$+$U_{pp}$.}
\label{table:structFe}
\begin{ruledtabular}
\begin{tabular}{rcccc}
\multicolumn{1}{c}{} & \multicolumn{4}{c}{\textbf{LaFeO$_3$}} \\
\multicolumn{1}{c}{} & \multicolumn{1}{c}{Expt.} & \multicolumn{1}{c}{PBE} & \multicolumn{1}{c}{PBE+$U_{dd}^{\text{Lit.}}$} & \multicolumn{1}{c}{ACBN0} \\ \cline{2-5}
V (\AA$^3$) & 242.88 & 247.51 & 252.38 & 245.13 \\
 &  & \textit{1.91} & \textit{3.91} & \textit{0.93} \\
$a$ (\AA) & 5.5549 & 5.558 & 5.595 & 5.547 \\
 &  & \textit{0.06} & \textit{0.72} & \textit{0.15} \\
$b$ (\AA) & 5.5663 & 5.653 & 5.679 & 5.617 \\
 &  & \textit{1.56} & \textit{2.02} & \textit{0.92} \\
$c$ (\AA) & 7.8549 & 7.877 & 7.944 & 7.867 \\
 &  & \textit{0.29} & \textit{1.13} & \textit{0.16} \\
M--$\text{O}_1$ (\AA) & 2.010 & 2.022 & 2.046 & 2.019 \\
 &  & \textit{0.60} & \textit{1.77} & \textit{0.41} \\
M--$\text{O}_{2,1}$ (\AA) & 2.019 & 2.048 & 2.055 & 2.028 \\
 &  & \textit{1.46} & \textit{1.82} & \textit{0.46} \\
M--$\text{O}_{2,2}$ (\AA) & 1.990 & 2.021 & 2.044 & 2.018 \\
 &  & \textit{1.56} & \textit{2.71} & \textit{1.43} \\
$\theta_{1}$ ($^{\circ}$) & 155.26 & 153.70 & 152.17 & 154.00 \\
 &  & \textit{1.01} & \textit{1.99} & \textit{0.81} \\
$\theta_{2}$ ($^{\circ}$) & 157.57 & 153.89 & 153.00 & 154.61 \\
 &  & \textit{2.33} & \textit{2.90} & \textit{1.88} \\
MARE (\%) &  & \textbf{1.20} & \textbf{2.11} & \textbf{0.79} \\
\end{tabular}
\end{ruledtabular}
\end{table}

\subsubsection{LaCoO$_{\text{3}}$}
Due to the smaller ionic radius of Co$^{3+}$, LaCoO$_3$ crystallizes in a rhombohedral structure with space group $R\bar{3}c$, with slight GFO-type octahedral distortions. Structural parameters and errors are listed in Table~\ref{table:structCo}. PBE and the smaller PBE+$U_{dd}^{\text{Lit.}}$ value of 4.2 eV (from cluster-CI calculations fit to experimental spectra\cite{nohara2009}) perform similarly, with MAREs of 1.20\% and 1.28\%, respectively. Increasing to a larger, LR $U_{dd}=8.5$ eV\cite{hsu2009}, the error increases significantly to 3.60\%. This provides yet another illustration of the pitfalls of choosing $U_{dd}$ uncritically, since the original work used LSDA+$U$; LSDA tends to overbind and shorten bond lengths, and the addition of a $U$ correction may increase the bond lengths closer to the experimental value. ACBN0 provides the highest structural accuracy for non-magnetic LaCoO$_3$, with a MARE of 0.12\%, which compares very favorably to the HSE-25 value of 0.42\% and the HSE-Opt value of 0.44\%. ACBN0 and hybrid functionals are the only methods reported here that decrease the over-estimated unit cell volume of PBE--applying a $U$ correction only to the $d$ electrons results in an increased cell volume.

\begin{table}[!tpb]
\small
\centering
\caption{Structural parameters for NM LaCoO$_3$. Experimental data measured at 4.2 K is taken from Thornton \emph{et al.}\cite{thornton1986} $U_{dd}^{\text{Lit.}}=4.2$ eV\cite{nohara2009} and 8.5 eV\cite{hsu2009}. Relative absolute error is shown shaded in gray (in \%), with the mean absolute relative error (MARE) listed at the bottom of the table. $\theta_1$ and $\theta_2$ describe O--$\hat{\text{Co}}$--O and Co--$\hat{\text{O}}$--Co angles, respectively. ACBN0 calculations are PBE+$U_{dd}$+$U_{pp}$.}
\label{table:structCo}
\begin{ruledtabular}
\begin{tabular}{rccccc}
\multicolumn{1}{c}{} & \multicolumn{5}{c}{\textbf{LaCoO$_3$}} \\
\multicolumn{1}{c}{} & \multicolumn{1}{c}{Expt.} & \multicolumn{1}{c}{PBE} & \multicolumn{2}{c}{PBE+$U_{dd}^{\text{Lit.}}$} & \multicolumn{1}{c}{ACBN0} \\
\multicolumn{1}{c}{} & \multicolumn{1}{c}{} & \multicolumn{1}{c}{} & \multicolumn{1}{c}{4.2 eV} & \multicolumn{1}{c}{8.5 eV} & \multicolumn{1}{c}{} \\ \cline{2-6}
V (\AA$^3$) & 110.17 & 112.43 & 112.73 & 113.31 & 110.19 \\
 &  & \textit{2.04} & \textit{2.32} & \textit{12.83} & \textit{0.02} \\
$a$ (\AA) & 5.3416 & 5.360 & 5.367 & 5.380 & 5.342 \\
 &  & \textit{0.35} & \textit{0.47} & \textit{3.67} & \textit{0.01} \\
$\alpha$ ($^{\circ}$) & 60.99 & 61.43 & 61.40 & 61.30 & 60.99 \\
 &  & \textit{0.73} & \textit{0.68} & \textit{0.95} & \textit{0.00} \\
M--$\text{O}_1$ (\AA) & 1.924 & 1.947 & 1.949 & 1.952 & 1.926 \\
 &  & \textit{1.18} & \textit{1.28} & \textit{1.43} & \textit{0.10} \\
$\theta_{1}$ ($^{\circ}$) & 88.56 & 87.91 & 87.93 & 88.02 & 88.49 \\
 &  & \textit{0.73} & \textit{0.71} & \textit{0.61} & \textit{0.09} \\
$\theta_{2}$ ($^{\circ}$) & 163.10 & 159.58 & 159.51 & 159.64 & 162.25 \\
 &  & \textit{2.16} & \textit{2.20} & \textit{2.12} & \textit{0.52} \\
MARE (\%) &  & \textbf{1.20} & \textbf{1.28} & \textbf{3.60} & \textbf{0.12} \\
\end{tabular}
\end{ruledtabular}
\end{table}

\begin{table}[!tpb]
\small
\centering
\caption{Structural parameters for NM LaNiO$_3$. Experimental data measured at 1.5 K is taken from Garc{\'{i}}a-Mu{\~{n}}oz \emph{et al.}\cite{garciamunoz1992} $U_{dd}^{\text{Lit.}}=5.7$ eV\cite{nohara2009}. Relative absolute error is shown shaded in gray (in \%), with the mean absolute relative error (MARE) listed at the bottom of the table. $\theta_1$ and $\theta_2$ describe O--$\hat{\text{Ni}}$--O and Ni--$\hat{\text{O}}$--Ni angles, respectively. ACBN0 calculations are PBE+$U_{dd}$+$U_{pp}$.}
\label{table:structNi}
\begin{ruledtabular}
\begin{tabular}{rcccc}
\multicolumn{1}{c}{} & \multicolumn{4}{c}{\textbf{LaNiO$_3$}} \\
\multicolumn{1}{c}{} & \multicolumn{1}{c}{Expt.} & \multicolumn{1}{c}{PBE} & \multicolumn{1}{c}{PBE+$U_{dd}^{\text{Lit.}}$} & \multicolumn{1}{c}{ACBN0} \\ \cline{2-5}
V (\AA$^3$) & 112.48 & 114.19 & 114.12 & 111.33 \\
 &  & \textit{1.53} & \textit{1.46} & \textit{1.02} \\
$a$ (\AA) & 5.3837 & 5.397 & 5.397 & 5.370 \\
 &  & \textit{0.25} & \textit{0.24} & \textit{0.26} \\
$\alpha$ ($^{\circ}$) & 60.86 & 61.21 & 61.19 & 60.75 \\
 &  & \textit{0.58} & \textit{0.54} & \textit{0.18} \\
M--$\text{O}_1$ (\AA) & 1.933 & 1.950 & 1.949 & 1.925 \\
 &  & \textit{0.88} & \textit{0.83} & \textit{0.41} \\
$\theta_{1}$ ($^{\circ}$) & 88.78 & 88.28 & 88.32 & 88.91 \\
 &  & \textit{0.55} & \textit{0.51} & \textit{0.15} \\
$\theta_{2}$ ($^{\circ}$) & 164.82 & 161.97 & 162.20 & 165.47 \\
 &  & \textit{1.73} & \textit{1.59} & \textit{0.39} \\
MARE (\%) &  & \textbf{0.92} & \textbf{0.86} & \textbf{0.40} \\
\end{tabular}
\end{ruledtabular}
\end{table}

\subsubsection{LaNiO$_{\text{3}}$}
LaNiO$_3$, similarly to LaCoO$_3$, has $R\bar{3}c$ symmetry with GFO-type octahedral tilting. Structural parameters and errors are listed in Table~\ref{table:structNi}. PBE provides a fairly accurate picture of the structure but also similarly to LaCoO$_3$, overestimates the unit cell volume. PBE+$U_{dd}^{\text{Lit.}}=5.7$ eV\cite{nohara2009} provides very marginal improvement in the structure, with a MARE value of 0.86\%. LDA+$U$ results from Gou \emph{et al.} optimized LaNiO$_3$ with an estimated MARE of 0.3\% and an empirical $U$ of 6 eV, thought it should be noted that plain PBE resulted in the best agreement with experimental Raman-active lattice modes and the large value of $U$ destabilized the lattice by introducing imaginary phonon modes\cite{gou2011}. ACBN0 improves the picture without significantly introducing larger errors to any of the structure parameters and yields a MARE of 0.40\%. HSE-25 (HSE-Opt is zero mixing fraction, or plain PBE for this material) yields additional improvement with a MARE of 0.19\%. While the geometry improves with increasing mixing fraction (up to HSE-35 with MARE of 0.1\%), the treatment of the electronic properties worsens, as discussed in the next section.

\begin{figure*}
   \centering
   \includegraphics{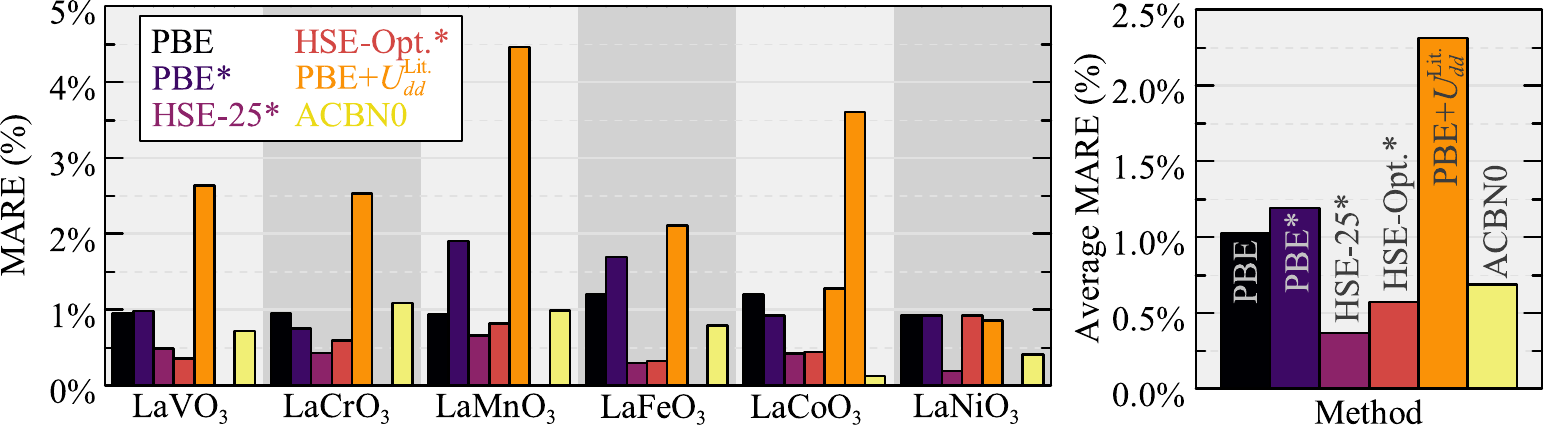}
   \caption{Mean absolute relative error (MARE) of perovskite structural parameters for PBE, PBE+$U_{dd}^{\text{Lit.}}$, HSE (from He and Franchini\cite{he2012}) and ACBN0 (PBE+$U_{dd}$+$U_{pp}$). HSE-25 refers to an exact exchange mixing fraction of 0.25, and HSE-Opt. is an empirically-optimized value to balance the description of both structural and electronic properties.}\label{fig:MAREstruc}
\end{figure*}

\begin{figure*}
   \centering
   \includegraphics{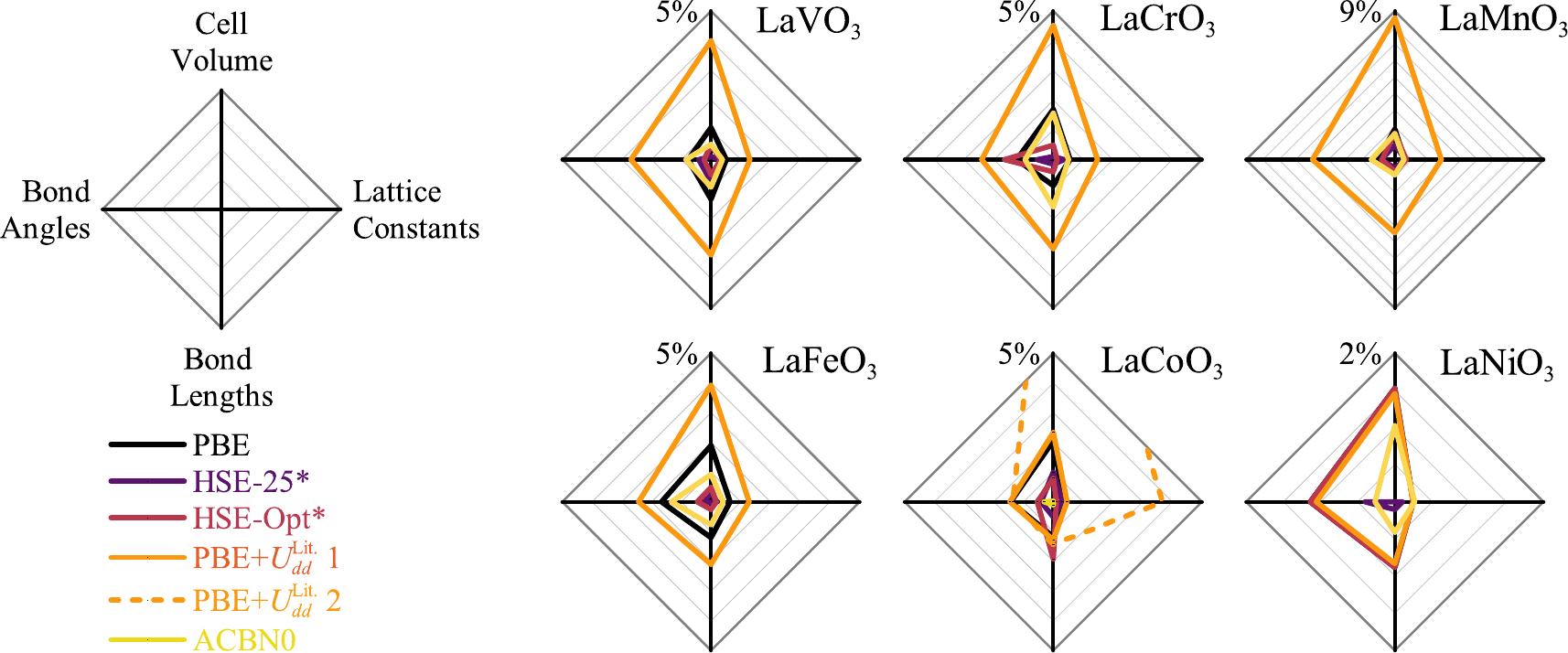}
   \caption{Relative absolute error of detailed subgroups of pervoskite structural parameters for PBE, PBE+$U_{dd}^{\text{Lit.}}$, HSE (from He and Franchini\cite{he2012}) and ACBN0 (PBE+$U_{dd}$+$U_{pp}$). HSE-25 refers to an exact exchange mixing fraction of 0.25, and HSE-Opt. is an empirically-optimized value to balance the description of both structural and electronic properties. For lattice constants, bond lengths, and bond angles, and average error for each structure is used. Cell volume error for LaCoO$_3$ goes beyond the axis limits to improve clarity (12.83\%).}\label{fig:MAREradar}
\end{figure*}

Figures~\ref{fig:MAREstruc} and~\ref{fig:MAREradar} illustrate the results of this section, showing the MARE values for each material and the average MARE for each method, along with detailed radar plots for each material that show each method's relative absolute errors for cell volume, lattice parameters, bond lengths and bond angles. The PBE results agree fairly well with the previously reported PBE calculations of He and Franchini\cite{he2012}, and describe the structures of the 3$d$ LaBO$_3$ perovskites fairly well with an average MARE of around 1\%. Applying the values of $U$ from the literature usually results in a poorly described structure (average MARE 2.3\%), with the exceptions of LaCoO$_3$ and LaNiO$_3$, where accuracy near the level of PBE is obtained. ACBN0 however, applying self-consistent values of $U$ to both metal $3d$ and oxygen $2p$ states, significantly improves the predicted structures with an average MARE of less than 0.7\%. There still are shortcomings when describing some of the early transition metal compounds that have significant JT distortion and orbital ordering, since the structure and electronic properties are so closely intertwined. While the previously-reported HSE-25 and HSE-Opt result in improved structural parameters vs. PBE (average MARE of 0.4\% and 0.6\% respectively), we will see in the next section that this does not necessarily translate to an improved overall picture including electronic properties, which is where the approach of applying $U_{pp}$ along with $U_{dd}$ in a self-consistent approach, like with ACBN0, clearly has benefits.

\subsection{\label{ssec:elec}Electronic Structure}
Bulk electronic structure properties for both experimental and optimized structures are presented in the following sections for each perovskite. Band gaps are determined from the KS density of states (DOS) and compared to experimental values where appropriate (i.e. not in the case of metallic LaNiO$_3$). Magnetic moments (from unit cell absolute magnetization divided by the number of metal cations) are also compared with literature values for magnetically ordered structures. The energies of AFM-A, AFM-C, AFM-G, FM and NM are tabulated relative to whichever magnetic structure is the experimentally-observed ground state. We present projected DOS (PDOS) for our calculations using PBE, ACBN0, and PBE+$U_{dd}^{\text{Lit.}}$. For brevity, only band structures for PBE and ACBN0 are compared; this is in order to present the dispersion of the bands, which is not visible from the PDOS.

\subsubsection{LaVO$_{\text{3}}$}
\begin{figure}[!tpb]
   \centering
   \includegraphics{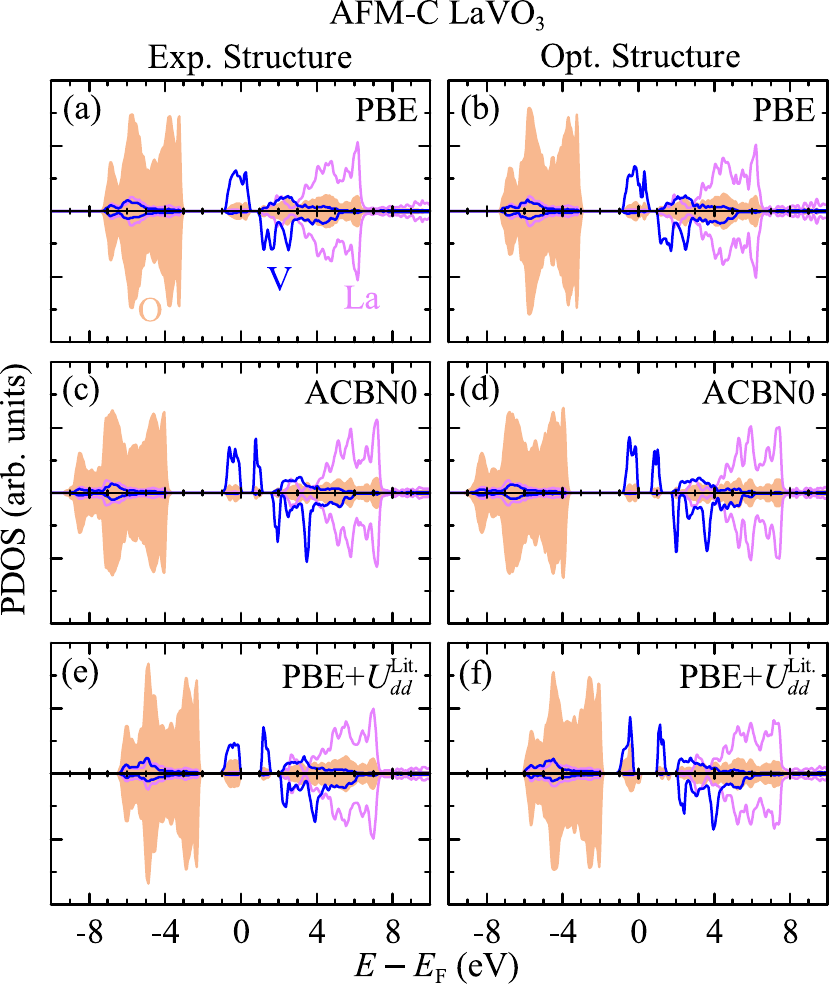}
   \caption{Projected density of states for AFM-C LaVO$_3$ (on the O, V, and La states); \textbf{a.} experimental structure with PBE; \textbf{b.} optimized structure with PBE; \textbf{c.} experimental structure with ACBN0 (PBE+$U_{dd}$+$U_{pp}$); \textbf{d.} optimized structure with ACBN0 (PBE+$U_{dd}$+$U_{pp}$); \textbf{e.} experimental structure with PBE+$U_{dd}^{\text{Lit.}}=3.0$ eV\cite{nohara2009}; \textbf{f.} optimized structure with PBE+$U_{dd}^{\text{Lit.}}=3.0$ eV\cite{nohara2009}.}\label{fig:VPDOS}
\end{figure}

The Mott insulator LaVO$_3$ is not correctly described by plain PBE DFT, which in this work predicts it as a AFM-A metal after geometry optimization. There is also a type-G $t_{2g}$ orbital ordering\cite{sage2007,varignon2015}, which will not be investigated here but may be included in future work. Table~\ref{table:VElec} presents electronic structure parameters for LaVO$_3$, including band gap and magnetic moment compared with experimental values, as well as the relative DFT-calculated energies of several possible magnetic orderings compared to the experimentally-observed AFM-C order\cite{miyasaka2003}. Even with the correct AFM-C ordering, PBE predicts a metallic ground state, as shown in Fig.~\ref{fig:VPDOS}a-b.

\begin{table}[!tpb]
\small
\centering
\caption{Parameters obtained from the electronic structure of LaVO$_3$, including band gap $E_{\text{g}}$, magnetic moment per V cation $\mu$ and the energy difference $\Delta E$ between various calculated magnetic ordering states for PBE, PBE+$U_{dd}^{\text{Lit.}}=3.0$ eV\cite{nohara2009}, and ACBN0 (PBE+$U_{dd}$+$U_{pp}$). Experimental values for band gap and magnetic moment are also provided. }
\label{table:VElec}
\begin{ruledtabular}
\begin{tabular}{rcccc}
 & \multicolumn{4}{c}{\textbf{LaVO}$_{\textbf{3}}$} \\
 & \multicolumn{4}{c}{AFM-C Optimized Structure} \\
 & Expt. & PBE & PBE+$U_{dd}^{\text{Lit.}}$ & ACBN0 \\ \cline{2-5}
$E_{\text{g}}$ (eV) & 1.1\cite{arima1993} & 0.5 & 0.8 & 0.8 \\
$\mu$ ($\mu_{\text{B}}$/V) & 1.3\cite{nguyen1995} & 1.86 & 2.10 & 1.98 \\ \cline{2-5}
 &  & \multicolumn{3}{c}{Relative Energy vs. AFM-C} \\
 &  & \multicolumn{3}{c}{Experimental Structure} \\
 &  & PBE & PBE+$U_{dd}^{\text{Lit.}}$ & ACBN0 \\ \cline{3-5} 
$\Delta E$ (meV) & AFM-A & 42 & -167 & 110 \\
 & AFM-G & 313 & -74 & 215 \\
 & FM & 59 & 94 & 324 \\
 & NM & 1547 & 5146 & 4264 \\ \cline{3-5}
 &  & \multicolumn{3}{c}{Optimized Structure} \\
 &  & PBE & PBE+$U_{dd}^{\text{Lit.}}$ & ACBN0 \\ \cline{3-5}
$\Delta E$ (meV) & AFM-A & -66 & 97 & 94 \\
 & AFM-G & 298 & 69 & 46 \\
 & FM & 21 & 20 & 44 \\
 & NM & 1326 & 5457 & 4096
\end{tabular}
\end{ruledtabular}
\end{table}

\begin{figure}[!tpb]
   \centering
   \includegraphics{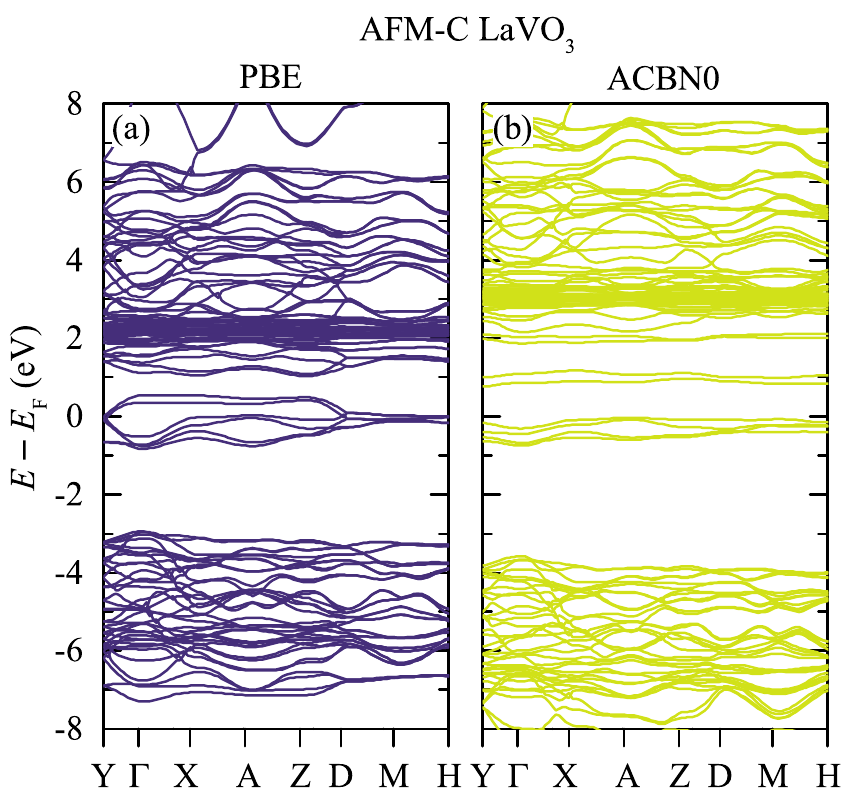}
   \caption{Band structure of AFM-C LaVO$_3$; \textbf{a.} PBE optimized structure; \textbf{b.} ACBN0 (PBE+$U_{dd}$+$U_{pp}$) optimized structure.}\label{fig:VBands}
\end{figure}

ACBN0 predicts the correct AFM-C ground state and also provides a very good estimate of the experimentally-observed band gap: a predicted 0.8 eV compared to the observed 1.1 eV\cite{arima1993}, introducing a gap between the occupied and unoccupied $t_{2g}$ states. This can also easily be seen in Fig.~\ref{fig:VPDOS}c-d and in the band structure of Fig.~\ref{fig:VBands}. The PBE+$U_{dd}^{\text{Lit.}}$ result and the HSE results of He and Franchini also result in a correct ground state, with the latter giving a slightly larger estimate of the band gap for HSE-Opt (1.46 eV). HSE-25 predicted a rather large value of 2.43 eV. Magnetic moments for ACBN0 and PBE+$U_{dd}^{\text{Lit.}}$ slightly overestimate the moment compared to PBE, which is also larger than experiment. This is a common error in hybrid functionals as well. Another important feature to notice is the charge transfer (CT) gap, or the difference between the predominately oxygen-derived lower valence band and the unoccupied conduction band of mostly $d$ parentage. Experimentally the value is reported to be 4.0 eV\cite{arima1993}, but PBE+$U_{dd}^{\text{Lit.}}$ predicts a smaller gap of approximately 3 eV and a higher mixing of O $2p$ and V $3d$ in the valence band vs ACBN0 and the previously reported HSE results. ACBN0 predicts a value near 4.2 eV, while HSE-Opt overestimates the experimental value, giving 4.9 eV. An additional empirical adjustment, HSE-10, can reduce this to 4.4 eV and gives a Mott-Hubbard (MH) gap of 0.89 eV. While both PBE+$U_{dd}^{\text{Lit.}}$ and ACBN0 give similar band gaps, the additional push of valence band oxygen $2p$ states to lower energy from the $U_{pp}$ term in ACBN0 opens up the CT gap to a value agreeing closer to experiment and widens the band dispersion of the valence band.

\subsubsection{LaCrO$_{\text{3}}$}
\begin{figure}[!tpb]
   \centering
   \includegraphics{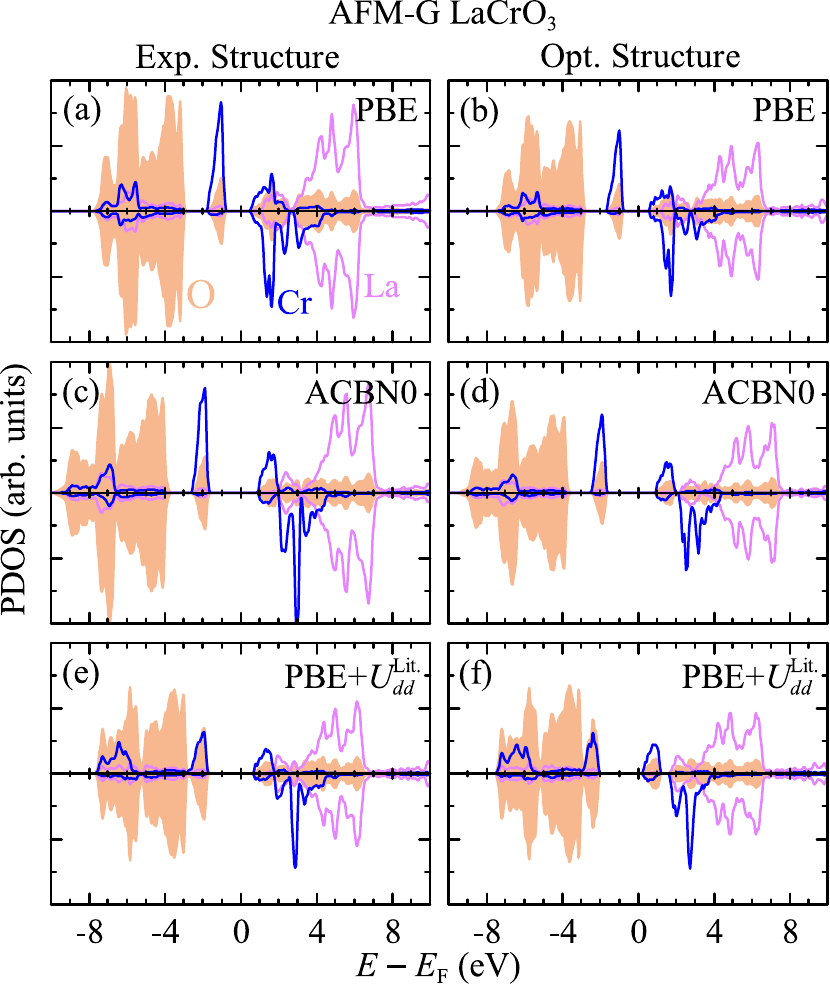}
   \caption{Projected density of states for AFM-G LaCrO$_3$ (on the O, Cr, and La states); \textbf{a.} experimental structure with PBE; \textbf{b.} optimized structure with PBE; \textbf{c.} experimental structure with ACBN0 (PBE+$U_{dd}$+$U_{pp}$); \textbf{d.} optimized structure with ACBN0 (PBE+$U_{dd}$+$U_{pp}$); \textbf{e.} experimental structure with PBE+$U_{dd}^{\text{Lit.}}=4.1$ eV\cite{nohara2009}; \textbf{f.} optimized structure with PBE+$U_{dd}^{\text{Lit.}}=4.1$ eV\cite{nohara2009}.}\label{fig:CrPDOS}
\end{figure}

Table~\ref{table:CrElec} presents electronic structure parameters of AFM-G LaCrO$_3$, an AFM insulator with an optical band gap of 3.4 eV as reported by Arima \emph{et al}\cite{arima1993}. They note in this early work that the weaker MH transition is completely indiscernible due to the stronger CT transition, meaning the two gaps are nearly equal in width or correspond to the same gap, with significant Cr $3d$--O $2p$ hybridization in the valence band. From the PBE+$U_{dd}^{\text{Lit.}}=4.1$ eV\cite{nohara2009} calculations shown in Fig.~\ref{fig:CrPDOS}f, this would seem to be a reasonable picture. Large values of exact exchange ($<$ 0.25) also lead to increased Cr--O hybridization in the valence band, although within this early interpretation of the optical data HSE-15 still provided the best overall picture of LaCrO$_3$\cite{he2012}, with a MH gap near 3.0 eV. However, this does not account for an important experimental observation--the green color of LaCrO$_3$, which would require a gap in the optical range. 

\begin{table}[!tpb]
\small
\centering
\caption{Parameters obtained from the electronic structure of LaCrO$_3$, including band gap $E_{\text{g}}$, magnetic moment per Cr cation $\mu$ and the energy difference $\Delta E$ between various calculated magnetic ordering states for PBE, PBE+$U_{dd}^{\text{Lit.}}=4.1$ eV\cite{nohara2009}, and ACBN0 (PBE+$U_{dd}$+$U_{pp}$). Experimental values for band gap and magnetic moment are also provided. The band gap in brackets corresponds to a more recent interpretation of optical data\cite{sushko2013}.}
\label{table:CrElec}
\begin{ruledtabular}
\begin{tabular}{rcccc}
 & \multicolumn{4}{c}{\textbf{LaCrO}$_{\textbf{3}}$} \\
 & \multicolumn{4}{c}{AFM-G Optimized Structure} \\
 & Expt. & PBE & PBE+$U_{dd}^{\text{Lit.}}$ & ACBN0 \\ \cline{2-5}
$E_{\text{g}}$ (eV) & 3.4\cite{arima1993} (2.4)\cite{sushko2013} & 1.5 & 2.3 & 2.7 \\
$\mu$ ($\mu_{\text{B}}$/Cr) & 2.45-2.8\cite{koehler1957,bertaut1966,sakai1996} & 2.86 & 3.17 & 3.04 \\ \cline{2-5}
 &  & \multicolumn{3}{c}{Relative Energy vs. AFM-G} \\
 &  & \multicolumn{3}{c}{Experimental Structure} \\
 &  & PBE & PBE+$U_{dd}^{\text{Lit.}}$ & ACBN0 \\ \cline{3-5} 
$\Delta E$ (meV) & AFM-A & 324 & 157 & 147 \\
 & AFM-C & 146 & 76 & 71 \\
 & FM & 519 & 250 & 233 \\
 & NM & 4839 & 10979 & 10641 \\ \cline{3-5}
 &  & \multicolumn{3}{c}{Optimized Structure} \\
 &  & PBE & PBE+$U_{dd}^{\text{Lit.}}$ & ACBN0 \\ \cline{3-5}
$\Delta E$ (meV) & AFM-A & 247 & 60 & 97 \\
 & AFM-C & 128 & 30 & 45 \\
 & FM & 388 & 90 & 155 \\
 & NM & 4818 & 11179 & 10379
\end{tabular}
\end{ruledtabular}
\end{table}

\begin{figure}[!tpb]
   \centering
   \includegraphics{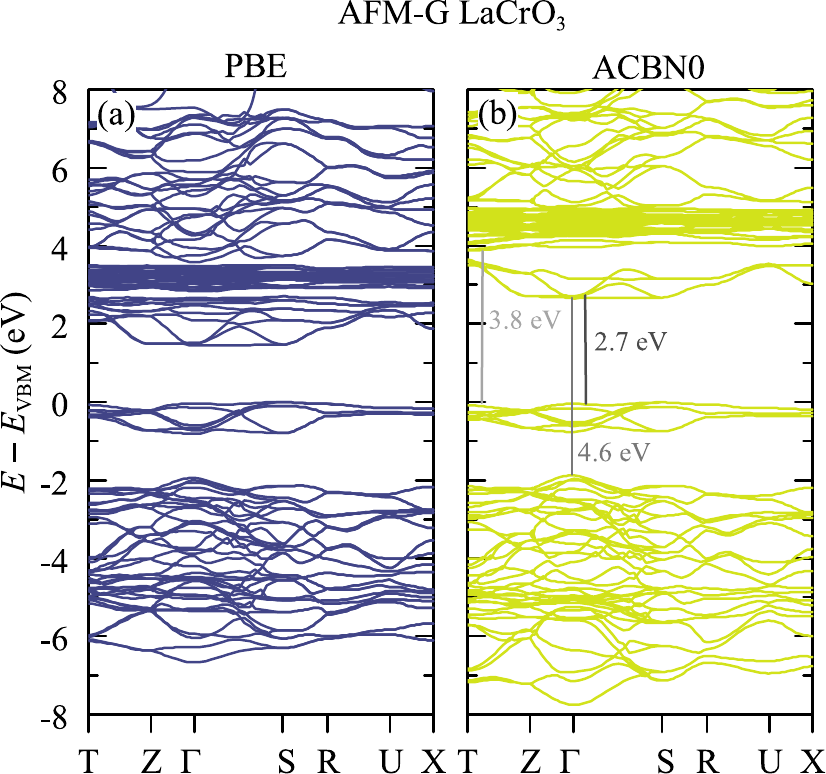}
   \caption{Band structure of AFM-G LaCrO$_3$; \textbf{a.} PBE optimized structure; \textbf{b.} ACBN0 (PBE+$U_{dd}$+$U_{pp}$) optimized structure.}\label{fig:CrBands}
\end{figure}

He and Franchini also mention a study by Ong \emph{et al.}\cite{ong2008} that interprets the electronic structure in a different way. They applied empirical $U_{dd}$ corrections of 2.72, 5.44 and 8.16 eV (the ACBN0-calculated value of $U_{dd}$ is 2.77 eV for Cr, as shown in the Supplemental Material\cite{supp}) to compare the simulated valence band PDOS to experimental X-ray photoemission spectroscopy (XPS) spectra. They found that applying any $U_{dd}$ correction both worsened their comparison with the experimental XPS spectra, and resulted in no features in the optical range near green light in simulated reflectivity spectra. The implication is that the CT and MH gaps remain distinct, and two separate transitions are present: the larger CT gap of 3.4 eV is responsible for the previous experimental measurements, while the smaller MH gap near 2.2 eV explains the green color of LaCrO$_3$ and the corresponding peaks in reflectivity measurements. Ong \emph{et al.} conclude that there are no strong electronic correlations in LaCrO$_3$, and that plain GGA is the most appropriate for describing this material. If one takes this interpretation as correct, it would appear at first glance that ACBN0, PBE+$U_{dd}^{\text{Lit.}}$ and HSE all do not describe this material correctly. ACBN0 and HSE result in CT gaps near 4.5 eV, while PBE results in an CT gap of near 3.4 eV. However, ACBN0 still results in a MH gap of near 2.6 eV, while HSE and PBE result in MH gaps of 3 eV and 1.4 eV, respectively (see Figs.~\ref{fig:CrPDOS} and \ref{fig:CrBands}). Ong \emph{et al.} mentioned that previous experimental spectra did not include features in the optical range, and suggested further experimental studies to find their predicted green light optical absorption.

At the time there was no additional experimental evidence clarifying the electronic structure of LaCrO$_3$, but in 2013 Sushko \emph{et al.}\cite{sushko2013} reported experimental measurements coupled with embedded cluster time-dependent DFT that discerned the multiple optical transitions present in this material. Spectroscopic ellipsometry revealed onset of absorption features near 2.3 eV and 3.2 eV, occurring before a large 5 eV optical absorption onset. They attributed the absorption features to families of $t_{2g}$--$e_g$, $t_{2g}$--$t_{2g}$, and Cr $3d$--O $2p$ transitions and conclude that the true CT gap is near $\sim$5 eV, not 3.4 eV like Ong \emph{et al.} suggested, while the green absorption feature (onset at $\sim$2.4 eV, centered near $\sim$2.7 eV) is due to $t_{2g}$--$e_g$ fundamental gap transitions and the previously-reported 3.4 eV gap is due to inter-Cr $t_{2g}$--$t_{2g}$ transitions. The band structure in Fig.~\ref{fig:CrBands} illustrates these transitions with lines of appropriate energy superimposed over the band structure. This is more in line with trends in the CT gap from X-ray spectroscopy experiments\cite{hong2015}, where the gaps are quite large since they are calculated from peak positions rather than band edges ($\sim$7.2 eV for LaCrO$_3$, and larger than the MH band gap) and generally decrease with increasing $d$ occupation. It is worth mentioning that for ACBN0, the spacing of the spin-down $t_{2g}$ peak and O $2p$ valence band peak is quite close to 7 eV, owing in part to $U_{pp}$ forcing the oxygen $2p$ states lower in energy. Further comparison of our calculations with experimental spectra are available in the Supplemental Material\cite{supp}. While ground state DFT strictly does not describe transition energies, the ACBN0 results generally support this picture in terms of the gaps and types of PDOS features present, in contrast to those of PBE+$U_{dd}^{\text{Lit.}}$, HSE-25 and HSE-Opt (HSE-10 provides a fairly similar picture to ACBN0). This alternative picture significantly affects the band gap error, shown in Fig.~\ref{fig:bandMARE}, bringing it more in line with the rest of the perovskites.

As shown in Table~\ref{table:CrElec}, all the methods used in this study, as well as the HSE results from He and Franchini, correctly predict the AFM-G magnetic ordering for LaCrO$_3$\cite{bertaut1966}. Magnetic moments are overestimated slightly by PBE, and further overestimated by ACBN0 and PBE+$U_{dd}^{\text{Lit.}}$ (although ACBN0 does to a lesser degree). Band structures for the PBE and ACBN0 optimized structures are shown in Fig.~\ref{fig:CrBands}.

\subsubsection{LaMnO$_{\text{3}}$}
LaMnO$_3$ is an type-A AFM MH insulator with significant JT distortions and $e_g$ orbital ordering\cite{murakami1998}. All the methods used in this work incorrectly predict a metallic FM ground state when the geometry and unit cell are optimized; in addition, only PBE predicts the correct AFM-A ground state when the experimental structure is used (summarized in Table~\ref{table:MnElec}). This illustrates the particular importance of the JT distortions in the existence of a band gap in this material. While there have been reports of DFT+$U$ both improving and worsening\cite{hashimoto2010} the structural and electronic properties of LaMnO$_3$, it is clear that in an orbitally-ordered material and/or where the $e_g$ and $t_{2g}$ bands exhibit markedly different localized or itinerant behavior, that the averaging used in both calculating and applying $U$ corrections in most commonly used implementations is likely inappropriate, and improvements from such treatments are fortuitous. This is especially true for the widely-used simplified rotationally-invariant implementation of DFT+$U$ which considers the exchange interaction $J$ as isotropic\cite{dudarev1998}. 

\begin{figure}[!tpb]
   \centering
   \includegraphics{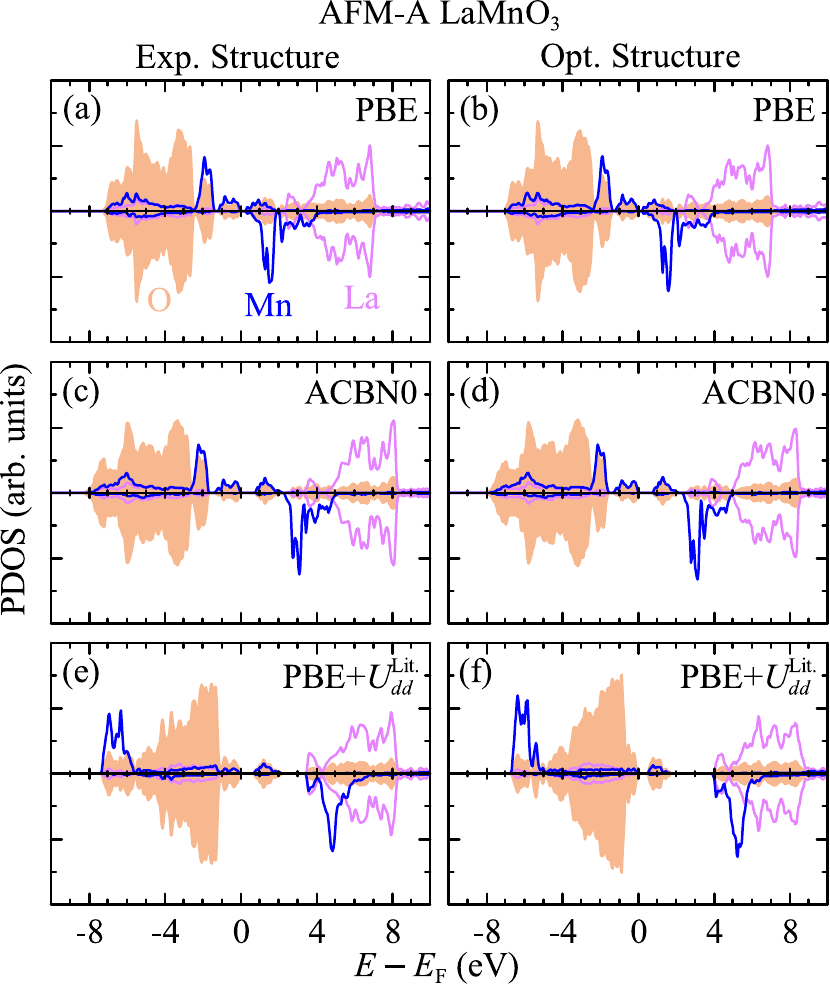}
   \caption{Projected density of states for AFM-A LaMnO$_3$ (on the O, Mn, and La states); \textbf{a.} experimental structure with PBE; \textbf{b.} optimized structure with PBE; \textbf{c.} experimental structure with ACBN0 (PBE+$U_{dd}$+$U_{pp}$); \textbf{d.} optimized structure with ACBN0 (PBE+$U_{dd}$+$U_{pp}$); \textbf{e.} experimental structure with PBE+$U_{dd}^{\text{Lit.}}=6.4$ eV\cite{nohara2009}; \textbf{f.} optimized structure with PBE+$U_{dd}^{\text{Lit.}}=6.4$ eV\cite{nohara2009}.}\label{fig:MnPDOS1}
\end{figure}

\begin{figure}[!tpb]
   \centering
   \includegraphics{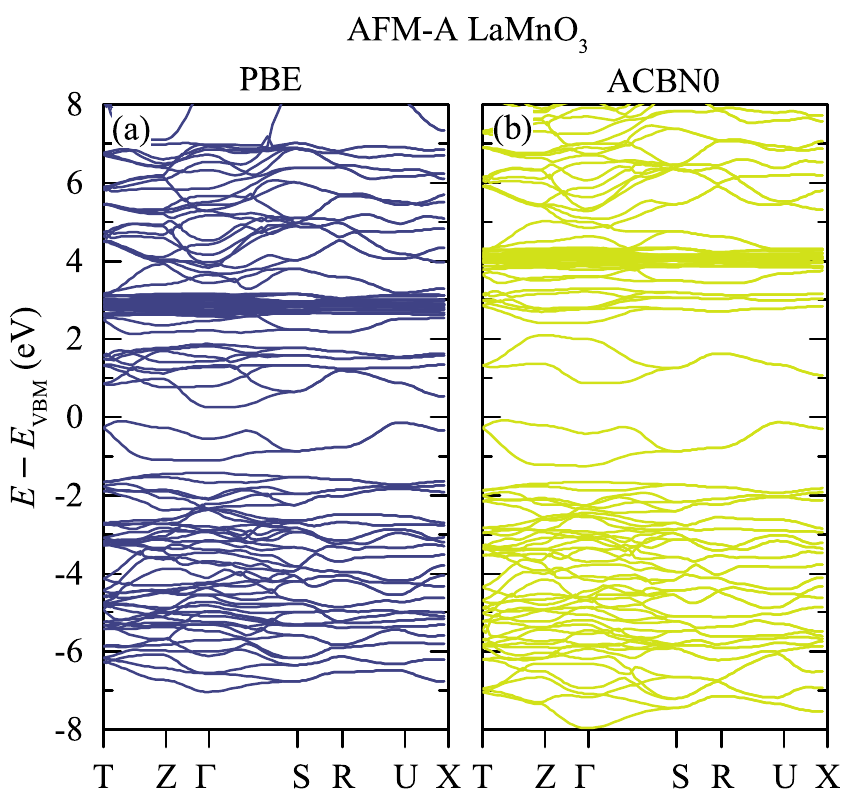}
   \caption{Band structure of AFM-A LaMnO$_3$; \textbf{a.} PBE optimized structure; \textbf{b.} ACBN0 (PBE+$U_{dd}$+$U_{pp}$) optimized structure.}\label{fig:MnBands1}
\end{figure}

\begin{figure}[!tpb]
   \centering
   \includegraphics{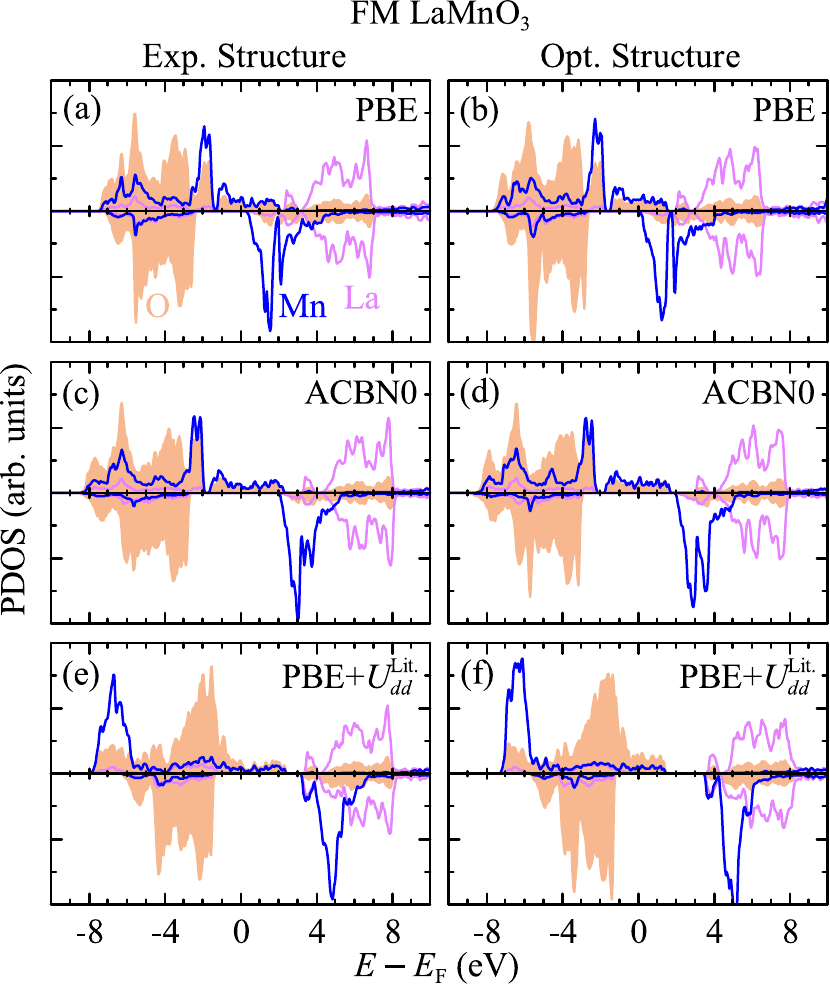}
   \caption{Projected density of states for FM LaMnO$_3$ (on the O, Mn, and La states); \textbf{a.} experimental structure with PBE; \textbf{b.} optimized structure with PBE; \textbf{c.} experimental structure with ACBN0 (PBE+$U_{dd}$+$U_{pp}$); \textbf{d.} optimized structure with ACBN0 (PBE+$U_{dd}$+$U_{pp}$); \textbf{e.} experimental structure with PBE+$U_{dd}^{\text{Lit.}}=6.4$ eV\cite{nohara2009}; \textbf{f.} optimized structure with PBE+$U_{dd}^{\text{Lit.}}=6.4$ eV\cite{nohara2009}.}\label{fig:MnPDOS2}
\end{figure}

\begin{figure}[!tpb]
   \centering
   \includegraphics{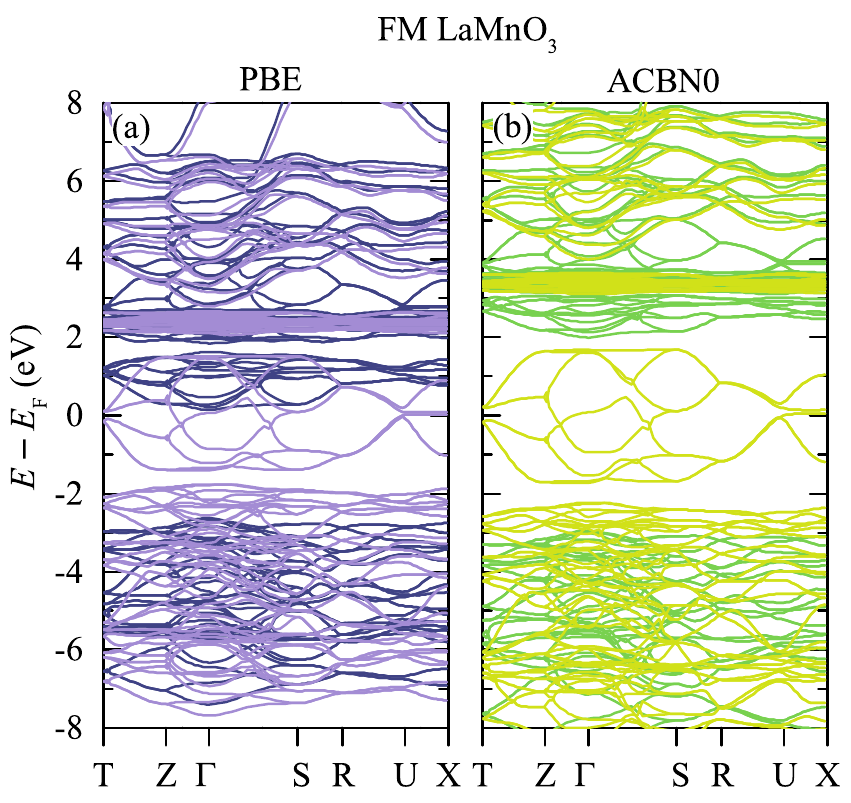}
   \caption{Band structure of FM LaMnO$_3$; \textbf{a.} PBE optimized structure; \textbf{b.} ACBN0 (PBE+$U_{dd}$+$U_{pp}$) optimized structure. Lighter colors correspond to spin up, while darker colors correspond to spin down.}\label{fig:MnBands2}
\end{figure}

Although in the DFT+$U$ implementation used in this work ACBN0 does not predict the correct ground state, for the AFM-A state it yields an accurate band gap of 1.0 eV, with the $e_g$ bands being isolated from the other bands (see Fig.~\ref{fig:MnBands1}), as reported in the HSE study of He and Franchini. Compare this with the ferromagnetic (FM) band structure presented in Fig.~\ref{fig:MnBands2}, where the most striking difference is the change in these bands near the Fermi level. For AFM-A LaMnO$_3$, HSE-25 grossly overestimates the band gap (2.47 eV) and HSE-Opt gives a reasonable value of 1.63 eV. PBE+$U_{dd}^{\text{Lit.}}$ highlights the previously mentioned failures of the simplified DFT+$U$ implementation for LaMnO$_3$ by giving a band gap of only 0.6 eV for $U=6.4$ eV\cite{nohara2009}. More notably, the spin-up $t_{2g}$ states are pushed down below the oxygen valence band, in contrast to ACBN0 (see Fig.~\ref{fig:MnPDOS1}) and the hybrid functional results (for all mixing fractions). The use of $U_{pp}$ in ACBN0 allows the opening of the MH gap to experimental values without especially large values of $U_{dd}$, and the lowering of O $2p$ energy preserves the general picture of bonding given by PBE and HSE calculations. These trends are also true in the FM case, shown in Fig.~\ref{fig:MnPDOS2}. The magnetic moment is again slightly overestimated by ACBN0 and PBE+$U_{dd}^{\text{Lit.}}$, while PBE gives a value at the upper end of the experimental range. LaMnO$_3$ is a widely-studied material and further discussion can be found in the literature\cite{franchini2012,mellan2015,sawada1998,he2012b}.

Owing to the difficulty in comparing not only self-consistent values of $U$ but also the results of calculations that use different computational parameters (pseudopotential, XC functional, DFT+$U$ implementation, etc.) it is challenging to make definitive statements regarding the treatment of the ground state of LaMnO$_3$ within DFT and DFT+$U$. It has been shown that explicit inclusion of orbital-dependent $J$ corrections, as in the original rotationally-invariant scheme by Liechtenstein \emph{et al.}\cite{liechtenstein1995}, is necessary for stabilizing the experimentally observed AFM-A magnetic ordering and reproducing the $e_g$ orbital ordering in LaMnO$_3$, at least when using LDA and PBEsol (GGA) functionals and computational parameters utilized in that work\cite{mellan2015}. Empirical values of $U=8$ eV and $J=2$ eV in that work yielded a good overall description of correct ground state, band gap, and structural properties. Other work using using PBE with explicit $U=2.7$ eV and $J=1.0$ (giving $U_{\textrm{eff}}=1.7$ eV, close in value to the ACBN0 values calculated in this work for Mn) has also reproduced this trend\cite{lee2013}. On the other hand, similar calculations using different DFT code with a simplified $U_{\textrm{eff}}=2.0$ eV may still give the correct AFM-A ground state in some circumstances\cite{hashimoto2010}.

As an aside, we perform some simple test calculations with explicit $U$ and $J$ on $d$ and $p$ states using the ACBN0-calculated values for both Mn and O (still calculated with the Dudarev implementation: $U = 3.62$ eV, $J = 1.18$ eV for Mn $3d$ and $U=12.185$ eV, $J=6.10$ eV for O $2p$) yields an energy difference of 0.0002 eV, compared to the value of 0.030 eV in Table~\ref{table:MnElec}. If we keep the ACBN0 correction on oxygen and increase $U$ and $J$ on Mn to 6.0 eV and 2.0 eV, respectively, closer to the values of Mellan \emph{et al.}\cite{mellan2015}, the AFM-A ordering is stabilized with the DFT+$U$ correction, with only a subtle push of Mn $d$ states deeper into the O $2p$ deep valence band with the larger values of $U$ and $J$ (shown in Supplemental Material\cite{supp}). Applying these same $U$ values within the simplified scheme does not stabilize the correct AFM-A ground state. Unfortunately, unit cell stress and pressure are not easily implemented in the generalized DFT+$U$ scheme, so only the calculations using the experimental structure has been performed for this additional comparison. Therefore, the inability of ACBN0 to improve geometry and electronic structure in this work may potentially be determined by the implementation of DFT+$U$ rather than the ACBN0 approach itself, leaving room for future improvement.

\begin{table}[!tpb]
\small
\centering
\caption{Parameters obtained from the electronic structure of LaMnO$_3$, including band gap $E_{\text{g}}$, magnetic moment per Mn cation $\mu$ and the energy difference $\Delta E$ between various calculated magnetic ordering states for PBE, PBE+$U_{dd}^{\text{Lit.}}=6.4$ eV\cite{nohara2009}, and ACBN0 (PBE+$U_{dd}$+$U_{pp}$). Experimental values for band gap and magnetic moment are also provided.}
\label{table:MnElec}
\begin{ruledtabular}
\begin{tabular}{rcccc}
 & \multicolumn{4}{c}{\textbf{LaMnO}$_{\textbf{3}}$} \\
 & \multicolumn{4}{c}{AFM-A Optimized Structure} \\
 & Expt. & PBE & PBE+$U_{dd}^{\text{Lit.}}$ & ACBN0 \\ \cline{2-5}
$E_{\text{g}}$ (eV) & 1.1-2.0\cite{arima1993,saitoh1995,jung1997,jung1998,kruger2004} & 0.4 & 0.6 & 1.0 \\
$\mu$ ($\mu_{\text{B}}$/Mn) & 3.4-3.9\cite{elemans1971,hauback1996,moussa1996} & 3.94 & 4.72 & 4.11 \\ \cline{2-5}
 &  & \multicolumn{3}{c}{Relative Energy vs. AFM-A} \\
 &  & \multicolumn{3}{c}{Experimental Structure} \\
 &  & PBE & PBE+$U_{dd}^{\text{Lit.}}$ & ACBN0 \\ \cline{3-5} 
$\Delta E$ (meV) & AFM-C & 277 & 415 & 309 \\
 & AFM-G & 293 & 597 & 391 \\
 & FM & 54 & -169 & -30 \\
 & NM & 6400 & 14172 & 12285 \\ \cline{3-5}
 &  & \multicolumn{3}{c}{Optimized Structure} \\
 &  & PBE & PBE+$U_{dd}^{\text{Lit.}}$ & ACBN0 \\ \cline{3-5}
$\Delta E$ (meV) & AFM-C & 213 & 279 & 241 \\
 & AFM-G & 168 & 422 & 268 \\
 & FM & -175 & -369 & -199 \\
 & NM & 5079 & 18053 & 10614
\end{tabular}
\end{ruledtabular}
\end{table}

\subsubsection{LaFeO$_{\text{3}}$}
\begin{figure}[!tpb]
   \centering
   \includegraphics{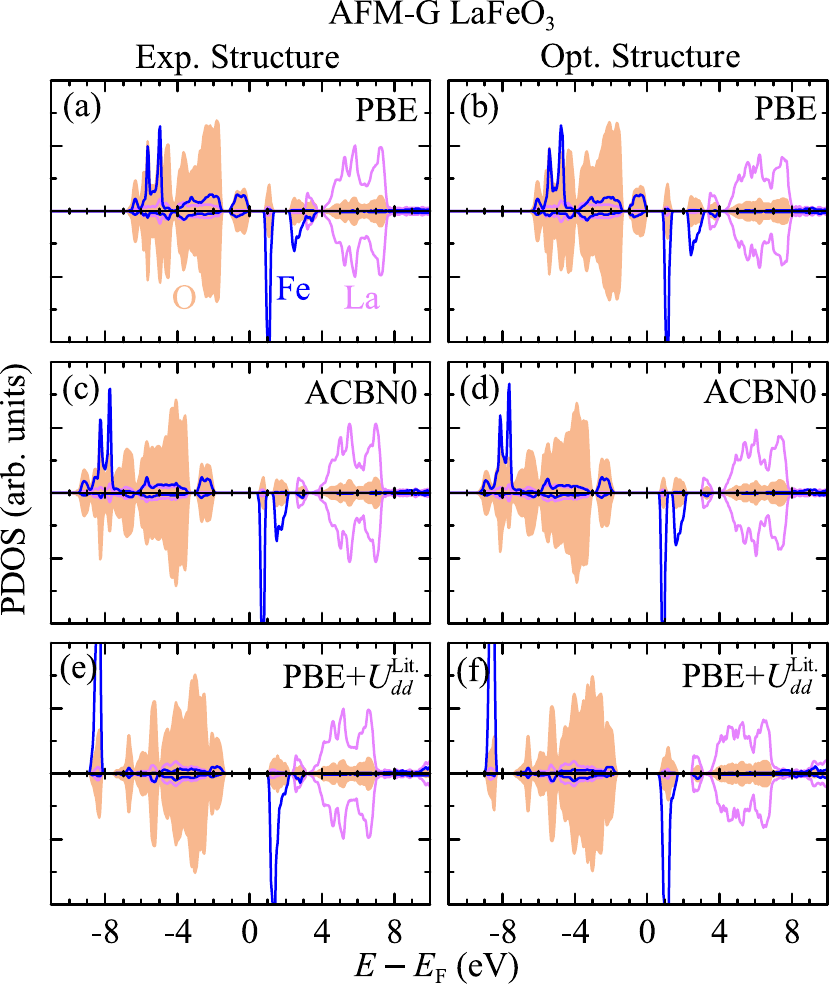}
   \caption{Projected density of states for AFM-G LaFeO$_3$ (on the O, Fe, and La states); \textbf{a.} experimental structure with PBE; \textbf{b.} optimized structure with PBE; \textbf{c.} experimental structure with ACBN0 (PBE+$U_{dd}$+$U_{pp}$); \textbf{d.} optimized structure with ACBN0 (PBE+$U_{dd}$+$U_{pp}$); \textbf{e.} experimental structure with PBE+$U_{dd}^{\text{Lit.}}=4.8$ eV\cite{nohara2009}; \textbf{f.} optimized structure with PBE+$U_{dd}^{\text{Lit.}}=4.8$ eV\cite{nohara2009}.}\label{fig:FePDOS}
\end{figure}

\begin{figure}[!tpb]
   \centering
   \includegraphics{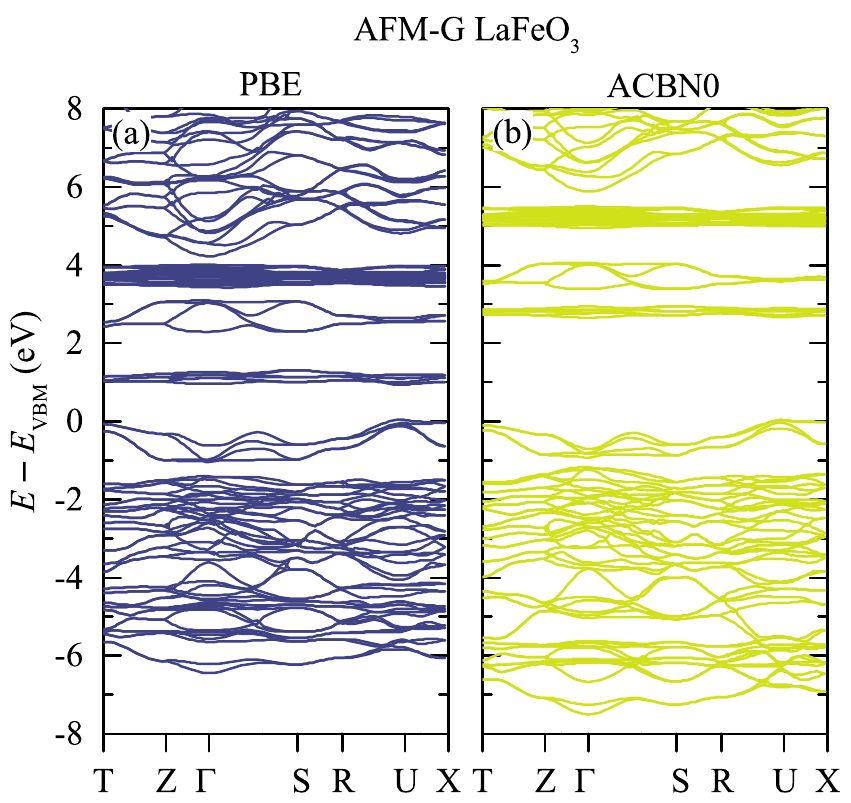}
   \caption{Band structure of AFM-G LaFeO$_3$; \textbf{a.} PBE optimized structure; \textbf{b.} ACBN0 (PBE+$U_{dd}$+$U_{pp}$) optimized structure.}\label{fig:FeBands}
\end{figure}

LaFeO$_3$ is often considered an intermediate CT/MH insulator\cite{arima1993}, owing to the considerable O $2p$ character in the $e_g$ valence band. This material exhibits AFM-G magnetic ordering\cite{koehler1957} with a band gap of 2.3 eV\cite{may2015,scafetta2014,comes2016}. All methods used in this work correctly predict the AFM-G ground state, which is much lower in energy than the other magnetic orderings listed in Table~\ref{table:FeElec}. The projected densities of states for PBE, ACBN0 and PBE+$U_{dd}^{\text{Lit.}}=4.8$ eV\cite{nohara2009} in both experimental and optimized structures are shown in Fig.~\ref{fig:FePDOS}, with band structures for the optimized structures of PBE and ACBN0 shown in Fig.~\ref{fig:FeBands}. 

\begin{table}[!tpb]
\small
\centering
\caption{Parameters obtained from the electronic structure of LaFeO$_3$, including band gap $E_{\text{g}}$, magnetic moment per Fe cation $\mu$ and the energy difference $\Delta E$ between various calculated magnetic ordering states for PBE, PBE+$U_{dd}^{\text{Lit.}}=4.8$ eV\cite{nohara2009}, and ACBN0 (PBE+$U_{dd}$+$U_{pp}$). Experimental values for band gap and magnetic moment are also provided.}
\label{table:FeElec}
\begin{ruledtabular}
\begin{tabular}{rcccc}
 & \multicolumn{4}{c}{\textbf{LaFeO}$_{\textbf{3}}$} \\
 & \multicolumn{4}{c}{AFM-G Optimized Structure} \\
 & Expt. & PBE & PBE+$U_{dd}^{\text{Lit.}}$ & ACBN0 \\ \cline{2-5}
$E_{\text{g}}$ (eV) & 2.3\cite{may2015} & 0.9 & 2.5 & 2.6 \\
$\mu$ ($\mu_{\text{B}}$/Fe) & 3.9, 4.6\cite{koehler1957,zhou2006,bellakki2010} & 4.10 & 4.34 & 4.43 \\ \cline{2-5}
 &  & \multicolumn{3}{c}{Relative Energy vs. AFM-G} \\
 &  & \multicolumn{3}{c}{Experimental Structure} \\
 &  & PBE & PBE+$U_{dd}^{\text{Lit.}}$ & ACBN0 \\ \cline{3-5} 
$\Delta E$ (meV) & AFM-A & 698 & 853 & 687 \\
 & AFM-C & 419 & 363 & 300 \\
 & FM & 838 & 1319 & 1064 \\
 & NM & 4418 & 10352 & 10010 \\ \cline{3-5}
 &  & \multicolumn{3}{c}{Optimized Structure} \\
 &  & PBE & PBE+$U_{dd}^{\text{Lit.}}$ & ACBN0 \\ \cline{3-5}
$\Delta E$ (meV) & AFM-A & 715 & 686 & 602 \\
 & AFM-C & 385 & 303 & 284 \\
 & FM & 921 & 1059 & 929 \\
 & NM & 3866 & 10121 & 9439
\end{tabular}
\end{ruledtabular}
\end{table}

PBE gives a qualitatively correct picture of the electronic structure but underestimates the band gap at 0.9 eV. ACBN0 and PBE+$U_{dd}^{\text{Lit.}}$ both give band gaps much closer to experiment at 2.6 and 2.5 eV, respectively. The band structures of PBE and ACBN0 are both qualitatively similar, with the states above the valence band maximum being more or less simply shifted upwards in energy. It is important to note the differences in the PDOS of ACBN0 and PBE+$U_{dd}^{\text{Lit.}}$. ACBN0 produces a picture similar to that of PBE, except for the band gap; the valence band retains significant Fe--O hybridization and remains separate from the deeper O $2p$ valence band, also giving a similar picture to the HSE-Opt results of He and Franchini\cite{he2012} both quantitatively and qualitatively. This also puts it in good agreement with the photoemission data of Wadati \emph{et al.}\cite{wadati2005} which was compared with the HSE results (see Supplemental Material\ref{supp}). In contrast, despite the fairly accurate band gap, the PBE+$U_{dd}^{\text{Lit.}}$ calculation in this work results in an electronic structure with significantly reduced hybridization, similar to the higher mixing fraction HSE calculations (HSE-35) by He and Franchini. The Fe $e_g$ parentage of the valence band is reduced, the valence band merges with the larger oxygen-derived valence band and occupied $t_{2g}$ states are pushed outside the band width of the oxygen valence band, leading to a much more localized, ionic picture that does not agree with the aforementioned experimental spectroscopic data. While the same trend of increasing magnetic moment with $U$ correction (also with hybrid functionals) continues with LaFeO$_3$, the larger variation in the reported experimental values in Table~\ref{table:FeElec} makes it difficult to make any claims about their accuracy.

\subsubsection{LaCoO$_{\text{3}}$}
A diamagnetic insulator (low-spin Co), LaCoO$_3$ is not well-described by plain DFT, which predicts a ferromagnetic metallic ground state. There still is no conclusive understanding of the higher temperature magnetic behavior of LaCoO$_3$, and a discussion of that topic is beyond the scope of this work. It has been reported in the literature that DFT+$U$ is at least \emph{capable} of stabilizing the correct low-spin insulating state, with $U$ values either being varied empirically\cite{yang1999} or calculated from first principles\cite{hsu2009,laref2010a,laref2010b,laref2012,ritzmann2014}. The values of $U$ themselves include an empirical $U_{\text{eff}}=6.5-0.65=5.85$ eV\cite{yang1999}, $U=3.3$ eV from empirical fitting to enthalpy of formation, $U=4.2$ eV from cluster-CI calculations fitted to experimental X-ray spectra\cite{nohara2009}, LR $U$ typically in the range of 7.8-8.5 eV\cite{hsu2009,laref2010a}, $U_{\text{eff}}=6.88,6.96$ eV calculated from constrained LDA\cite{solovyev1996,nohara2009}, and a renormalized $U=4.0$ eV from unrestricted HF (uHF) orbitals\cite{ritzmann2014} (a method from which ACBN0 takes inspiration). The ACBN0 $U_{dd}$ value on Co of $\sim$3.4 eV is in best agreement with the screened GW, cluster-CI and explicit Coulomb/exchange integrals from uHF, and PBE+$U_{dd}^{\text{Lit.}}$ values of both 4.2 eV\cite{nohara2009} and 8.5 eV\cite{hsu2009} are used for comparison.

\begin{table}[!tpb]
\small
\centering
\caption{Parameters obtained from the electronic structure of LaCoO$_3$, including band gap $E_{\text{g}}$ and the energy difference $\Delta E$ between various calculated magnetic ordering states for PBE, PBE+$U_{dd}^{\text{Lit.}}$ values of 4.2 eV\cite{nohara2009} and 8.5 eV\cite{hsu2009}, and ACBN0 (PBE+$U_{dd}$+$U_{pp}$). Experimental values for band gap are also provided. Gap of ``m'' refers to a metallic system.}
\label{table:CoElec}
\begin{ruledtabular}
\begin{tabular}{rccccc}
 & \multicolumn{5}{c}{\textbf{LaCoO}$_{\textbf{3}}$} \\
 & \multicolumn{5}{c}{NM Optimized Structure} \\
 & Expt. & PBE & \multicolumn{2}{c}{PBE+$U_{dd}^{\text{Lit.}}$} & ACBN0 \\
 &  &  & 4.2 eV & 8.5 eV & \\ \cline{2-6}
$E_{\text{g}}$ (eV) & 0.3\cite{arima1993} & m & 0.8 & 1.2 & 0.8 \\ \cline{2-6}
 &  & \multicolumn{4}{c}{Relative Energy vs. NM} \\
 &  & \multicolumn{4}{c}{Experimental Structure} \\
 &  & PBE & \multicolumn{2}{c}{PBE+$U_{dd}^{\text{Lit.}}$} & ACBN0 \\ 
 &  &  & 4.2 eV & 8.5 eV &  \\ \cline{3-6}
$\Delta E$ (meV) & FM & -102 & -313 & -1551 & -331 \\ \cline{3-6}
 &  & \multicolumn{4}{c}{Optimized Structure} \\
 &  & PBE & \multicolumn{2}{c}{PBE+$U_{dd}^{\text{Lit.}}$} & ACBN0 \\ 
 &  &  & 4.2 eV & 8.5 eV &  \\ \cline{3-6} 
$\Delta E$ (meV) & FM & -128 & -618 & -2334 & -476
\end{tabular}
\end{ruledtabular}
\end{table}

\begin{figure}[!tpb]
   \centering
   \includegraphics{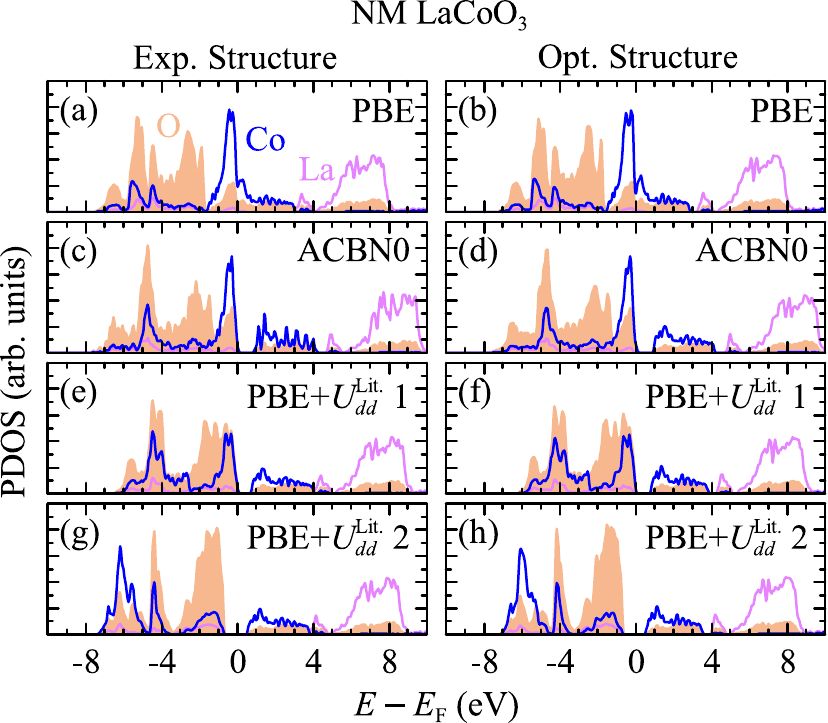}
   \caption{Projected density of states for NM LaCoO$_3$ (on the O, Co, and La states); \textbf{a.} experimental structure with PBE; \textbf{b.} optimized structure with PBE; \textbf{c.} experimental structure with ACBN0 (PBE+$U_{dd}$+$U_{pp}$); \textbf{d.} optimized structure with ACBN0 (PBE+$U_{dd}$+$U_{pp}$); \textbf{e.} experimental structure with PBE+$U_{dd}^{\text{Lit.}}=4.2$ eV\cite{nohara2009}; \textbf{f.} optimized structure with PBE+$U_{dd}^{\text{Lit.}}=4.2$ eV\cite{nohara2009}; \textbf{g.} experimental structure with PBE+$U_{dd}^{\text{Lit.}}=8.5$ eV\cite{nohara2009}; \textbf{h.} optimized structure with PBE+$U_{dd}^{\text{Lit.}}=8.5$ eV\cite{nohara2009}}\label{fig:CoPDOS1}
\end{figure}

\begin{figure}[!tpb]
   \centering
   \includegraphics{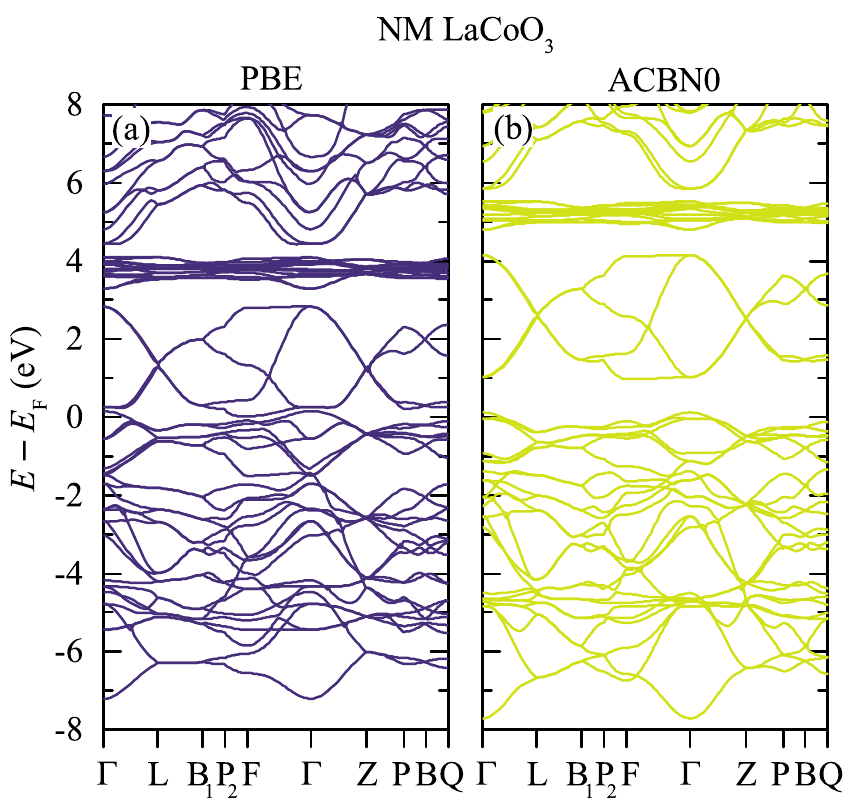}
   \caption{Band structure of NM LaCoO$_3$; \textbf{a.} PBE optimized structure; \textbf{b.} ACBN0 (PBE+$U_{dd}$+$U_{pp}$) optimized structure.}\label{fig:CoBands1}
\end{figure}

\begin{figure}[!tpb]
   \centering
   \includegraphics{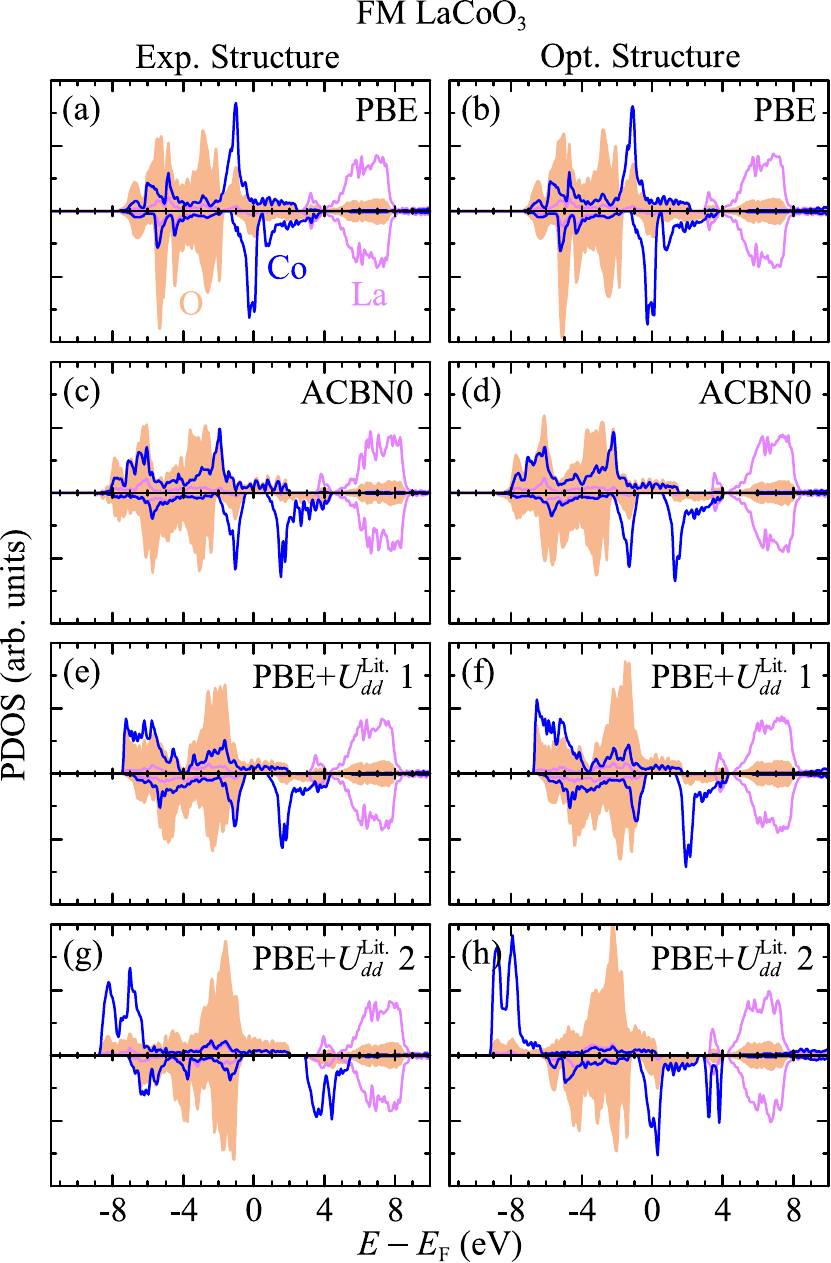}
   \caption{Projected density of states for FM LaCoO$_3$ (on the O, Co, and La states); \textbf{a.} experimental structure with PBE; \textbf{b.} optimized structure with PBE; \textbf{c.} experimental structure with ACBN0 (PBE+$U_{dd}$+$U_{pp}$) \textbf{d.} optimized structure with ACBN0 (PBE+$U_{dd}$+$U_{pp}$); \textbf{e.} experimental structure with PBE+$U_{dd}^{\text{Lit.}}=4.2$ eV\cite{nohara2009}; \textbf{f.} optimized structure with PBE+$U_{dd}^{\text{Lit.}}=4.2$ eV\cite{nohara2009}; \textbf{g.} experimental structure with PBE+$U_{dd}^{\text{Lit.}}=8.5$ eV\cite{nohara2009}; \textbf{h.} optimized structure with PBE+$U_{dd}^{\text{Lit.}}=8.5$ eV\cite{nohara2009}}\label{fig:CoPDOS2}
\end{figure}

\begin{figure}[!tpb]
   \centering
   \includegraphics{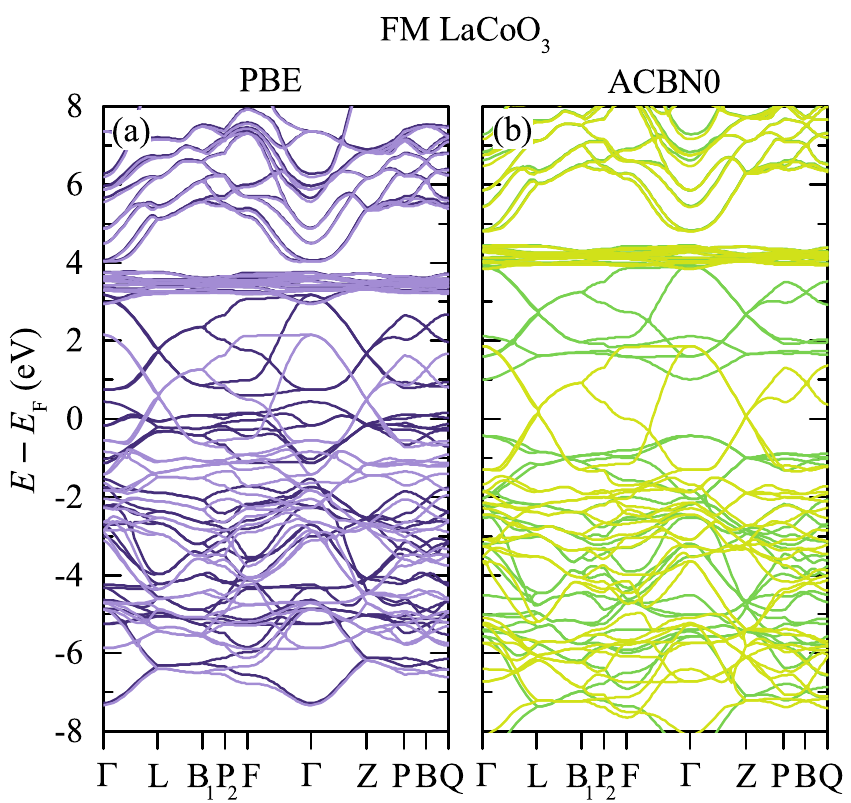}
   \caption{Band structure of FM LaCoO$_3$; \textbf{a.} PBE optimized structure; \textbf{b.} ACBN0 (PBE+$U_{dd}$+$U_{pp}$) optimized structure. Lighter colors correspond to spin up, while darker colors correspond to spin down.}\label{fig:CoBands2}
\end{figure}

Both ACBN0 and PBE+$U_{dd}=4.2$ eV calculations yield similar descriptions of the electronic structure, with the PDOS shown in Figs.~\ref{fig:CoPDOS1} and \ref{fig:CoPDOS2} for the non-magnetic (NM) and FM states, respectively. The overall picture of the electronic structure is similar to that of PBE, except that a gap is opened of approximately 0.8 eV. The PBE+$U_{dd}=8.5$ eV data reduce the Co--O mixing in the valence band and push occupied Co states further to the bottom of the valence band and only slightly increase the gap to 1.2 eV. All the resulting DFT+$U$ gaps are fairly large compared to experiment (0.3 eV), as shown in Table~\ref{table:CoElec}. However, all the results fail to predict the correct NM low-$T$ ground state. This may be attributed to the fact that all the previous DFT+$U$ studies mentioned used LDA+$U$ as opposed to the GGA+$U$ used here, and our results are consistent with those reported by Ritzmann \emph{et al.}\cite{ritzmann2014}. The tendency of larger $U_{dd}$ to extend bond lengths in this material combined with the tendency of LDA to overbind could explain why using the same value of $U$ for a GGA+$U$ calculation fails to improve the description, and illustrates the non-transferability of $U$. The HSE results of He and Franchini\cite{he2012} of course use PBE as the base for mixing exact exchange, with HSE-25 giving a very large gap of 2.4 eV; the value of mixing for HSE-Opt is very small at 0.05, but yields a band gap of 0.1 eV and also provides the best description of the structure in that work. It should be mentioned that despite the larger gap, ACBN0 provides a very similar picture of hybridization to that of HSE-Opt, which also agrees with the CT-like nature of the optical gap\cite{arima1993} and makes sense given the similarity of the former to the PBE result and the very low exchange mixing fraction of the latter. As mentioned in the previous section, ACBN0 provides an excellent description of the structure of LaCoO$_3$. The larger, LR $U$ of 8.5 eV significantly reduces the $d$ character of the valence band and thus departs from the picture provided by PBE, ACBN0 and HSE. The band structures of PBE and ACBN0 Fig.~\ref{fig:CoBands1} and \ref{fig:CoBands2} further illustrate the electronic structure as a simple shifting of the $e_g$ manifold higher in energy from the $t_{2g}$ manifold, resulting in a band gap in the NM case (the FM state remains metallic).

\subsubsection{LaNiO$_{\text{3}}$}
\begin{figure}[!tpb]
   \centering
   \includegraphics{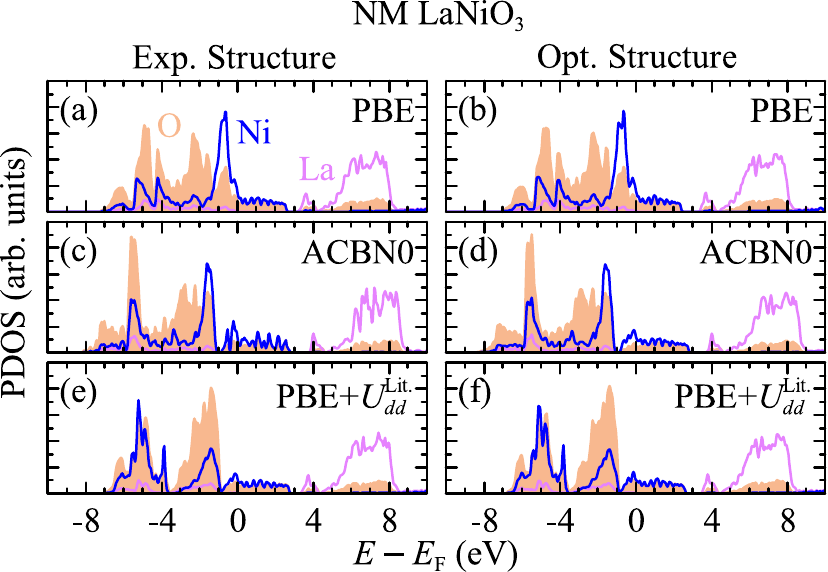}
   \caption{Projected density of states for NM LaNiO$_3$ (on the O, Ni, and La states); \textbf{a.} experimental structure with PBE; \textbf{b.} optimized structure with PBE; \textbf{c.} experimental structure with ACBN0 (PBE+$U_{dd}$+$U_{pp}$); \textbf{d.} optimized structure with ACBN0 (PBE+$U_{dd}$+$U_{pp}$); \textbf{e.} experimental structure with PBE+$U_{dd}^{\text{Lit.}}=5.7$ eV\cite{nohara2009}; \textbf{f.} optimized structure with PBE+$U_{dd}^{\text{Lit.}}=5.7$ eV\cite{nohara2009}.}\label{fig:NiPDOS1}
\end{figure}

\begin{figure}[!tpb]
   \centering
   \includegraphics{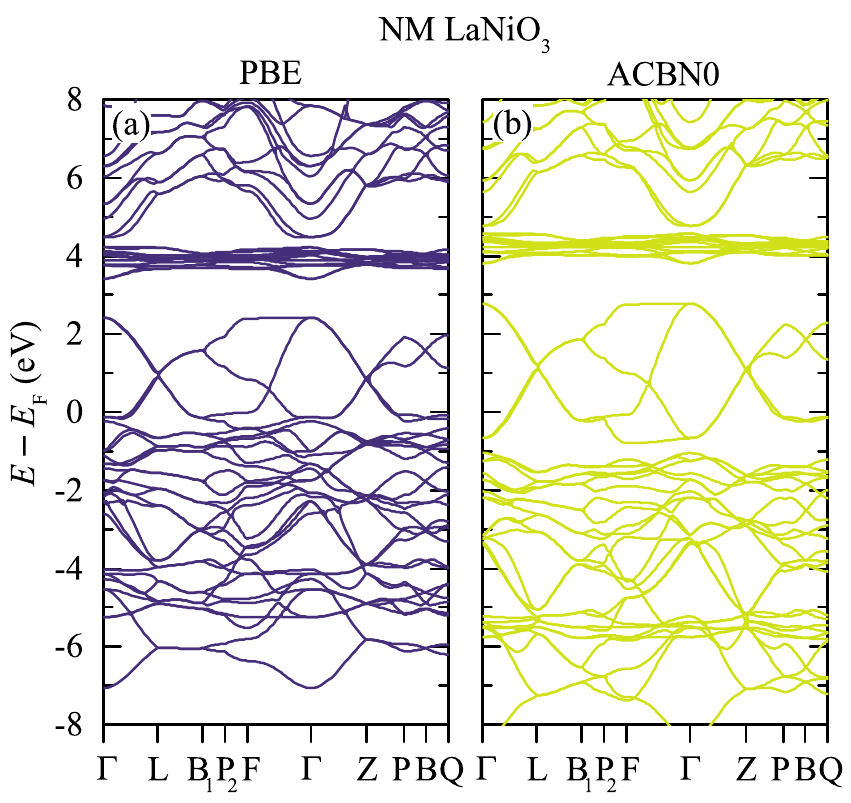}
   \caption{Band structure of NM LaNiO$_3$; \textbf{a.} PBE optimized structure; \textbf{b.} ACBN0 (PBE+$U_{dd}$+$U_{pp}$) optimized structure.}\label{fig:NiBands1}
\end{figure}

\begin{figure}[!tpb]
   \centering
   \includegraphics{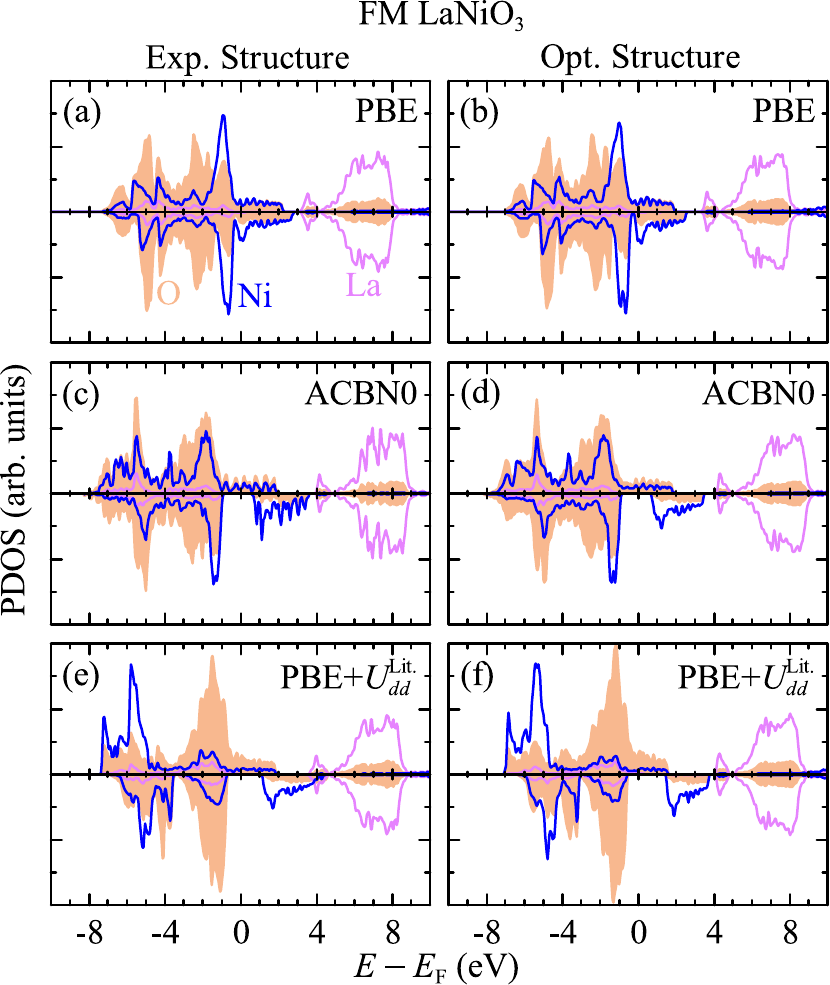}
   \caption{Projected density of states for FM LaNiO$_3$ (on the O, Ni, and La states); \textbf{a.} experimental structure with PBE; \textbf{b.} optimized structure with PBE; \textbf{c.} experimental structure with ACBN0 (PBE+$U_{dd}$+$U_{pp}$); \textbf{d.} optimized structure with ACBN0 (PBE+$U_{dd}$+$U_{pp}$); \textbf{e.} experimental structure with PBE+$U_{dd}^{\text{Lit.}}=5.7$ eV\cite{nohara2009}; \textbf{f.} optimized structure with PBE+$U_{dd}^{\text{Lit.}}=5.7$ eV\cite{nohara2009}}\label{fig:NiPDOS2}
\end{figure}

\begin{figure}[!tpb]
   \centering
   \includegraphics{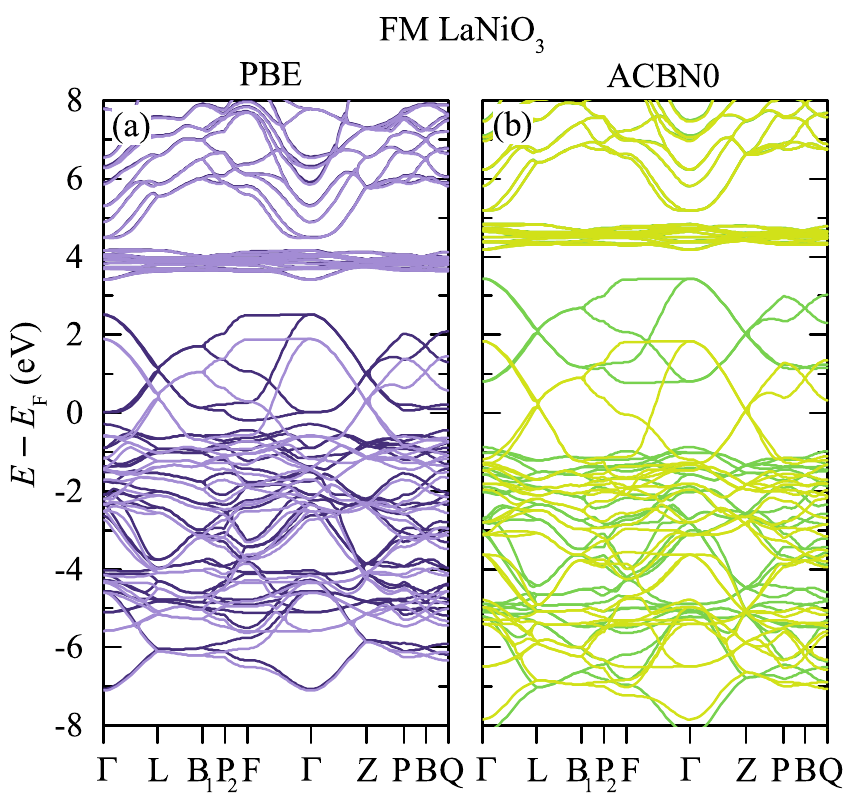}
   \caption{Band structure of FM LaNiO$_3$; \textbf{a.} PBE optimized structure; \textbf{b.} ACBN0 (PBE+$U_{dd}$+$U_{pp}$) optimized structure. Lighter colors correspond to spin up, while darker colors correspond to spin down.}\label{fig:NiBands2}
\end{figure}

The last material to be studied is LaNiO$_{3}$, where the strong covalent interaction between Ni and O screen results in itinerant electrons that screen correlation to a degree and result in a paramagnetic (PM) metal, albeit still one with important electron-electron interactions as revealed by the $T^2$ dependence of resistivity and heat capacity\cite{gou2011,garciamunoz1992,medarde1997}. The electronic parameters of LaNiO$_3$ are summarized in Table~\ref{table:NiElec}, with PDOS for both NM and FM states (for PBE, ACBN0 and PBE+$U_{dd}^{\text{Lit.}}$) shown in Fig.~\ref{fig:NiPDOS1} and \ref{fig:NiPDOS2}, respectively; and band structures for NM and FM states (for PBE and ACBN0) shown in Fig.~\ref{fig:NiBands1} and \ref{fig:NiBands2}, respectively.

\begin{table}[!tpb]
\small
\centering
\caption{Parameters obtained from the electronic structure of LaNiO$_3$, including band gap $E_{\text{g}}$ and the energy difference $\Delta E$ between various calculated magnetic ordering states for PBE, PBE+$U_{dd}^{\text{Lit.}}=5.7$ eV\cite{nohara2009}, and ACBN0 (PBE+$U_{dd}$+$U_{pp}$). Gap of ``m'' refers to a metallic system.}
\label{table:NiElec}
\begin{ruledtabular}
\begin{tabular}{rcccc}
 & \multicolumn{4}{c}{\textbf{LaNiO}$_{\textbf{3}}$} \\
 & \multicolumn{4}{c}{NM Optimized Structure} \\
 & Expt. & PBE & PBE+$U_{dd}^{\text{Lit.}}$ & ACBN0 \\ \cline{2-5}
$E_{\text{g}}$ (eV) & m\cite{garciamunoz1992} & m & m & m \\ \cline{2-5}
 &  & \multicolumn{3}{c}{Relative Energy vs. NM} \\
 &  & \multicolumn{3}{c}{Experimental Structure} \\
 &  & PBE & PBE+$U_{dd}^{\text{Lit.}}$ & ACBN0 \\ \cline{3-5} 
$\Delta E$ (meV) & FM & 1 & -839 & -596 \\ \cline{3-5}
 &  & \multicolumn{3}{c}{Optimized Structure} \\
 &  & PBE & PBE+$U_{dd}^{\text{Lit.}}$ & ACBN0 \\ \cline{3-5}
$\Delta E$ (meV) & FM & -33 & -1000 & -604
\end{tabular}
\end{ruledtabular}
\end{table}

PBE stabilizes a NM state with the experimental structure. This is in contrast to Gou \emph{et al.} but in agreement with He and Franchini\cite{gou2011,he2012}. The absolute energy difference vs. the FM state is extremely small at 1 meV, about two orders of magnitude smaller than that reported by the latter study. ACBN0 and PBE+$U_{dd}^{\text{Lit.}}=5.7$ eV\cite{nohara2009} both stabilize FM ordering, similarly to previously-reported LSDA+$U$\cite{gou2011} and hybrid functional results\cite{gou2011,he2012}. The PBE+$U_{dd}^{\text{Lit.}}$ calculation, similar to the aforementioned LSDA+$U$ study, suppresses the contribution of Ni states near the Fermi level and pushes them to the bottom of the valence band, yielding a qualitatively incorrect picture of the electronic structure. Aside from the fact that ACBN0 incorrectly stabilizes FM ordering in the bulk, the deviations from the PBE result are less extreme. Hybrid functionals of increasing mixing fraction have a similar trend as when increasing $U$, but their description of valence band spectra is significantly worse than LDA or DFT+$U$. However, they also describe bound core states more accurately\cite{gou2011} where DFT+$U$ does not (since the correction is only applied to the valence states). Further study is needed to determine how ACBN0 performs in comparison with experimental spectra.

\begin{figure*}[!htpb]
   \centering
   \includegraphics{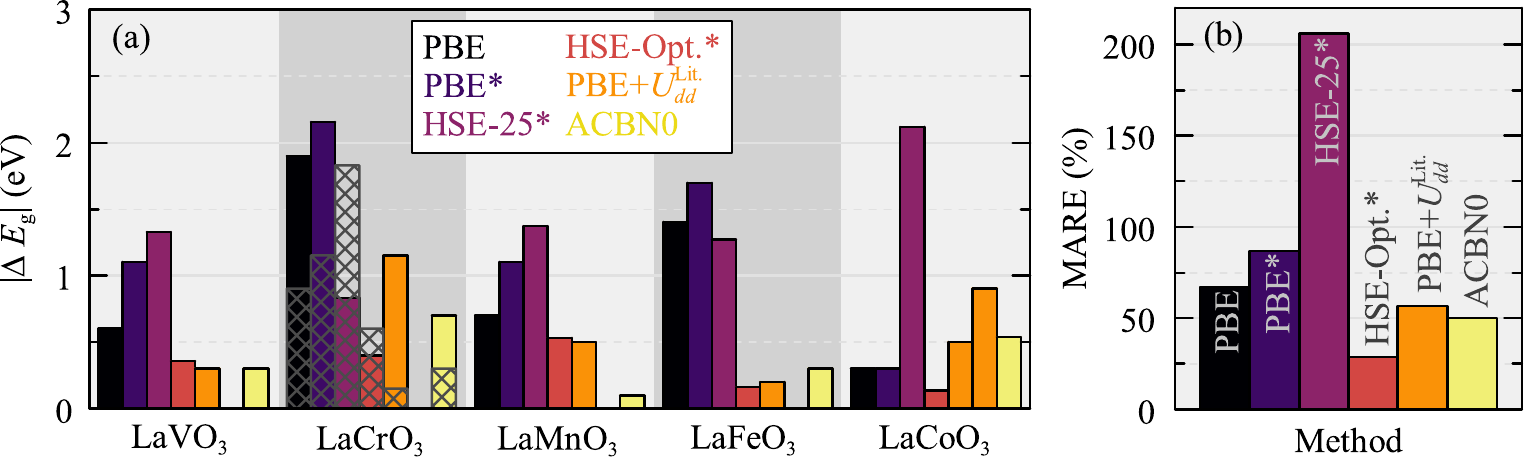}
   \caption{The absolute band gap error for PBE and several corrective methods, and the MARE of the band gap predictions for each method. Asterisks denote data from He and Franchini\cite{he2012}. ACBN0 calculations are PBE+$U_{dd}$+$U_{pp}$. Bars with gray hashed lines represent errors using an alternative interpretation of the electronic structure (see main text).}\label{fig:bandMARE}
\end{figure*}

It should be mentioned that all the reported DFT, DFT+$U$ and hybrid functional calculations are fundamentally incapable of describing the electronic structure of LaNiO$_3$ accurately. The delocalized states lead to screened correlation effects that are not captured accurately with approximate XC functionals\cite{gou2011}. Corrections such as hybrid functionals and DFT+$U$ are intended to correct self-interaction error arising from inexact exchange, and strictly speaking do not treat correlation. Many-body methods such as dynamical mean field theory (DMFT) are necessary to treat these correlation effects in a meaningful way\cite{hansmann2010,deng2012}.

The results of this section are summarized in Fig.~\ref{fig:bandMARE}. The absolute error is significantly improved using ACBN0 vs. PBE (the PBE error in LaCoO$_3$ is due to predicting a metallic state). The more stringent \% MARE (since \% errors for small gaps can be very high) demonstrates that ACBN0 is still improved vs. PBE, PBE+$U_{dd}^{\text{Lit.}}$ and HSE-25. Only HSE-Opt performs better on average, but as can be seen by the absolute errors, ACBN0 still outperforms HSE in several cases, most notably LaVO$_3$, LaCrO$_3$, and LaMnO$_3$. However, one must be careful not to attribute too much meaning to just the band gap, as simply applying an empirical $U_{dd}$ can result in a correct band gap but incorrect picture of bonding, worsening structural description and other properties. The Supplemental Material has some comparison to experimental X-ray photoemission and X-ray emission/absorption spectra that may be illustrative of this point\cite{supp}. While ACBN0 can predict band gaps fairly well, the overall failure to predict the correct ground state of several of these oxides may be reflective of a need for more sophistication in applying $U$ corrections. Many materials may require separate corrections for $t_{2g}$, $e_g$, etc. as has been noted in many early literature works. Hybrid functionals do not have this problem, and as mentioned in Sec.~\ref{sec:intro}, can utilize first-principles self-consistent mixing fractions, but still suffer from high computational cost. Further developments in the implementation of DFT+$U$ may help bridge this gap.

\subsubsection{\label{subsec:Uox}Effect of \emph{U} applied to oxygen 2\emph{p} states}
For most the perovskites studied in this work, the values of $U_{dd}$ predicted by ACBN0 are significantly smaller in magnitude compared to the corresponding values of $U$ referenced from in the literature (see Supplemental Material\cite{supp}). However, regardless of the transferability of $U$, these reference values of $U_{dd}$ were not calculated or fitted with a non-zero value of $U_{pp}$ applied to the oxygen $2p$ states. This brings up the effect of $U_{pp}$ as an important point for discussion.

The use of non-zero $U_{pp}$ has been reported extensively on several transition metal oxides. In general, $U_{pp}$ can be expected to be on the same order of magnitude as $U_{dd}$, based on both cRPA calculations\cite{sakuma2013} and on extensive experimental spectroscopy results\cite{sawatzky1979,ghijsen1988,bar-deroma1992,chainani1992,chainani1993,sarma1994,chainani2017}. One notable case is that of ZnO, where additional Hubbard corrections beyond $U_{dd}$ are necessary in order to open the band gap to the experimental value. Even large values of $U_{dd}$ are not successful in this regard\cite{walsh2008}. Recent studies on bulk \cite{ma2013} and nanowire\cite{sheetz2009} ZnO have found that empirical values of $U_{dd}\approx 10$ eV and $U_{pp}\approx 7$ eV provide the best overall description of lattice geoemetry, band gap, and defect formation energies. Another study attempted to empirically fit $U_{dd}$ and $U_{pp}$ to match experimental XPS spectra, which resulted in values of 9.3 and 18.4 eV, respectively; based on the calculated dielectric tensor, the authors concluded XPS spectra are not representative of the ground state and are unsuitable for direct comparison with calculated DOS\cite{bashyal2018}. Interestingly, ACBN0 results in $U_{dd}=\sim 13$ eV and $U_{pp}=\sim 5.5$ eV, which is very close to the values chosen empirically\cite{gopal2015,agapito2015}. Several other papers have investigated TiO$_2$, where the use of both $U_{dd}$ and $U_{pp}$ allows for the simultaneous improvement of both band gap and unit cell parameters. Using LR for both Hubbard correction values yielded $U_{dd}=3.5$ eV and $U_{pp}=10.65$ eV for rutile TiO$_2$\cite{mattioli2010}. However, an empirical $U_{pp}=5.0$ eV was used to give a correct band gap. ACBN0 yielded $U{dd}=0.15$ eV (owing to the nearly empty $d$ bands) and $U_{pp}=7.6$ eV, closer to the empirical $U_{pp}$ used to give the correct band gap\cite{agapito2015}. Empirical tuning of the Hubbard parameters on both titanium $3d$ and oxygen $2p$ has also been investigated\cite{park2010,ataei2016}. In these studies, while $U_{pp}$ preserves the hybridized character of the bonds, the lattice geometry is often worse compared to LDA or GGA\cite{park2010}. Molecular organometallic Ni magnets\cite{cao2008} and cobalt oxyhydroxide catalysts\cite{mattioli2013} have also been studied using $U_{pp}$. In many cases, justifications based on the experimental observation of large Coulombic interactions between O $2p$ states, as well as preservation of strong $d$--$p$ hybridization--strongly modified by applying only $U_{dd}$--have been put forward, but many of these studies still suffer from difficulty in applying the $U$ correction self-consistently.

The simultaneous, self-consistent determination of both $U_{dd}$ and $U_{pp}$ in this study has the effect of maintaining the metal-oxygen covalency present in the PBE calculation (which is also the case for the HSE-Opt and HSE-25 cases when using hybrid functionals\cite{he2012}) while also improving the description of lattice geometry and the band gap. While $U_{dd}$ can sometimes be tuned to give a more accurate description of certain properties (usually band gap), by examining the pDOS in this work, we can see that the bonding character is often significantly modified, especially for larger values of $U_{dd}$. Metal-oxygen covalency is reduced, with the metal $d$ states being pushed to lower energy with narrower DOS features. In contrast, ACBN0 shifts both the O $2p$ and metal $3d$ bands; by not drastically changing the bonding character or introducing spurious charge transfer that can occur when only $U_{dd}$ is applied\cite{mattioli2010}, the accuracy of several material properties can be improved simultaneously, which is encouraging for researchers desiring to take a less empirical approach in their calculations. In addition, the nature of the renormalization procedure in ACBN0 will reduce the magnitude of $U$ corrections when the KS states are not well-represented by a basis of localized orbitals. Building on the arguments put forward by other studies that utilize a $U_{pp}$ correction, we believe the results presented in this paper make a strong case for the incorporation of Hubbard correction terms on oxygen $2p$ states in many transition metal oxides, when calculated self-consistently from first principles. 

\section{Conclusions}
This work has demonstrated that ACBN0 improves the description of the first row transition metal perovskites compared with PBE and a na\"{i}ve choice of $U$ taken from the literature. ACBN0 also compares favorably with the hybrid functional HSE, offering improved descriptions of band gaps vs. HSE-25 and performing competitively to an \emph{empirically optimized} HSE-Opt for both structure and to a lesser degree, band gap, from completely first-principles values of $U$ that directly depend on the charge density.

Simply choosing a value of $U$ from the literature is insufficient when trying to obtain an overall picture of material properties. In addition, values of $U$ can vary across functionals, approaches to calculating $U$, and implementation of the DFT+$U$ method itself, leading to results at odds with other published calculations in the literature. We have also demonstrated the importance of explicit $U$ and $J$ values in some orbitally ordered materials, which can also be easily performed with ACBN0. Overall, there still remains potential room for improvement in using and verifying ACBN0 that is mainly limited by currently-available implementation in software. The use of unique values of $U$ for specific subsets of orbitals such as $e_g$ and $t_{2g}$ may yet offer improved descriptions of materials such as LaMnO$_3$, in addition to the necessity of using explicit $U$ and $J$. ACBN0 should also be applicable to the DFT+U+V approach\cite{campo2010} that offers improved descriptions of covalent materials. If these developments are fruitful, this method holds promise not only in high-throughput applications but also in treating a wide variety of complex materials with first-principles site-specific $U$ values at a reasonable computational cost.

\begin{acknowledgments}
We wish to thank Prof. Marco Buongiorno Nardelli (University of North Texas) for sharing an early ACBN0 implementation, as well as Dr. Priya Gopal (University of North Texas) and Dr. Andrew Supka (Central Michigan University) for helpful discussion. This work made use of computational resources from the National Energy Research Scientific Computing Center (NERSC), a DOE Office of Science User Facility supported by the Office of Science of the U.S. Department of Energy under Contract No. DE-AC02-05CH11231, as well as computational resources from the Texas Advanced Computing Center (TACC) at The University of Texas at Austin. Financial support was received from the Skoltech-MIT Center for Electrochemical Energy Storage. K.J.M. was partially supported by a doctoral postgraduate scholarship (PGS-D) from the Natural Sciences and Engineering Research Council of Canada (NSERC).
\end{acknowledgments}

\bibliography{arxivmain}
\clearpage
\onecolumngrid
\begin{center}
\textbf{\large Supplemental Material}
\end{center}

\twocolumngrid

\setcounter{equation}{0}
\setcounter{figure}{0}
\setcounter{table}{0}
\setcounter{page}{1}
\makeatletter
\renewcommand{\theequation}{S-\arabic{equation}}
\renewcommand{\thefigure}{S-\arabic{figure}}
\renewcommand{\thetable}{S-\Roman{table}}

\renewcommand{\citenumfont}[1]{S#1}
\begin{table}[ht!]
\centering
\caption{ACBN0 calculations of $U$, $J$ and $U_{\text{eff.}}=U-J$ for V $3d$ and O $2p$ in LaVO$_3$. Average values for energetically-competitive phases are bolded.}
\label{table:UvalsV}
\begin{ruledtabular}
\begin{tabular}{rccccccc}
\multicolumn{1}{r}{} & \multicolumn{7}{c}{LaVO$_3$ ACBN0 $U$ Values (eV)} \\
\multicolumn{1}{r}{\begin{tabular}[c]{@{}c@{}}Magnetic\\ State\end{tabular}} & \multicolumn{1}{r}{\begin{tabular}[c]{@{}c@{}}Quantity\end{tabular}} & \multicolumn{3}{c}{\begin{tabular}[c]{@{}c@{}}Experimental\\ Structure\end{tabular}} & \multicolumn{3}{c}{\begin{tabular}[c]{@{}c@{}}Optimized\\ Structure\end{tabular}} \\ \cline{3-8}
 & & V$_1$ & V$_2$ & O & V$_1$ & V$_2$ & O \\ \cline{1-8}
NM & $U$ & \multicolumn{2}{c}{6.16} & 15.48 & \multicolumn{2}{c}{6.25} & 15.10 \\
   & $J$ & \multicolumn{2}{c}{3.03} & 7.64 & \multicolumn{2}{c}{5.86} & 7.46 \\
   & $U_{\text{eff.}}$ & \multicolumn{2}{c}{3.13} & 7.84 & \multicolumn{2}{c}{0.39} & 7.64 \\
FM & $U$ & \multicolumn{2}{c}{6.36} & 15.23 & \multicolumn{2}{c}{6.24} & 15.09 \\
   & $J$  & \multicolumn{2}{c}{4.70} & 7.53 & \multicolumn{2}{c}{4.60} & 7.46 \\
   & $U_{\text{eff.}}$ & \multicolumn{2}{c}{1.66} & 7.70 & \multicolumn{2}{c}{1.64} & 7.63 \\
AFM-A & $U$ & 6.27 & 6.43 & 15.21 & 6.40 & 6.38 & 15.09 \\
      & $J$ & 4.74 & 4.84 & 7.51 & 4.75 & 4.74 & 7.45 \\
      & $U_{\text{eff.}}$ & 1.53 & 1.60 & 7.70 & 1.65 & 1.64 & 7.64 \\
AFM-C & $U$ & 6.29 & 6.32 & 15.26 & 6.28 & 6.29 & 15.11 \\
      & $J$ & 4.75 & 4.77 & 7.53 & 4.77 & 4.78 & 7.46 \\
      & $U_{\text{eff.}}$ & 1.54 & 1.56 & 7.72 & 1.51 & 1.52 & 7.65 \\
AFM-G & $U$ & 6.42 & 6.42 & 15.25 & 6.46 & 6.46 & 15.11 \\
      & $J$ & 5.01 & 5.01 & 7.53 & 5.01 & 5.01 & 7.46 \\
      & $U_{\text{eff.}}$ & 1.41 & 1.41 & 7.72 & 1.45 & 1.45 & 7.65 \\
\textbf{AVG} & $\textbf{\textit{U}}_{\textbf{eff.}}$ & \textbf{1.54} & \textbf{1.56} & \textbf{7.71} & \textbf{1.56} & \textbf{1.56} & \textbf{7.64} \\
\end{tabular}
\end{ruledtabular}
\end{table}

\begin{table}[ht!]
\centering
\caption{ACBN0 calculations of $U$, $J$ and $U_{\text{eff.}}=U-J$ for Cr $3d$ and O $2p$ in LaCrO$_3$. Average values for energetically-competitive phases are bolded.}
\label{table:UvalsCr}
\begin{ruledtabular}
\begin{tabular}{rccccccc}
\multicolumn{1}{r}{} & \multicolumn{7}{c}{LaCrO$_3$ ACBN0 $U$ Values (eV)} \\
\multicolumn{1}{r}{\begin{tabular}[c]{@{}c@{}}Magnetic\\ State\end{tabular}} & \multicolumn{1}{r}{\begin{tabular}[c]{@{}c@{}}Quantity\end{tabular}} & \multicolumn{3}{c}{\begin{tabular}[c]{@{}c@{}}Experimental\\ Structure\end{tabular}} & \multicolumn{3}{c}{\begin{tabular}[c]{@{}c@{}}Optimized\\ Structure\end{tabular}} \\ \cline{3-8}
 & & Cr$_1$ & Cr$_2$ & O & Cr$_1$ & Cr$_2$ & O \\ \cline{1-8}
NM & $U$ & \multicolumn{2}{c}{7.69} & 14.53 & \multicolumn{2}{c}{7.11} & 14.19 \\
   & $J$  & \multicolumn{2}{c}{3.18} & 7.17 & \multicolumn{2}{c}{3.76} & 7.00 \\
   & $U_{\text{eff.}}$ & \multicolumn{2}{c}{4.51} & 7.35 & \multicolumn{2}{c}{3.35} & 7.18 \\
FM & $U$ & \multicolumn{2}{c}{5.52} & 13.87 & \multicolumn{2}{c}{5.27} & 13.67 \\
   & $J$  & \multicolumn{2}{c}{2.70} & 6.90 & \multicolumn{2}{c}{2.54} & 6.81 \\
   & $U_{\text{eff.}}$ & \multicolumn{2}{c}{2.82} & 6.97 & \multicolumn{2}{c}{2.74} & 6.87 \\
AFM-A & $U$ & 5.38 & 5.38 & 13.72 & 5.15 & 5.15 & 13.52 \\
      & $J$ & 2.64 & 2.64 & 6.78 & 2.50 & 2.49 & 6.68 \\
      & $U_{\text{eff.}}$ & 2.74 & 2.74 & 6.94 & 2.66 & 2.65 & 6.84 \\
AFM-C & $U$ & 5.44 & 5.44 & 13.93 & 5.24 & 5.24 & 13.76 \\
      & $J$ & 2.68 & 2.68 & 6.89 & 2.55 & 2.54 & 6.81 \\
      & $U_{\text{eff.}}$ & 2.76 & 2.76 & 7.04 & 2.70 & 2.69 & 6.95 \\
AFM-G & $U$ & 5.45 & 5.45 & 13.86 & 5.30 & 5.30 & 13.75 \\
      & $J$ & 2.68 & 2.68 & 6.85 & 2.58 & 2.58 & 6.80 \\
      & $U_{\text{eff.}}$ & 2.77 & 2.77 & 7.01 & 2.71 & 2.71 & 6.95 \\
\textbf{AVG} & $\textbf{\textit{U}}_{\textbf{eff.}}$ & \textbf{2.77} & \textbf{2.77} & \textbf{6.99} & \textbf{2.70} & \textbf{2.70} & \textbf{6.90} \\
\end{tabular}
\end{ruledtabular}
\end{table}

\begin{table}[ht!]
\centering
\caption{ACBN0 calculations of $U$, $J$ and $U_{\text{eff.}}=U-J$ for Mn $3d$ and O $2p$ in LaMnO$_3$. Average values for energetically-competitive phases are bolded.}
\label{table:UvalsMn}
\begin{ruledtabular}
\begin{tabular}{rccccccc}
\multicolumn{1}{r}{} & \multicolumn{7}{c}{LaMnO$_3$ ACBN0 $U$ Values (eV)} \\
\multicolumn{1}{r}{\begin{tabular}[c]{@{}c@{}}Magnetic\\ State\end{tabular}} & \multicolumn{1}{r}{\begin{tabular}[c]{@{}c@{}}Quantity\end{tabular}} & \multicolumn{3}{c}{\begin{tabular}[c]{@{}c@{}}Experimental\\ Structure\end{tabular}} & \multicolumn{3}{c}{\begin{tabular}[c]{@{}c@{}}Optimized\\ Structure\end{tabular}} \\ \cline{3-8}
 & & Mn$_1$ & Mn$_2$ & O & Mn$_1$ & Mn$_2$ & O \\ \cline{1-8}
NM & $U$ & \multicolumn{2}{c}{7.63} & 13.28 & \multicolumn{2}{c}{8.14} & 13.51 \\
   & $J$ & \multicolumn{2}{c}{4.31} & 6.57 & \multicolumn{2}{c}{3.42} & 6.67 \\
   & $U_{\text{eff.}}$ & \multicolumn{2}{c}{3.33} & 6.71 & \multicolumn{2}{c}{4.72} & 6.84 \\
FM & $U$ & \multicolumn{2}{c}{3.88} & 12.38 & \multicolumn{2}{c}{4.10} & 12.48 \\
   & $J$ & \multicolumn{2}{c}{1.27} & 6.28 & \multicolumn{2}{c}{1.26} & 6.35 \\
   & $U_{\text{eff.}}$ & \multicolumn{2}{c}{2.61} & 6.11 & \multicolumn{2}{c}{2.83} & 6.13 \\
AFM-A & $U$ & 3.36 & 3.36 & 11.99 & 3.14 & 3.14 & 11.79 \\
      & $J$ & 1.09 & 1.09 & 5.93 & 1.02 & 1.02 & 5.83 \\
      & $U_{\text{eff.}}$ & 2.26 & 2.26 & 6.06 & 2.12 & 2.12 & 5.95 \\
AFM-C & $U$ & 3.39 & 3.39 & 12.07 & 3.03 & 3.03 & 11.76 \\
      & $J$ & 1.14 & 1.14 & 6.00 & 1.01 & 1.01 & 5.84 \\
      & $U_{\text{eff.}}$ & 2.25 & 2.25 & 6.07 & 2.01 & 2.01 & 5.92 \\
AFM-G & $U$ & 3.42 & 3.42 & 12.04 & 3.05 & 3.05 & 11.68 \\
      & $J$ & 1.17 & 1.17 & 5.96 & 1.03 & 1.03 & 5.78 \\
      & $U_{\text{eff.}}$ & 2.25 & 2.25 & 6.08 & 2.02 & 2.02 & 5.89 \\
\textbf{AVG} & $\textbf{\textit{U}}_{\textbf{eff.}}$ & \textbf{2.34} & \textbf{2.34} & \textbf{6.08} & \textbf{2.25} & \textbf{2.25} & \textbf{5.97} \\
\end{tabular}
\end{ruledtabular}
\end{table}

\begin{table}[ht!]
\centering
\caption{ACBN0 calculations of $U$, $J$ and $U_{\text{eff.}}=U-J$ for Fe $3d$ and O $2p$ in LaFeO$_3$. Average values for energetically-competitive phases are bolded.}
\label{table:UvalsFe}
\begin{ruledtabular}
\begin{tabular}{rccccccc}
\multicolumn{1}{r}{} & \multicolumn{7}{c}{LaFeO$_3$ ACBN0 $U$ Values (eV)} \\
\multicolumn{1}{r}{\begin{tabular}[c]{@{}c@{}}Magnetic\\ State\end{tabular}} & \multicolumn{1}{r}{\begin{tabular}[c]{@{}c@{}}Quantity\end{tabular}} & \multicolumn{3}{c}{\begin{tabular}[c]{@{}c@{}}Experimental\\ Structure\end{tabular}} & \multicolumn{3}{c}{\begin{tabular}[c]{@{}c@{}}Optimized\\ Structure\end{tabular}} \\ \cline{3-8}
 & & Fe$_1$ & Fe$_2$ & O & Fe$_1$ & Fe$_2$ & O \\ \cline{1-8}
NM & $U$ & \multicolumn{2}{c}{6.59} & 12.14 & \multicolumn{2}{c}{6.91} & 12.19 \\
   & $J$ & \multicolumn{2}{c}{2.79} & 6.00 & \multicolumn{2}{c}{2.95} & 6.02 \\
   & $U_{\text{eff.}}$ & \multicolumn{2}{c}{3.80} & 6.14 & \multicolumn{2}{c}{3.96} & 6.17 \\
FM & $U$ & \multicolumn{2}{c}{5.47} & 13.44 & \multicolumn{2}{c}{8.07} & 14.58 \\
   & $J$ & \multicolumn{2}{c}{1.75} & 6.65 & \multicolumn{2}{c}{2.57} & 7.20 \\
   & $U_{\text{eff.}}$ & \multicolumn{2}{c}{3.71} & 6.80 & \multicolumn{2}{c}{5.50} & 7.38 \\
AFM-A & $U$ & 2.99 & 2.99 & 11.84 & 3.66 & 3.66 & 12.30 \\
      & $J$ & 0.91 & 0.91 & 5.86 & 1.14 & 1.44 & 6.09 \\
      & $U_{\text{eff.}}$ & 2.08 & 2.08 & 5.97 & 2.52 & 2.52 & 6.20 \\
AFM-C & $U$ & 3.19 & 3.19 & 12.19 & 3.69 & 3.69 & 12.55 \\
      & $J$ & 0.97 & 0.97 & 6.02 & 1.15 & 1.15 & 6.20 \\
      & $U_{\text{eff.}}$ & 2.22 & 2.22 & 6.17 & 2.54 & 2.54 & 6.35 \\
AFM-G & $U$ & 3.81 & 3.81 & 12.35 & 4.13 & 4.13 & 12.56 \\
      & $J$ & 1.17 & 1.17 & 6.11 & 1.28 & 1.28 & 6.22 \\
      & $U_{\text{eff.}}$ & 2.63 & 2.63 & 6.24 & 2.84 & 2.84 & 6.34 \\
\textbf{AVG} & $\textbf{\textit{U}}_{\textbf{eff.}}$ & \textbf{2.31} & \textbf{2.31} & \textbf{6.13} & \textbf{2.63} & \textbf{2.63} & \textbf{6.30} \\
\end{tabular}
\end{ruledtabular}
\end{table}
\clearpage
\begin{table}[ht!]
\centering
\caption{ACBN0 calculations of $U$, $J$ and $U_{\text{eff.}}=U-J$ for Co $3d$ and O $2p$ in LaCoO$_3$. Average values for energetically-competitive phases are bolded.}
\label{table:UvalsCo}
\begin{ruledtabular}
\begin{tabular}{rccccc}
\multicolumn{1}{r}{} & \multicolumn{5}{c}{LaCoO$_3$ ACBN0 $U$ Values (eV)} \\
\multicolumn{1}{r}{\begin{tabular}[c]{@{}c@{}}Magnetic\\ State\end{tabular}} &
\multicolumn{1}{r}{\begin{tabular}[c]{@{}c@{}}Quantity\end{tabular}} & \multicolumn{2}{c}{\begin{tabular}[c]{@{}c@{}}Experimental\\ Structure\end{tabular}} & \multicolumn{2}{c}{\begin{tabular}[c]{@{}c@{}}Optimized\\ Structure\end{tabular}} \\ \cline{3-6}
 & & Co & O & Co & O \\ \cline{1-6}
NM & $U$ & 5.50 & 10.49 & 5.47 & 10.45 \\
   & $J$ & 2.10 & 5.19 & 2.09 & 5.17 \\
   & $U_{\text{eff.}}$ & 3.39 & 5.30 & 3.38 & 5.28 \\
FM & $U$ & 5.08 & 10.68 & 4.96 & 10.68 \\
   & $J$ & 1.79 & 5.36 & 1.74 & 5.37 \\
   & $U_{\text{eff.}}$ & 3.29 & 5.32 & 3.22 & 5.31  \\
\textbf{AVG} & $\textbf{\textit{U}}_{\textbf{eff.}}$ & \textbf{3.34} & \textbf{5.31} & \textbf{3.30} & \textbf{5.29}\\
\end{tabular}
\end{ruledtabular}
\end{table}

\begin{table}[ht!]
\centering
\caption{ACBN0 calculations of $U$, $J$ and $U_{\text{eff.}}=U-J$ for Ni $3d$ and O $2p$ in LaNiO$_3$. Average values for energetically-competitive phases are bolded.}
\label{table:UvalsNi}
\begin{ruledtabular}
\begin{tabular}{rccccc}
\multicolumn{1}{r}{} & \multicolumn{5}{c}{LaNiO$_3$ ACBN0 $U$ Values (eV)} \\
\multicolumn{1}{r}{\begin{tabular}[c]{@{}c@{}}Magnetic\\ State\end{tabular}} &
\multicolumn{1}{r}{\begin{tabular}[c]{@{}c@{}}Quantity\end{tabular}} & \multicolumn{2}{c}{\begin{tabular}[c]{@{}c@{}}Experimental\\ Structure\end{tabular}} & \multicolumn{2}{c}{\begin{tabular}[c]{@{}c@{}}Optimized\\ Structure\end{tabular}} \\ \cline{3-6}
 & & Ni & O & Ni & O \\ \cline{1-6}
NM & $U$ & 5.42 & 9.57 & 5.43 & 9.57 \\
   & $J$ & 1.64 & 4.76 & 1.64 & 4.76 \\
   & $U_{\text{eff.}}$ & 3.78 & 4.81 & 3.79 & 4.81 \\
FM & $U$ & 4.56 & 9.47 & 4.45 & 9.36  \\
   & $J$ & 1.47 & 4.72 & 1.43 & 4.66 \\
   & $U_{\text{eff.}}$ & 3.08 & 4.75 & 3.02 & 4.70  \\
\textbf{AVG} & $\textbf{\textit{U}}_{\textbf{eff.}}$ & \textbf{3.43} & \textbf{4.78} & \textbf{3.40} & \textbf{4.76}\\
\end{tabular}
\end{ruledtabular}
\end{table}

\begin{table*}[ht!]
\centering
\caption{Values of $U$ taken from the literature ("Lit. $U$"). Cluster-CI refers to cluster configuration interaction calculations, and those $U$ values are an effective term given by $U_{\text{eff}}=U-J$. Values used in this work are bolded. In LaCoO$_3$ two values were used in separate calculations.}
\label{table:LitU}
\begin{ruledtabular}
\begin{tabular}{cclll}
   B-site cation & \multicolumn{4}{c}{Literature $U$}                             \\
   & $U_{\text{eff}}$ (eV)      & \multicolumn{1}{c}{Method} & \multicolumn{1}{c}{Underlying XC Functional} & \multicolumn{1}{c}{Code Used}     \\ \cline{1-5}
V  & 2.49          & cRPA ($t_{2g}$-$t_{2g}$ model)\cite{kim2018} & PBE (GGA) & VASP\\
   & \textbf{3.0}  & Cluster-CI\cite{nohara2009}                  & N/A & N/A \\
   & 3.0           & Fit to $E_{\text{g}}$\cite{fang2004}          & LDA & Not available\\
   & 3.16          & cRPA ($t_{2g}$-$t_{2g},p$ model)\cite{kim2018} & PBE (GGA) & VASP\\
   & 3.6           & Fit to optical absorption spectra\cite{wang2015} & PBE (GGA) & VASP\\
   & 3.85          & Fit to $E_{\text{g}}$\cite{kumari2017}         & PBE (GGA) & The ELK Code (FP-LAPW method)\\
   & 6.76          & Constrained DFT\cite{nohara2009}               & LSDA & Not available (LMTO-ASA method)\\
Cr & 1.97          & cRPA ($t_{2g}$-$t_{2g}$ model)\cite{kim2018}   & PBE (GGA) & VASP\\
   & 2.66          & cRPA ($t_{2g}$-$t_{2g},p$ model)\cite{kim2018} & PBE (GGA) & VASP\\
   & 3.5           & Fit to Cr$_2$O$_3$ enthalpy of formation\cite{wang2006} & PBE (GGA) & VASP\\
   & 3.8           & Fit to HSE calculation\cite{hong2012}          & PBEsol (GGA) & VASP\\
   & \textbf{4.1}  & Cluster-CI\cite{nohara2009}                    & N/A & N/A\\
   & 4.5           & Fit to $E_{\text{g}}$ and $\mu$\cite{yang1999} & LSDA & Not available (LMTO-ASA method)\\
   & 6.96          & Constrained DFT\cite{nohara2009}               & LSDA & Not available (LMTO-ASA method)\\
Mn & 3.3           & cRPA ($d$-$dp$ model)\cite{jang2018}           & PBE (GGA) & openMX\\
   & 4.0           & Fit to enthalpy of formation\cite{lee2009}     & PW91 (GGA) & VASP\\
   & 5.0           & Fit to $E_{\text{g}}$ and $\mu$\cite{yang1999} & LSDA & Not available (LMTO-ASA method)\\
   & \textbf{6.4}  & Cluster-CI\cite{nohara2009}                    & N/A & N/A\\
   & 7.1           & Constrained DFT\cite{nohara2009}               & LSDA & Not available (LMTO-ASA method)\\
   & 8.0           & Fit to $E_{\text{g}}$\cite{yin2006}            & LSDA & Not available\\
Fe & 3.7           & cRPA ($d$-$dp$ model)\cite{jana2019}           & PBE (GGA) & WIEN2K (FP-LAPW method) \\
   & 4.0           & Fit to enthalpy of formation\cite{lee2009}     & PW91 (GGA) & VASP\\
   & 4.0           & Fit to $E_{\text{g}}$\cite{javaid2014}         & PBE (GGA) & QUANTUM ESPRESSO\\
   & \textbf{4.8}  & Cluster-CI\cite{nohara2009}                    & N/A & N/A\\
   & 5.1           & Fit to HSE calculation\cite{hong2012}          & PBEsol (GGA) & VASP\\
   & 5.4           & Fit to $E_{\text{g}}$ and $\mu$\cite{yang1999} & LSDA & Not available (LMTO-ASA method)\\
   & 7.43          & Constrained DFT\cite{nohara2009}               & LSDA & Not available (LMTO-ASA method)\\
Co & 3.3           & Fit to enthalpy of formation\cite{lee2009}     & PW91 (GGA) & VASP\\
   & 4.0           & Unrestricted Hartree-Fock\cite{ritzmann2014}   & N/A & GAMESS\\
   & \textbf{4.2}  & Cluster-CI\cite{nohara2009}                    & N/A & N/A\\
   & 5.85          & Fit to $E_{\text{g}}$ and $\mu$\cite{yang1999} & LDA & Not available (LMTO-ASA method)\\
   & 6.9           & Constrained DFT\cite{dutta2018}                & LDA & WIEN2K, ELK (FP-LAPW method) \\
   & 6.96          & Constrained DFT\cite{nohara2009}               & LDA & Not available (LMTO-ASA method)\\
   & \textbf{8.46} & Linear response\cite{hsu2009}                  & LDA & QUANTUM ESPRESSO\\
Ni & 1.1           & cRPA ($e_g$-$e_g$ model)\cite{hampel2019}      & PBE (GGA) & VASP\\
   & 5.64          & Linear response\cite{gou2011}                  & LSDA & QUANTUMESPRESSO, VASP\\
   & \textbf{5.7}  & Cluster-CI\cite{nohara2009}                    & N/A & N/A\\
   & 6.35          & Fit to $E_{\text{g}}$ and $\mu$\cite{yang1999} & LSDA & Not available (LMTO-ASA method)\\
   & 6.4           & Fit to enthalpy of formation\cite{lee2009}     & PW91 (GGA) & VASP\\
   & 7.57          & Constrained DFT\cite{nohara2009}               & LSDA & Not available (LMTO-ASA method)\\
\end{tabular}
\end{ruledtabular}
\end{table*}

\begin{figure*}[ht!]
   \centering
   \includegraphics{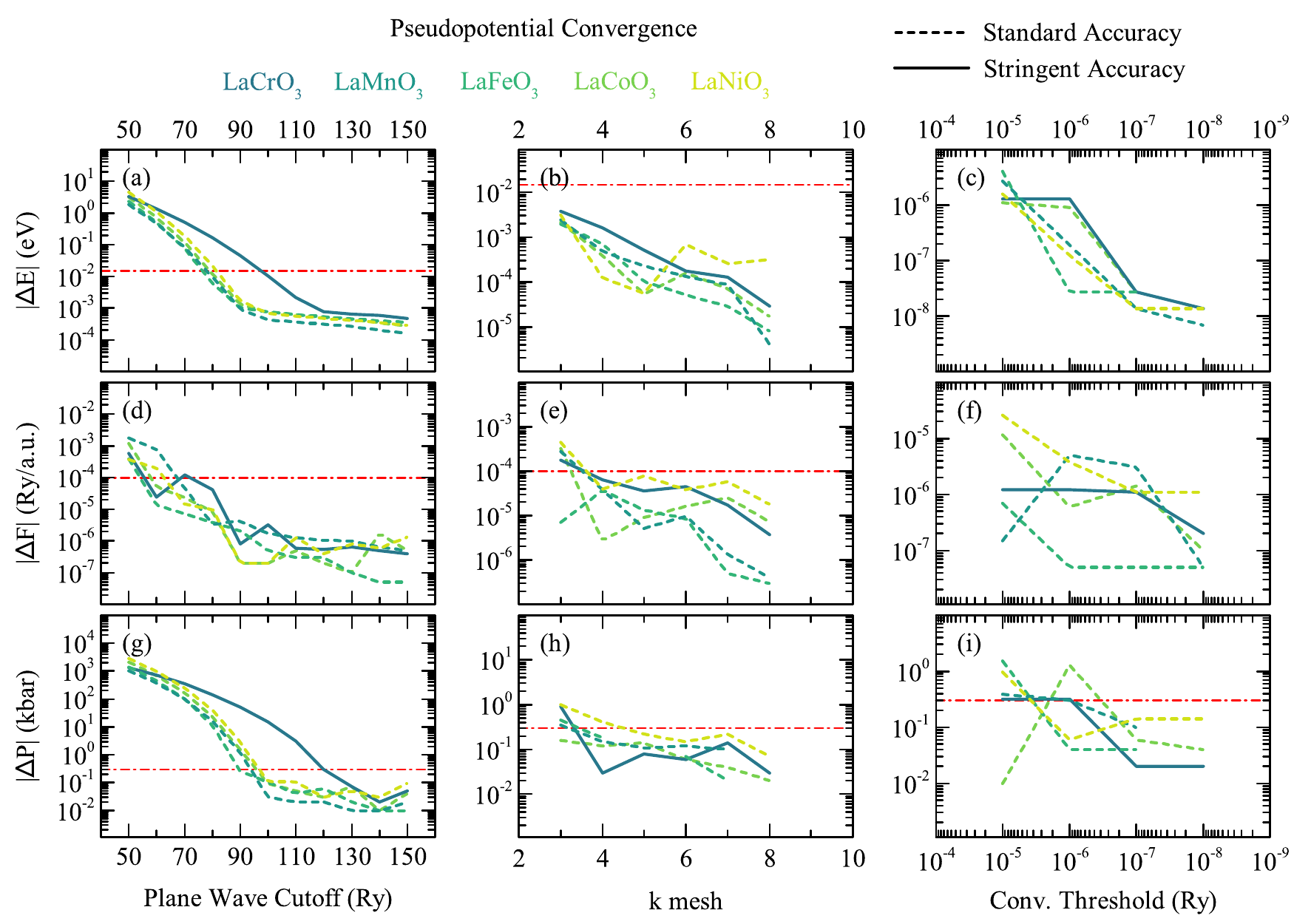}
   \caption{Pseudopotential convergence tests; total energy per atom vs. (a) plane wave cutoff, (b) k-point mesh, (c) scf convergence threshold; total force per atom vs. (d) plane wave cutoff, (e) k-point mesh, (f) scf convergence threshold; total cell pressure vs. (g) plane wave cutoff, (h) k-point mesh, (i) scf convergence threshold. The quantity "k mesh" $n$ refers to the dimensions of the k-point mesh, which is approximately $n\times n\times 0.75n$ for orthorhombic cells and $n\times n\times n$ for rhombohedral cells. All quantities are referenced to a well-converged calculation with two of the three paramters fixed at 250 Ry plane wave cutoff, $9\times 9\times 9$ k-point grid, and convergence threshold of $10^{-9}$ Ry.}\label{fig:convtest}
\end{figure*}

\begin{figure}[ht!]
   \centering
   \includegraphics{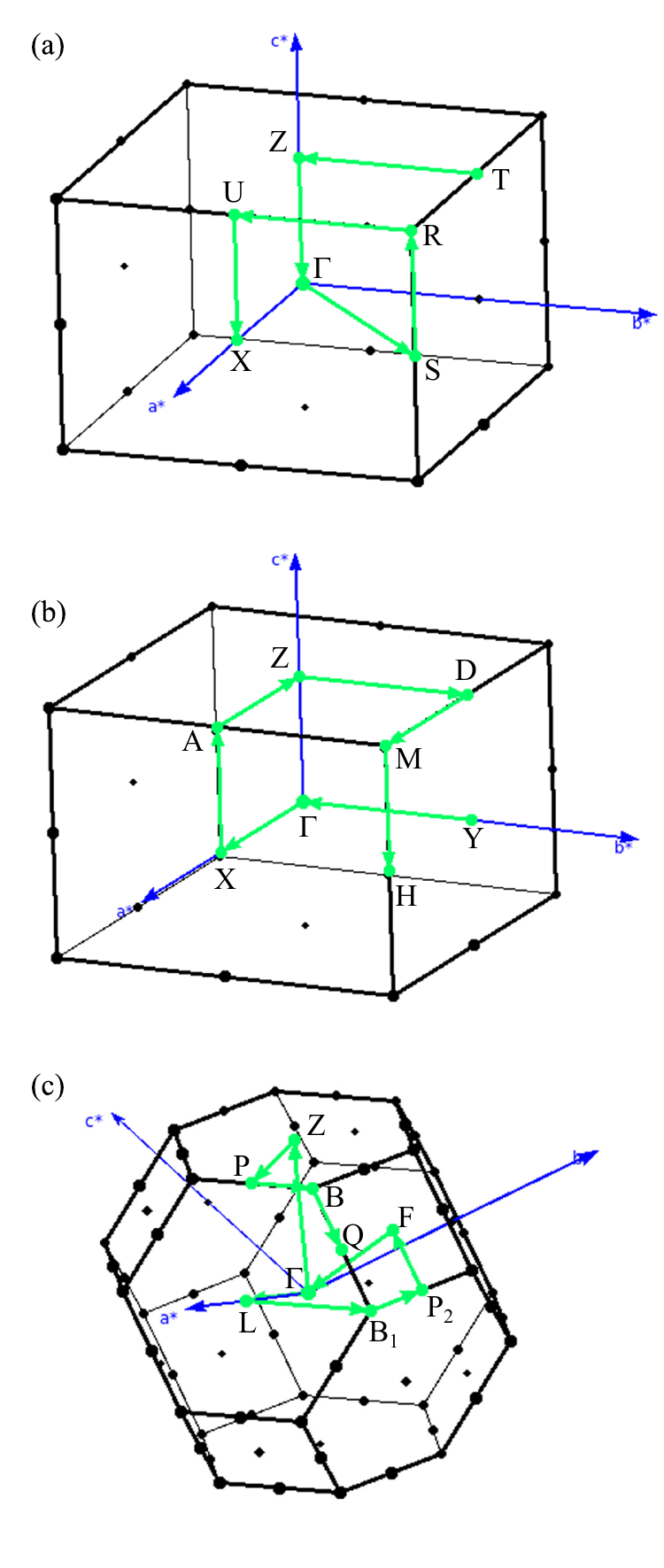}
   \caption{Paths through k-space for generating the band structures presented in the main text; (a) $Pbnm$; (b) $P2_1/b$; (c) $R\bar{3}c$. Images generated using XCrySDen\cite{kokalj2003}.}\label{fig:kpts}
\end{figure}

\begin{figure}[ht!]
   \centering
   \includegraphics{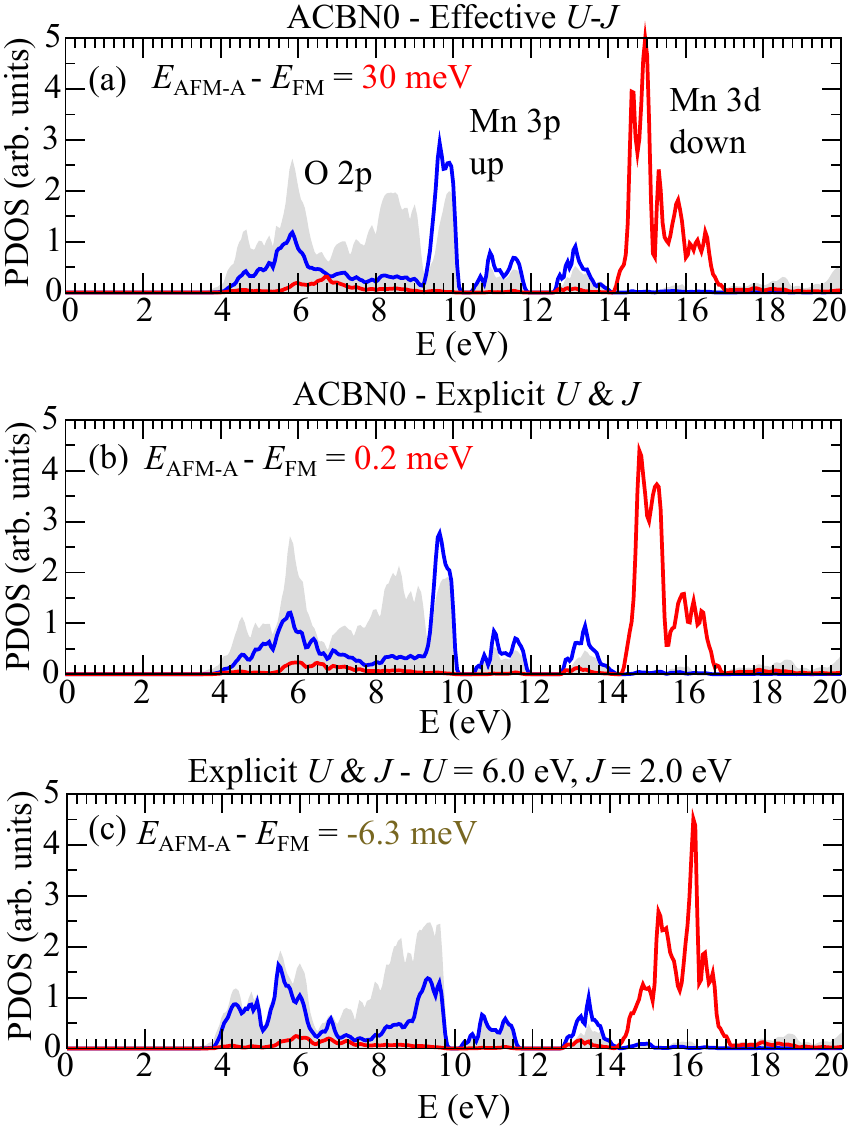}
   \caption{Projected density of states for LaMnO$_3$; (a) ACBN0 with $U_{\textrm{eff}}=U-J$ from a simplified DFT+$U$ scheme (see table \ref{table:UvalsMn}); (b) the same values of $U$ and $J$ but applied explicitly in a generalized rotationally-invariant DFT+$U$ implementation; (c) the same calculation as panel (b) but with $U$ and $J$ on Mn increased to 6.0 and 2.0 eV, respectively.}\label{fig:U-J}
\end{figure}

\begin{figure}[ht!]
   \centering
   \includegraphics{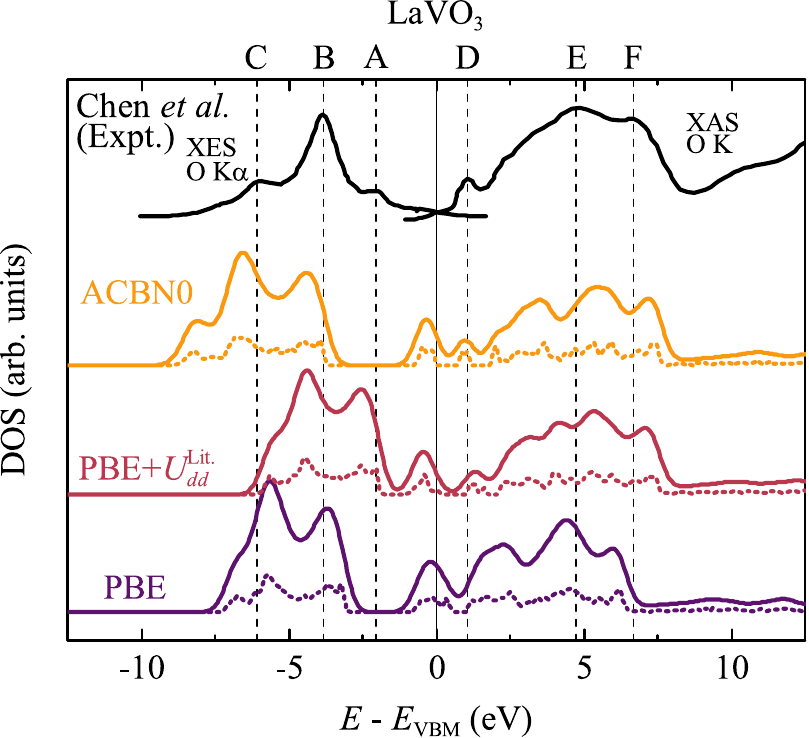}
   \caption{Total density of states for LaVO$_3$ (dotted lines), and with Gaussian broadening applied (solid lines, 0.6 eV), comparing PBE, PBE+$U_{dd}$, and ACBN0 (PBE+$U_{dd}$+$U_{pp}$). For comparison is experimental spectra (solid lines, top of figure) from Chen \emph{et al.}\cite{chen2015}}\label{fig:VSpectra}
\end{figure}

\begin{figure}[ht!]
   \centering
   \includegraphics{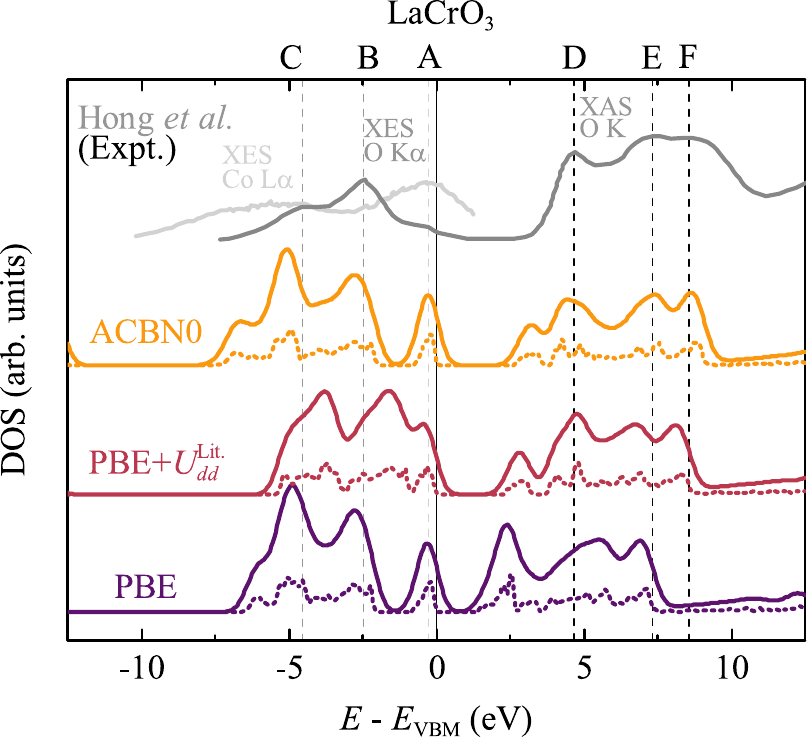}
   \caption{Total density of states for LaCrO$_3$ (dotted lines), and with Gaussian broadening applied (solid lines, 0.6 eV), comparing PBE, PBE+$U_{dd}$, and ACBN0 (PBE+$U_{dd}$+$U_{pp}$). For comparison is experimental spectra (solid lines, top of figure) from Hong \emph{et al.}\cite{hong2015}.}\label{fig:CrSpectra}
\end{figure}

\begin{figure}[ht!]
   \centering
   \includegraphics{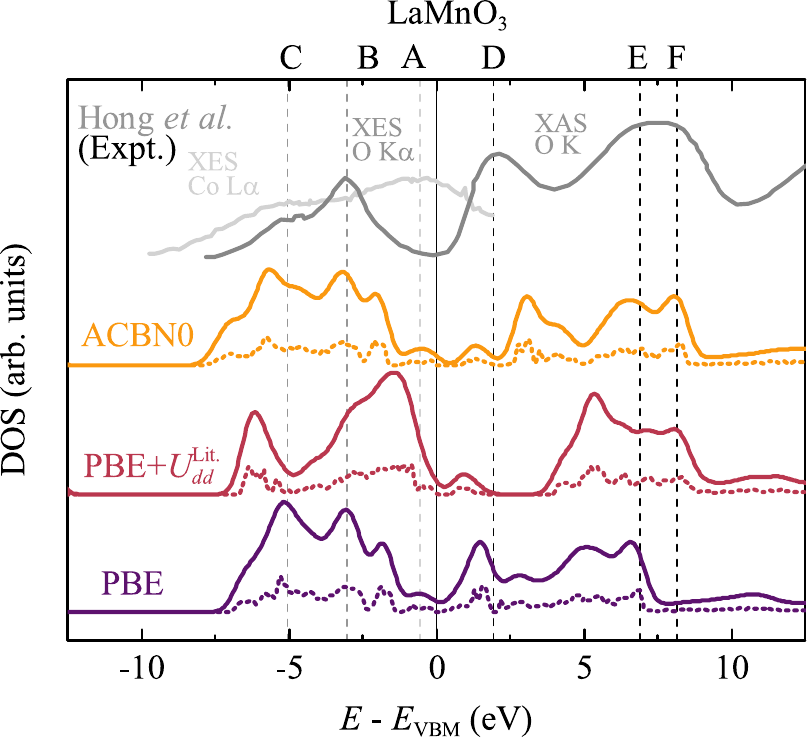}
   \caption{Total density of states for LaMnO$_3$ (dotted lines), and with Gaussian broadening applied (solid lines, 0.6 eV), comparing PBE, PBE+$U_{dd}$, and ACBN0 (PBE+$U_{dd}$+$U_{pp}$). For comparison is experimental spectra (solid lines, top of figure) from Hong \emph{et al.}\cite{hong2015}.}\label{fig:MnSpectra}
\end{figure}

\begin{figure}[ht!]
   \centering
   \includegraphics{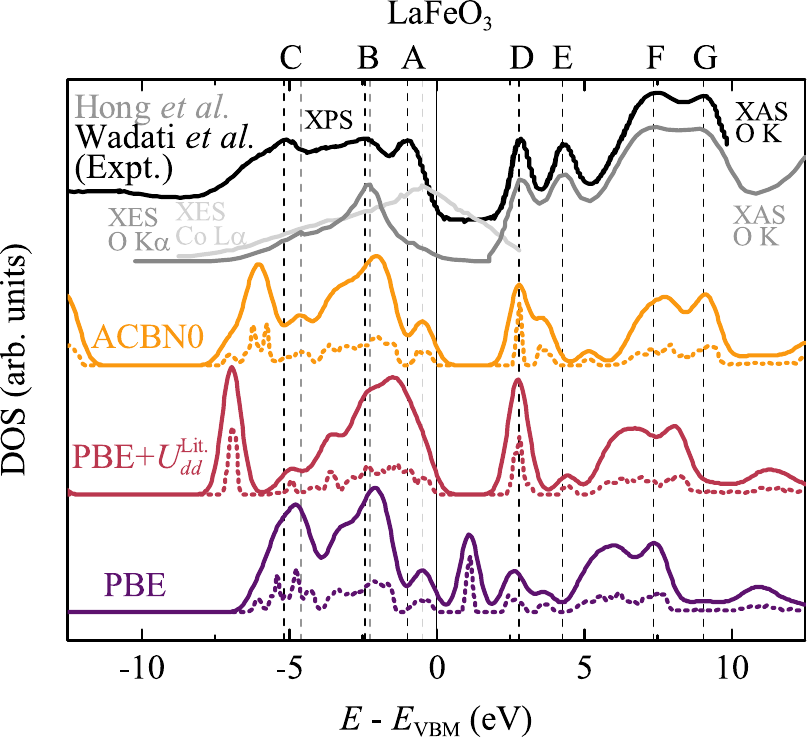}
   \caption{Total density of states for LaFeO$_3$ (dotted lines), and with Gaussian broadening applied (solid lines, 0.6 eV), comparing PBE, PBE+$U_{dd}$, and ACBN0 (PBE+$U_{dd}$+$U_{pp}$). For comparison is experimental spectra (solid lines, top of figure) from Hong \emph{et al.} and Wadati \emph{et al.}\cite{wadati2005,hong2015}}\label{fig:FeSpectra}
\end{figure}

\begin{figure}[ht!]
   \centering
   \includegraphics{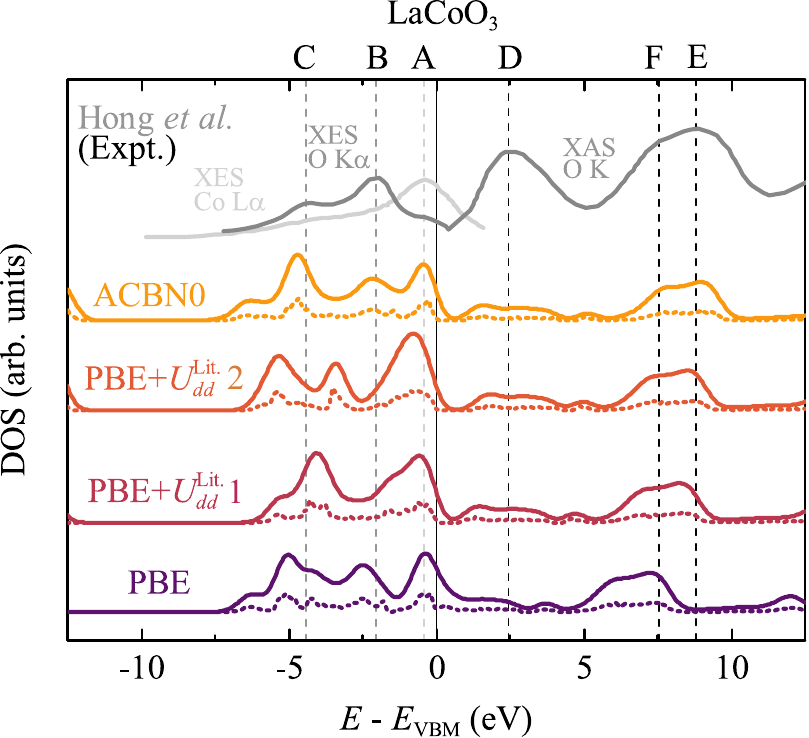}
   \caption{Total density of states for LaCoO$_3$ (dotted lines), and with Gaussian broadening applied (solid lines, 0.6 eV), comparing PBE, PBE+$U_{dd}$, and ACBN0 (PBE+$U_{dd}$+$U_{pp}$). For comparison is experimental spectra (solid lines, top of figure) from Hong \emph{et al.}\cite{hong2015}.}\label{fig:CoSpectra}
\end{figure}

\begin{figure}[ht!]
   \centering
   \includegraphics{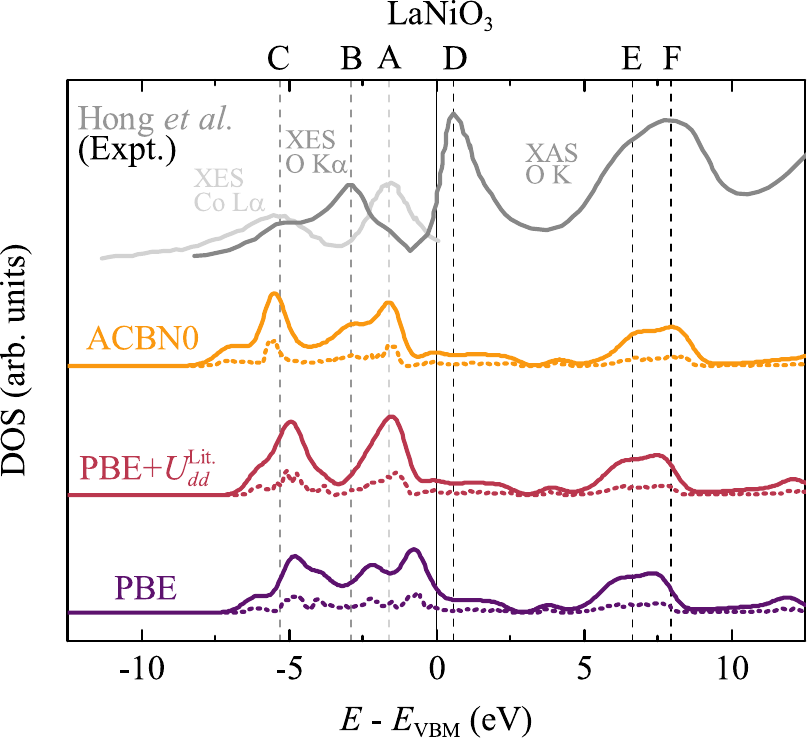}
   \caption{Total density of states for LaNiO$_3$ (dotted lines), and with Gaussian broadening applied (solid lines, 0.6 eV), comparing PBE, PBE+$U_{dd}$, and ACBN0 (PBE+$U_{dd}$+$U_{pp}$). For comparison is experimental spectra (solid lines, top of figure) from Hong \emph{et al.}\cite{hong2015}.}\label{fig:NiSpectra}
\end{figure}

\clearpage
\end{document}